%% file: Paper.tex
\documentclass[a4paper, 10pt]{article}
\pdfoutput=1

\usepackage{jheppub}
\usepackage{amsthm}
\usepackage[numbers,sort&compress]{natbib}
\usepackage{dsfont}
\usepackage{slashed}
\usepackage{color}
\usepackage{empheq}
\usepackage{adjustbox}
\usepackage{soul}
\usepackage[english]{babel}
\usepackage[utf8x]{inputenc}
\usepackage{graphicx,epsfig}
\usepackage{pstricks}
\usepackage{tikz}
\usepackage{bashful}
\usepackage{subfigure}
\usepackage{setspace}
\usepackage{booktabs}
\usepackage{float}
\usepackage{nextpage}
\usepackage{pdfpages}
\usepackage{hypcap}
\usepackage[small,bf]{caption}
\usepackage{longtable}
\usepackage{fancyhdr}
\usepackage[makeroom]{cancel}
\usepackage{microtype}
\usepackage{feynmp-auto-mod}
\usepackage{xparse}
\usepackage{etoolbox}
\usepackage[subfigure]{tocloft}
\usepackage{listings}
\usepackage{multirow}
\usepackage{makecell}

\newcommand{\p}{\partial}

\newcommand{\<}{\langle}
\renewcommand{\>}{\rangle}

\renewcommand{\O}{\mathcal{O}}
\newcommand{\N}{\mathcal{N}}

\newcommand{\tr}{\mathrm{Tr}}
\newcommand{\M}{\mathcal{M}}

\renewcommand{\L}{\mathcal{L}}

\newcommand{\E}{\mathcal{E}}
\newcommand{\D}{\mathcal{D}}
\newcommand{\nn}{\nonumber\\}
\newcommand{\GeV}{\,\mathrm{GeV}}
\newcommand{\Tr}{\mathrm{Tr}}

\newcommand{\msbar}{$\overline{\text{MS}}$}

\NewDocumentCommand{\Op}{ m m O{} o }{
	\O^{\ifblank{#3}{}{#3,}#2 }_{\IfNoValueTF{#4}{#1}{\substack{#1\\#4}}}
}

\definecolor{darkgreen}{rgb}{0,0.5,0}
\definecolor{darkblue}{rgb}{0,0,0.5}
\definecolor{darkred}{rgb}{0.5,0,0}
\definecolor{beige}{rgb}{0.7,0.4,0.3}

\makeatletter
  \def\my@tag@font{\normalsize}
  \def\maketag@@@#1{\hbox{\m@th\normalfont\my@tag@font#1}}
  \let\amsmath@eqref\eqref
  \renewcommand\eqref[1]{{\let\my@tag@font\relax\amsmath@eqref{#1}}}
\makeatother

\newenvironment{myfmf}[1]
{\begin{fmffile}{#1}
\fmfset{curly_len}{2mm}
\fmfcmd{%
  style_def wboson expr p =
  cdraw (wiggly p);
  shrink (1);
  cfill (arrow p);
  endshrink;
  enddef;}
\fmfcmd{%
  style_def momins expr p =
  drawarrow p;
  enddef;}
\fmfcmd{%
  style_def marrowc expr p =
  drawarrow subpath (1/4, 3/4) of p withpen pencircle scaled 0.4;
  enddef;}
\fmfcmd{%
  style_def marrowd expr p =
  drawarrow subpath (1/4, 3/4) of p shifted 10 down withpen pencircle scaled 0.4;
  enddef;}
\fmfcmd{%
  style_def marrowu expr p =
  drawarrow subpath (1/4, 3/4) of p shifted 10 up withpen pencircle scaled 0.4;
  enddef;}
\fmfcmd{%
  style_def marrowl expr p =
  drawarrow subpath (1/4, 3/4) of p shifted 10 left withpen pencircle scaled 0.4;
  enddef;}
\fmfcmd{%
  style_def mlarrowd expr p =
  drawarrow subpath (1/8, 7/8) of p shifted 10 down withpen pencircle scaled 0.4;
  enddef;}
\fmfcmd{%
  style_def mlarrowu expr p =
  drawarrow subpath (1/8, 7/8) of p shifted 10 up withpen pencircle scaled 0.4;
  enddef;}
\fmfcmd{%
  style_def mlarrowl expr p =
  drawarrow subpath (0, 1.75) of p shifted 10 left withpen pencircle scaled 0.4;
  enddef;}
\fmfcmd{%
  style_def mlarrowr expr p =
  drawarrow subpath (0, 1.75) of p shifted 10 right withpen pencircle scaled 0.4;
  enddef;}
\fmfcmd{%
  style_def marrowr expr p =
  drawarrow subpath (1/4, 3/4) of p shifted 10 right withpen pencircle scaled 0.4;
  enddef;}
\fmfcmd{%
  style_def marrowdr expr p =
  drawarrow subpath (1/4, 3/4) of p shifted 10 right shifted 3 down withpen pencircle scaled 0.4;
  enddef;}
\fmfcmd{%
  style_def darrowd expr p =
  drawarrow subpath (1/4, 3/4) of p shifted 10 down dashed evenly withpen pencircle scaled 0.4;
  enddef;}
\fmfcmd{%
  style_def darrowu expr p =
  drawarrow subpath (1/4, 3/4) of p shifted 10 up dashed evenly withpen pencircle scaled 0.4;
  enddef;}
\fmfcmd{%
  style_def darrowl expr p =
  drawarrow subpath (1/4, 3/4) of p shifted 10 left dashed evenly withpen pencircle scaled 0.4;
  enddef;}
\fmfcmd{%
  style_def darrowr expr p =
  drawarrow subpath (1/4, 3/4) of p shifted 10 right dashed evenly withpen pencircle scaled 0.4;
  enddef;}
\fmfcmd{%
  style_def darrowdl expr p =
  drawarrow subpath (1/4, 3/4) of p shifted 7 left shifted 3 down dashed evenly withpen pencircle scaled 0.4;
  enddef;}
\fmfcmd{%
  style_def darrowdr expr p =
  drawarrow subpath (1/4, 3/4) of p shifted 10 right shifted 3 down dashed evenly withpen pencircle scaled 0.4;
  enddef;}
\fmfcmd{%
  style_def dlarrowdr expr p =
  drawarrow subpath (1/8, 7/8) of p shifted 10 right shifted 3 down dashed evenly withpen pencircle scaled 0.4;
  enddef;}

}
{
\end{fmffile}
}

\makeatletter
\renewcommand\paragraph{\@startsection{paragraph}{4}{\z@}%
  {-3.25ex\@plus -1ex \@minus -.2ex}%
  {1.5ex \@plus .2ex}%
  {\normalfont\normalsize\bfseries}}
\makeatother

\cftsetindents{section}{0em}{2em}
\cftsetindents{subsection}{2em}{3em}
\cftsetindents{subsubsection}{5em}{4em}

\allowdisplaybreaks[1]

\title{\boldmath One-loop matching of $CP$-odd four-quark operators to the gradient-flow scheme}

\author[a,b,c]{Jona B\"uhler,}

\author[b,c]{Peter Stoffer}

\emailAdd{jobuehle@student.ethz.ch}
\emailAdd{stoffer@physik.uzh.ch}

\affiliation[a]{Institute for Particle Physics and Astrophysics, ETH Z\"urich, 8093 Z\"urich, Switzerland}
\affiliation[b]{Physik-Institut, Universit\"at Z\"urich, Winterthurerstrasse 190, 8057 Z\"urich, Switzerland}
\affiliation[c]{Paul Scherrer Institut, 5232 Villigen PSI, Switzerland}

\abstract{
The translation of experimental limits on the neutron electric dipole moment into constraints on heavy $CP$-violating physics beyond the Standard Model requires knowledge about non-perturbative matrix elements of effective operators, which ideally should be computed in lattice QCD. However, this necessitates a matching calculation as an interface to the effective field theory framework, which is based on dimensional regularization and renormalization by minimal subtraction.
We calculate the one-loop matching between the gradient-flow and minimal-subtraction schemes for the $CP$-violating four-quark operators contributing to the neutron electric dipole moment. The gradient flow is a modern regularization-independent scheme amenable to lattice computations that promises, e.g., better control over power divergences than traditional momentum-subtraction schemes. Our results extend previous work on dimension-five operators and provide a necessary ingredient for future lattice-QCD computations of the contribution of four-quark operators to the neutron electric dipole moment.
}

\numberwithin{equation}{section}

\begin{document}

\preprint{
\mbox{}\hfill{} PSI-PR-23-8 \\
\mbox{}\hfill{} ZU-TH 16/23
}
	\maketitle


	\begin{myfmf}{diags/diags}

	\input{sections/Introduction}

	\input{sections/Operators}

	\input{sections/GradientFlow}

	\input{sections/Results}

	\input{sections/Conclusions}

	
	\section*{Acknowledgements}
	\addcontentsline{toc}{section}{\numberline{}Acknowledgements}

	We thank K.~Kirch, Ò.~Lara~Crosas, and L.~Naterop for useful discussions and E.~Mereghetti, C.~J.~Monahan, M.~D.~Rizik, and A.~Shindler for collaboration on closely related projects.
	Furthermore, we thank Ò.~Lara~Crosas and E.~Mereghetti for comments on the manuscript.
	Financial support by the Swiss National Science Foundation (Project No.~PCEFP2\_194272) is gratefully acknowledged.

	
	\appendix
	
	\input{sections/Conventions}

	\input{sections/FeynmanRules}

	\end{myfmf}

	\addcontentsline{toc}{section}{\numberline{}References}
	\bibliographystyle{utphysmod}
	\bibliography{Literature}
	
\end{document}

%% file: sections/Introduction.tex

\section{Introduction}
\label{sec:Introduction}

The search for sources of $CP$ violation beyond the Standard Model (SM) is primarily motivated by the baryon asymmetry of the universe and has resulted in a very active program adressing both leptonic~\cite{Muong-2:2008ebm,ACME:2018yjb,Adelmann:2021udj,Roussy:2022cmp} and hadronic electric dipole moments (EDMs), see Refs.~\cite{Chupp:2017rkp,Alarcon:2022ero} for reviews. The current best experimental bound on the neutron EDM (nEDM)~\cite{Abel:2020pzs}
\begin{align}
	\label{eq:nEDMExperimentalBound}
	|d_n| < 1.8 \times 10^{-26} \, e\, \mathrm{cm} \text{ (90\% C.L.)}
\end{align}
was obtained by the nEDM collaboration at PSI. Further improvements in the experimental sensitivities are expected in the near future~\cite{Ito:2017ywc,nEDM:2019qgk,Wurm:2019yfj,Martin:2020lbx,n2EDM:2021yah}.
The SM contribution to the nEDM due to the $CP$-violating phase in the CKM matrix is several orders of magnitude smaller than the current experimental bound~\cite{Shabalin:1979gh,Khriplovich:1981ca,Czarnecki:1997bu,Seng:2014lea}. Therefore, the measurement of the nEDM is an interesting probe of $CP$ violation beyond the SM.

Given the absence of clear direct signals of physics beyond the SM at the LHC, new particles need to be either very weakly coupled or very heavy, with masses well above the electroweak scale. In the second case, their indirect low-energy effects can be described in terms of effective field theories (EFTs), in particular the Standard Model EFT (SMEFT) above the electroweak scale~\cite{Buchmuller:1985jz,Grzadkowski:2010es} and the low-energy EFT (LEFT) below the weak scale~\cite{Jenkins:2017jig}. The matching of models for new physics to the SMEFT is currently being automated~\cite{Carmona:2021xtq,Fuentes-Martin:2022jrf} and the complete renormalization-group equations (RGEs) and the matching of the SMEFT and LEFT have been derived at one loop~\cite{Jenkins:2013zja,Jenkins:2013wua,Alonso:2013hga,Jenkins:2017dyc,Dekens:2019ept}, enabling a treatment that avoids large logarithms in each step of the calculation.

The EFT approach is ideal to obtain constraints on new physics at a high scale from low-energy precision observables, such as the nEDM. However, the calculation of the observable itself within the LEFT involves matrix elements of effective operators between neutron states, schematically\footnote{The sign $\sim$ indicates that $d_n$ is obtained from the projection of the electric dipole form factor at zero momentum transfer.}
\begin{align}
	\label{eq:nEDMSchematics}
	\nn[-0.25cm]
	d_n \; \sim \quad
		\begin{gathered}
			\begin{fmfgraph*}(60,25)
 			\fmfleft{in,i1}
 			\fmfright{out,o1}
 			\fmftop{phot}
			\fmflabel{$N$}{in}
			\fmflabel{$N$}{out}
			\fmflabel{$\gamma$}{phot}
 			\fmf{quark}{in,v1}
 			\fmf{quark}{v1,out}
 			\fmf{photon}{phot,v1}
 			\fmfblob{.25w}{v1}
 			\fmffreeze
			\fmfi{plain}{vpath (__in,__v1) shifted (thick*(-1,2))}
			\fmfi{plain}{vpath (__in,__v1) shifted (thick*(1,-2))}
			\fmffreeze
			\fmfi{plain}{vpath (__v1,__out) shifted (thick*(1,2))}
			\fmfi{plain}{vpath (__v1,__out) shifted (thick*(-1,-2))}
			\end{fmfgraph*}
		\end{gathered}
		\quad = \sum_i L_i(\mu) \< N | \O_i^\mathrm{MS} | N \gamma \> \, ,
\end{align}
where $L_i(\mu)$ denotes renormalized LEFT Wilson coefficients, and the sum runs over all renormalized operators $\O_i^\mathrm{MS}$ that are not excluded by symmetry principles: due to the nature of the strong interaction at low energies, the hadronic matrix elements are non-perturbative and all $CP$-odd and flavor-neutral operators contribute. Due to the running and mixing effects of the RGEs, it is not trivial to turn the experimental bound~\eqref{eq:nEDMExperimentalBound} into a strong constraint on heavy new physics: in order to avoid possible cancellations, the uncertainties on the non-perturbative operator matrix elements in Eq.~\eqref{eq:nEDMSchematics} should be reduced to a level of $10-25\%$~\cite{Chien:2015xha,Alarcon:2022ero}. In addition, disentangling different sources of $CP$ violation will require not only the nEDM as a single observable, but rather a whole portfolio of experiments~\cite{Chupp:2017rkp,Alarcon:2022ero}.

Ideally, the non-perturbative matrix elements of effective operators relevant for the nEDM should be obtained from lattice-QCD computations, which provide a first-principles approach with controlled systematic uncertainties, see Ref.~\cite{Shindler:2021bcx} for a recent review. Since the EFT framework is based on dimensional regularization and renormalization by (modified) minimal subtraction (MS or \msbar{}), the EFT description of the observable~\eqref{eq:nEDMSchematics} involves matrix elements of MS operators. Therefore, the use of lattice-QCD input necessitates a matching calculation to a different renormalization scheme. Traditional schemes amenable to lattice-QCD computations are momentum-subtraction schemes (MOM), and the matching between MOM and \msbar{} has been worked out for the dimension-five operators contributing to the nEDM~\cite{Bhattacharya:2015rsa}, as well as the dimension-six $CP$-odd three-gluon operator~\cite{Cirigliano:2020msr}. A more modern scheme is provided by the gradient flow~\cite{Luscher:2010iy,Luscher:2013cpa}, which promises a better control of power divergences~\cite{Rizik:2020naq,Kim:2021qae}. Recently, the one-loop matching between the gradient-flow and MS schemes was worked out at dimension five~\cite{Mereghetti:2021nkt}. In the present paper, we extend this work to dimension-six four-quark operators. The gradient-flow matching was previously performed for the left-chiral current-current operators arising in the Fermi theory of weak interactions~\cite{Suzuki:2020zue,Harlander:2022tgk}, using the naive dimensional regularization (NDR) and dimensional-reduction schemes. Here, we consider instead the $CP$-odd and flavor-neutral four-quark operators that contribute to the nEDM.

The article is structured as follows: in Sect.~\ref{sec:Operators}, we define our operator basis up to dimension six. In addition to the physical operators, we define the unphysical nuisance operators that appear in the off-shell matching calculation, as well as evanescent operators in two different schemes. In Sect.~\ref{sec:GradientFlow} we briefly review the gradient-flow formalism and comment on a modification of the flow equations in the presence of an electromagnetic field. In Sect.~\ref{sec:MatchingCoefficients}, we discuss the short-flow-time expansion and present our results for the one-loop matching coefficients of the flowed four-quark operators to MS operators up to dimension six. We conclude in Sect.~\ref{sec:Conclusions} and provide some details on conventions and Feynman rules in the appendices.

%% file: sections/Operators.tex

\section{Operator basis}
\label{sec:Operators}

The indirect effect of heavy physics beyond the SM on observables below the electroweak scale is described by the LEFT with the Lagrangian
\begin{align}
	\L_\mathrm{LEFT} = \L_\mathrm{QCD+QED} +\sum_{d \geq 5} \sum_{i=1}^{n_d} L_i^{(d)} \O_i^{(d)} \, ,
\end{align}
where the sum runs over all operators with mass dimension $d \geq 5$ that respect the $SU(3)_c \times U(1)_\mathrm{em}$ gauge symmetry. Although dipole operators in the LEFT appear at dimension five, their contribution beyond the SM arises in the SMEFT only at dimension six due to $SU(2)_L$ gauge invariance. Hence, in a scenario of heavy new physics where the LEFT is matched to the SMEFT at the weak scale, a consistent treatment should involve both dimension-five and -six effects in the LEFT.

We start by identifying the LEFT operators at the hadronic scale of a few GeV that can contribute to the neutron EDM up to dimension six. Having an accuracy goal for the hadronic matrix elements of about $10-25\%$ in mind~\cite{Alarcon:2022ero}, we are not interested in higher-order QED corrections and we treat the photon as a static external field. At the hadronic scale, we consider either three or four active quark flavors, collected in a vector $q=(u,d,s)$ or $q=(u,d,s,c)$. Due to non-perturbative effects, any operator with the right symmetry properties has to be taken into account: the operators need to be $CP$-odd and flavor neutral. 

We start by listing the physical operators in Sect.~\ref{sec:PhysicalOperators}. In a second step, we extend the operator basis to unphysical operators that appear in the loop calculation. In Sect.~\ref{sec:NuisanceOperators}, we list the nuisance operators, which vanish by the equations of motion (EOM). Evanescent operators related to dimensional regularization are discussed in Sect.~\ref{sec:EvanescentOperators}: their definition is part of the renormalization scheme.

\subsection{Physical operators}
\label{sec:PhysicalOperators}

From the LEFT operator basis up to dimension six as classified in Ref.~\cite{Jenkins:2017jig}, we select the $CP$-odd and flavor-neutral operators that can contribute to the nEDM at leading order in QED. We work in $D=4-2\varepsilon$ Euclidean space-time dimensions and largely follow the conventions of Ref.~\cite{Mereghetti:2021nkt}, see App.~\ref{sec:Conventions}. The Lagrangian of Euclidean QCD is given by
\begin{align}
	\label{eq:QCDLagrangian}
	\L_\mathrm{QCD} = \frac{1}{4g_0^2} G_{\mu\nu}^a G_{\mu\nu}^a + \bar q ( \slashed D + \M_q ) q + \L_\mathrm{GF} + \L_\mathrm{gh} \, ,
\end{align}
where the covariant derivative
\begin{align}
	D_\mu = \p_\mu + G_\mu + A_\mu \, , \quad G_\mu = t^a G_\mu^a
\end{align}
includes the external electromagnetic field $A_\mu$. The field-strength tensors are defined by
\begin{align}
	G_{\mu\nu} = \p_\mu G_\nu - \p_\nu G_\mu + [ G_\mu, G_\nu ] \, , \quad F_{\mu\nu} = \p_\mu A_\nu - \p_\nu A_\mu \, .
\end{align}
The quark-mass matrix is $\M_q = \mathrm{diag}(m_u, m_d, m_s)$ or $\M_q = \mathrm{diag}(m_u, m_d, m_s, m_c)$ for $N_f = 3$ or $N_f = 4$ active quark flavors, respectively. The gauge-fixing and ghost terms are
\begin{align}
	\label{eq:GaugeFixing}
	\L_\mathrm{GF} &= \frac{1}{2g_0^2 \xi} \left(\p_\mu G_\mu^a \right)^2 \, , \quad
	\L_\mathrm{gh} = \left(\p_\mu \bar{c}^a \right) D_\mu^{ac}c^c \, ,
\end{align}
with the covariant derivative in the adjoint representation
\begin{align}
	D_\mu^{ac} = \p_\mu \delta^{ac} + f^{abc} G_\mu^b \, .
\end{align}
We include a trivial leptonic Lagrangian
\begin{align}
	\L_\mathrm{lept} = \bar l (\slashed D + \M_l ) l
\end{align}
with $D_\mu = \p_\mu + A_\mu$, but we disregard dynamical photons, i.e., we approximate the full LEFT by the QCD and leptonic Lagrangian, supplemented by a tower of effective operators:
\begin{align}
	\label{eq:ApproximateLEFTLagrangian}
	\L = \L_\mathrm{QCD} + \L_\mathrm{lept} + \sum_i L_i \O_i \, .
\end{align}
Without dynamical photons, the leptonic interactions are restricted to the effective operators as well as the coupling to the external electromagnetic field.
In the context of the nEDM we focus on $CP$-odd and flavor-neutral operators up to dimension six that involve quarks or gluons. At dimension three, there is the pseudoscalar density, or $CP$-odd mass term
\begin{align}
	\label{eq:PseudoscalarDensity}
	\O^P_{p} = \bar q_p \gamma_5 q_p \, ,
\end{align}
where $p$ is a fixed quark-flavor index.
It is always possible to switch to a basis where the mass matrix is real and diagonal, which removes the $CP$-odd mass term. In general, this field redefinition involves an anomalous axial rotation, which affects the only $CP$-odd operator at dimension four, the theta term, or topological charge density:
\begin{align}
	\label{eq:ThetaTerm}
	\O_\theta = \tr[ G_{\mu\nu} \widetilde G_{\mu\nu} ] \, ,
\end{align}
where the dual field-strength tensor is $\widetilde G_{\mu\nu} = \frac{1}{2} \epsilon_{\mu\nu\alpha\beta} G_{\alpha\beta}$.

At dimension five, there are the electric and chromo-electric dipole operators,
\begin{align}
	\O^E_{p} &= \bar q_p \tilde\sigma_{\mu\nu}  F_{\mu\nu} q_p \, , \nn
	\O^{CE}_{p} &= \bar q_p \tilde\sigma_{\mu\nu} t^a  G^a_{\mu\nu} q_p \, ,
\end{align}
where we again only keep the flavor-diagonal contributions relevant for the nEDM and we do not sum over the flavor index $p$. We are using the definition~\cite{Mereghetti:2021nkt}
\begin{align}
	\label{eq:SigmaTilde}
	\tilde\sigma_{\mu\nu}^\mathrm{HV} = - \frac{1}{2} \epsilon_{\mu\nu\alpha\beta} \sigma_{\alpha\beta} \, , \quad \tilde\sigma_{\mu\nu}^\mathrm{NDR} = \sigma_{\mu\nu} \gamma_5 \, ,
\end{align}
depending on the scheme for $\gamma_5$, with $\sigma_{\mu\nu} = \frac{i}{2} [ \gamma_\mu, \gamma_\nu ]$.

At dimension six we encounter the $CP$-odd three-gluon operator
\begin{align}
	\O_{\widetilde G} = \frac{1}{g_0^2} \tr[ G_{\mu\nu} G_{\nu\lambda} \widetilde G_{\lambda\mu} ] \, ,
\end{align}
which will be left for future studies~\cite{CP3GOMatching}, as well as a larger number of four-fermion operators, which are the focus of this article. Leptonic four-fermion operators only contribute at higher orders in $\alpha_\mathrm{QED}$, hence we restrict ourselves to semileptonic and non-leptonic operators (baryon- and lepton-number-violating operators do not contribute to the nEDM at dimension six). Schematically, they have the form
\begin{align}
	\label{eq:FourFermiGeneric}
	\O^{\Gamma_1\Gamma_2}_{\substack{2\ell2q \\ prst}} &= (\bar l_p \Gamma_1 l_r)(\bar q_s \Gamma_2 q_t) \, , \nn
	\O^{\Gamma_1\Gamma_2,1}_{\substack{4q \\ prst}} &= (\bar q_p \Gamma_1 q_r)(\bar q_s \Gamma_2 q_t) \, , \nn
	\O^{\Gamma_1\Gamma_2,8}_{\substack{4q \\ prst}} &= (\bar q_p \Gamma_1 t^a q_r)(\bar q_s \Gamma_2 t^a q_t) \, ,
\end{align}
where $\Gamma_{1,2}$ denote Dirac structures.
The condition that the operators be flavor neutral means that in the case of semileptonic operators we are interested in flavor indices $(p=r) \wedge (s=t)$, while in the case of four-quark operators we need to consider either $(p=r) \wedge (s=t)$ or $(p=t) \wedge (s=r)$.

\subsubsection{Semileptonic operators}

To leading order in $\alpha_\mathrm{QED}$, the contribution of semileptonic operators to the nEDM can be written as
\begin{align}
	\label{eq:nEDMSemileptonic}
	\< N | (\bar l_p \Gamma_1 l_p)(\bar q_r \Gamma_2 q_r) | N \gamma \> &= \< 0 | \bar l_p \Gamma_1 l_p | \gamma \> \< N | \bar q_r \Gamma_2 q_r | N \> \nn
		&\quad + \< 0 | \bar l_p \Gamma_1 l_p | 0 \> \< N | \bar q_r \Gamma_2 q_r | N\gamma \> + \O(\alpha_\mathrm{QED}) \, ,
\end{align}
where $p$ and $r$ are fixed flavor indices. The nEDM is determined by the terms linear in the momentum of the external photon. Due to Lorentz and gauge invariance, to leading order in QED the nEDM receives a contribution from the first term only in the case of semileptonic tensor operators with $\Gamma_1 = \sigma_{\mu\nu}$ and $\Gamma_2 = \tilde\sigma_{\mu\nu}$ or vice versa. The non-perturbative hadronic matrix element is the matrix element of a tensor quark-bilinear operator. The matching of quark bilinears to the gradient-flow scheme is known~\cite{Hieda:2016lly} and we have reproduced these results.

The second contraction in Eq.~\eqref{eq:nEDMSemileptonic} vanishes unless $\Gamma_1 = 1$ and hence $\Gamma_2 = \gamma_5$. In this case, the semileptonic operator contributes as a renormalization of a $CP$-odd quark-mass term. Again, this contribution can be shifted into the theta term by an anomalous axial field redefinition and the calculation of the hadronic matrix element of the pseudoscalar density is equivalent to the one of the nEDM induced by the topological charge~\cite{Aoki:1990ix}.

\subsubsection{Four-quark operators}

The non-redundant set of $CP$-odd four-quark operators that contribute to the nEDM has been identified previously in Ref.~\cite{Khatsimovsky:1987fr}. We write them as
\begin{align}
	\label{eq:FourQuarkParityBasis}
	\O^{S1}_p &= (\bar{q}_p \gamma_5 q_p)(\bar{q}_p q_p) \, , \nn*
	\O^{S8}_p &= (\bar{q}_p \gamma_5 t^a q_p)(\bar{q}_p t^a q_p) \, , \nn
	\O^{S1}_{pr} &= (\bar{q}_p \gamma_5 q_p)(\bar{q}_r q_r) \, , \qquad p \neq r \, , \nn
	\O^{S8}_{pr} &= (\bar{q}_p \gamma_5 t^a q_p)(\bar{q}_r t^a q_r) \, , \qquad p \neq r \, , \nn
	\O^{T1}_{pr} &= \frac{1}{2} \Bigl[ (\bar{q}_p \tilde{\sigma}_{\mu \nu} q_p)(\bar{q}_r \sigma_{\mu \nu} q_r) + (\bar{q}_r \tilde{\sigma}_{\mu \nu} q_r)(\bar{q}_p \sigma_{\mu \nu} q_p) \Bigr] \, , \qquad p \neq r \, , \nn
	\O^{T8}_{pr} &= \frac{1}{2} \Bigl[ (\bar{q}_p \tilde{\sigma}_{\mu \nu} t^a q_p)(\bar{q}_r \sigma_{\mu \nu} t^a q_r) + (\bar{q}_r \tilde{\sigma}_{\mu \nu} t^a q_r)(\bar{q}_p \sigma_{\mu \nu} t^a q_p) \Bigr] \, , \qquad p \neq r \, ,
\end{align}
where no implicit sums over flavor indices are performed. The tensor operators are symmetric in the two flavor indices, in contrast to the scalar operators.
There are in total $n_q(3n_q-1)$ $CP$-odd and flavor-neutral four-quark operators, i.e., 24 operators for $n_q=3$ quark flavors or 44 for $n_q = 4$ quark flavors.
Tensor operators with identical quark flavors in the two bilinears have been reduced to scalar operators using the Fierz relations~\cite{Khatsimovsky:1987fr}. In the parity basis, the relevant Fierz relations in $D=4$ space-time dimensions read
\begin{align}
	\label{eq:FierzDiracParity}
	\left( \gamma_5 \right) \otimes \left[ \mathds{1} \right] &=  \frac{1}{4} \left[  \left( \mathds{1} \right] \otimes \left[ \gamma_5 \right) + \left( \gamma_5 \right] \otimes \left[ \mathds{1} \right) - \left( \gamma_\mu \gamma_5 \right] \otimes \left[ \gamma_\mu \right) + \left( \gamma_\mu \right] \otimes \left[ \gamma_\mu \gamma_5 \right) + \frac{1}{2} \left( \tilde{\sigma}_{\mu \nu} \right] \otimes [ \sigma_{\mu \nu} ) \right] \, , \nn
	\left( \gamma_\mu \gamma_5 \right) \otimes \left[ \gamma_\mu \right] &= \left( \mathds{1} \right] \otimes \left[ \gamma_5 \right) - \left( \gamma_5 \right] \otimes \left[ \mathds{1} \right) - \frac{1}{2}  \left( \gamma_\mu \gamma_5 \right] \otimes \left[ \gamma_\mu \right) - \frac{1}{2}  \left( \gamma_\mu \right] \otimes \left[ \gamma_\mu \gamma_5 \right) \, , \nn
	\left(\tilde\sigma_{\mu\nu} \right) \otimes \left[ \sigma_{\mu\nu} \right] &= 3 \left(\mathds{1} \right] \otimes \left[ \gamma_5 \right) + 3 \left( \gamma_5 \right] \otimes \left[ \mathds{1} \right) - \frac{1}{2} \left( \tilde\sigma_{\mu\nu} \right] \otimes \left[ \sigma_{\mu\nu} \right) \, ,
\end{align}
where round and square brackets denote Dirac indices and the sign from the anticommutation of fermion fields is not included. Together with the $SU(N_c)$ Fierz relation
\begin{align}
	\label{eq:SUNFierz}
	t^a_{\alpha \beta} t^a_{\gamma \delta} &= -\left( \frac{1}{2} \delta_{\alpha \delta} \delta_{\gamma \beta} - \frac{1}{2 N_c} \delta_{\alpha \beta} \delta_{\gamma \delta} \right)
\end{align}
for the anti-Hermitian $SU(N_c)$ generators $t^a$, the four-quark operators in Eq.~\eqref{eq:FourQuarkParityBasis} can be related to LEFT operators in the chiral basis as follows:
\begin{align}
	\label{eq:ParityChiralBasisChange}
	\O^{S1}_p &= (\bar q_{Lp} q_{Rp})(\bar q_{Lp} q_{Rp}) - (\bar q_{Rp} q_{Lp})(\bar q_{Rp} q_{Lp}) \, , \nn
	\O^{S8}_p &= (\bar q_{Lp} t^a q_{Rp})(\bar q_{Lp} t^a q_{Rp}) - (\bar q_{Rp} t^a q_{Lp})(\bar q_{Rp} t^a q_{Lp}) \, , \nn
	\O^{S1}_{pr} &= (\bar q_{Lp} q_{Rp})(\bar q_{Lr} q_{Rr}) - (\bar q_{Rp} q_{Lp})(\bar q_{Rr} q_{Lr}) \nn
		&\quad - \frac{1}{2N_c} (\bar q_{Lp} \gamma_\mu q_{Lr})(\bar q_{Rr} \gamma_\mu q_{Rp}) + \frac{1}{2N_c} (\bar q_{Lr} \gamma_\mu q_{Lp})(\bar q_{Rp} \gamma_\mu q_{Rr}) \nn
		&\quad + (\bar q_{Lp} \gamma_\mu t^a q_{Lr})(\bar q_{Rr} \gamma_\mu t^a q_{Rp}) - (\bar q_{Lr} \gamma_\mu t^a q_{Lp})(\bar q_{Rp} \gamma_\mu t^a q_{Rr}) \, , \nn
	\O^{S8}_{pr} &= (\bar q_{Lp} t^a q_{Rp})(\bar q_{Lr} t^a q_{Rr}) - (\bar q_{Rp} t^a q_{Lp})(\bar q_{Rr} t^a q_{Lr}) \nn
		&\quad + \frac{N_c^2-1}{4N_c^2} (\bar q_{Lp} \gamma_\mu q_{Lr})(\bar q_{Rr} \gamma_\mu q_{Rp}) - \frac{N_c^2-1}{4N_c^2} (\bar q_{Lr} \gamma_\mu q_{Lp})(\bar q_{Rp} \gamma_\mu q_{Rr}) \nn
		&\quad + \frac{1}{2N_c} (\bar q_{Lp} \gamma_\mu t^a q_{Lr})(\bar q_{Rr} \gamma_\mu t^a q_{Rp}) - \frac{1}{2N_c} (\bar q_{Lr} \gamma_\mu t^a q_{Lp})(\bar q_{Rp} \gamma_\mu t^a q_{Rr}) \, , \nn
	\O^{T1}_{pr} &= -4\Big[ (\bar q_{Lp} q_{Rp})(\bar q_{Lr} q_{Rr}) - (\bar q_{Rp} q_{Lp})(\bar q_{Rr} q_{Lr}) \nn*
		&\qquad + \frac{2}{N_c} (\bar q_{Lp} q_{Rr})(\bar q_{Lr} q_{Rp}) - \frac{2}{N_c} (\bar q_{Rp} q_{Lr})(\bar q_{Rr} q_{Lp}) \nn*
		&\qquad - 4 (\bar q_{Lp} t^a q_{Rr})(\bar q_{Lr} t^a q_{Rp}) + 4 (\bar q_{Rp} t^a q_{Lr})(\bar q_{Rr} t^a q_{Lp}) \Big] \, , \nn
	\O^{T8}_{pr} &= -4\Big[ (\bar q_{Lp} t^a q_{Rp})(\bar q_{Lr} t^a q_{Rr}) - (\bar q_{Rp} t^a q_{Lp})(\bar q_{Rr} t^a q_{Lr}) \nn
		&\qquad - \frac{N_c^2-1}{N_c^2} (\bar q_{Lp} q_{Rr})(\bar q_{Lr} q_{Rp}) + \frac{N_c^2-1}{N_c^2} (\bar q_{Rp} q_{Lr})(\bar q_{Rr} q_{Lp}) \nn
		&\qquad - \frac{2}{N_c} (\bar q_{Lp} t^a q_{Rr})(\bar q_{Lr} t^a q_{Rp}) + \frac{2}{N_c} (\bar q_{Rp} t^a q_{Lr})(\bar q_{Rr} t^a q_{Lp}) \Big] \, ,
\end{align}
where the chiral fields are defined by
\begin{align}
	\psi_{L,R} = P_{L,R} \psi \, , \quad \bar\psi_{L,R} = \bar\psi P_{R,L} \, , \quad P_L = \frac{1-\gamma_5}{2} \, , \quad P_R = \frac{1+\gamma_5}{2}  \, .
\end{align}
However, the relations~\eqref{eq:ParityChiralBasisChange} only hold in $D=4$ space-time dimensions and in order to do a proper matching to the LEFT operator basis of Ref.~\cite{Jenkins:2017jig}, one needs to take into account evanescent operators in the Fierz relations~\cite{Buras:1989xd,Dugan:1990df,Herrlich:1994kh,Aebischer:2022aze}. In the present context of the nEDM we find it convenient to work in the parity basis given in Eq.~\eqref{eq:FourQuarkParityBasis}.

\subsection{Nuisance operators}
\label{sec:NuisanceOperators}

Since we will perform an off-shell matching calculation, we will encounter not only physical operators, but in addition unphysical ``nuisance operators.'' They can be split into two classes~\cite{Dixon:1974ss,Kluberg-Stern:1975ebk,Joglekar:1975nu,Deans:1978wn,Collins:1984xc}: on the one hand, we need gauge-invariant operators that vanish by the classical EOM, known as class-IIa operators. On the other hand, additional gauge-variant nuisance operators appear as the solutions of the Ward--Slavnov--Taylor identities, known as class-IIb operators. They can be constructed as BRST variations of operators with ghost number $-1$.

\subsubsection{Equation-of-motion operators}

From the Euclidean QCD Lagrangian~\eqref{eq:QCDLagrangian}, we obtain the quark EOM
\begin{align}
	\label{eq:QuarkEOM}
	(\slashed D + \M_q ) q = 0
\end{align}
as well as the gluon EOM
\begin{align}
	\label{eq:GluonEOM}
	D_\mu^{ac} G_{\mu\nu}^c = g_0^2 \sum_p \bar{q}_p \gamma_\nu t^a q_p \, ,
\end{align}
where the index $p$ runs over the quark flavors.
Operators that are proportional to the classical EOM can be removed from the operator basis by field redefinitions that effectively shift their effects to higher orders in the EFT power counting. Furthermore, $S$-matrix elements of EOM operators vanish~\cite{Deans:1978wn}. However, in an off-shell matching of Green's functions unphysical operators appear as counterterm contributions.

The gluon EOM is not relevant if we restrict our attention to operators contributing to the nEDM: a $CP$-odd operator involving $D_\mu G_{\mu\nu}$ needs to have the form
\begin{align}
	\label{eq:GluonEOMOperator}
	\O^\mathrm{odd}_{pr} = (\bar q_p \gamma_\nu \gamma_5 t^a q_r)(D_\mu G_{\mu\nu})^a - (\bar q_r \gamma_\nu \gamma_5 t^a q_p)(D_\mu G_{\mu\nu})^a, 
\end{align}
and hence it vanishes for flavor-conserving operators with $p=r$.

For the quark EOM operators, it is convenient to define the fields
\begin{align}
	q_E &= (\slashed D + \M_q)q \, , \quad
	\bar q_E = \bar q (-\overleftarrow{\slashed{D}}+\M_q) \, ,
\end{align}
where
\begin{align}
	\overleftarrow{D}_\mu = \overleftarrow{\p}_\mu - G_\mu - A_\mu \, .
\end{align}
We work with the following set of EOM operators:
\begin{align}
	\label{eq:ClassIIaNuisance}
	\N_{p}^{1} &= \bar q_{Ep} \gamma_5 \slashed D q_p - \bar q_p \overleftarrow{\slashed D}  \gamma_5 q_{Ep} \, , \nn
	\N_{p}^{2} &= \left( \bar q_{Ep} \tilde\sigma_{\mu\nu} t^a q_p + \bar q_{p} \tilde\sigma_{\mu\nu} t^a q_{Ep} \right) G_{\mu\nu}^a \, , \nn
	\N_{p}^{3} &= \left( \bar q_{Ep} \tilde\sigma_{\mu\nu} q_p + \bar q_{p} \tilde\sigma_{\mu\nu} q_{Ep} \right) F_{\mu\nu} \, ,
\end{align}
where no implicit sum over flavor indices is understood.

\subsubsection{Gauge-variant operators}
\label{sec:ClassIIbNuisanceOperators}

The gauge-fixing term in Eq.~\eqref{eq:GaugeFixing}, which is introduced in perturbation theory, breaks $SU(3)_c$ gauge symmetry down to BRST invariance. This implies that we encounter not only gauge-invariant counterterms, but also nuisance operators that are only BRST exact but not gauge invariant~\mbox{\cite{Dixon:1974ss,Kluberg-Stern:1975ebk,Joglekar:1975nu,Deans:1978wn,Collins:1984xc}}. These nuisance operators could be avoided by working with the background-field method. A~background-field formulation of the gradient flow has been established in Ref.~\cite{Suzuki:2015bqa}, but in the present paper we will work with conventional $R_\xi$ gauge.

Since we do not consider dynamical photons, we do not need to fix the QED gauge. We will work in a formalism with manifest $U(1)_\mathrm{em}$ invariance, which however also requires flow equations that respect electromagnetic gauge invariance, as will be discussed in Sect.~\ref{sec:GradientFlow}. This allows us to include only class-IIb nuisance operators that are $U(1)_\mathrm{em}$ gauge invariant. They have been classified in Ref.~\cite{Cirigliano:2020msr} in the context of the three-gluon operator. Imposing momentum conservation, we drop total-derivative operators and relabel the remaining relevant nuisance operators from Ref.~\cite{Cirigliano:2020msr} as follows:\footnote{We omit two additional operators from Ref.~\cite{Cirigliano:2020msr}, which are not generated at one loop in our calculation.}
\begin{align}
	\label{eq:ClassIIbNuisance}
	\N_p^{4} &= (\bar q_{Ep} \gamma_\mu \gamma_5 t^a q_p - \bar q_p \gamma_\mu \gamma_5 t^a q_{Ep})G_\mu^a \, , \nn
	\N_p^{5} &= (\bar q_{Ep} \gamma_5 t^a q_p - \bar q_p \gamma_5 t^a q_{Ep})\p_\mu G_\mu^a \, , \nn
	\N_p^{6} &= (\bar q_{Ep} \gamma_5 t^a D_\mu q_p - \bar q_p \overleftarrow D_\mu \gamma_5 t^a q_{Ep}) G_\mu^a \, , \nn
	\N_p^{7} &= (\bar q_{Ep} \tilde\sigma_{\mu\nu} t^a q_p + \bar q_p \tilde\sigma_{\mu\nu} t^a q_{Ep})\p_\mu G_\nu^a \, , \nn
	\N^{8} &= G^a_{\mu\nu} \biggl[ \p_\lambda \Bigl( D_\rho G_{\rho\sigma}^a - g_0^2 \sum_p \bar q_p t^a \gamma_\sigma q_p + g_0^2 f^{abc}(\p_\sigma \bar c^b)c^c \Bigr) \biggr] \epsilon_{\mu \nu \lambda \sigma} \, .
\end{align}
Because the four-quark operators are not singlets under chiral transformations, we find it convenient not to include explicit factors of the quark masses in the operators~\eqref{eq:ClassIIaNuisance} and~\eqref{eq:ClassIIbNuisance}, apart from the quark EOM fields $q_E$ and $\bar q_E$.

\subsection[Evanescent operators and $\gamma_5$ schemes]{\boldmath Evanescent operators and $\gamma_5$ schemes}
\label{sec:EvanescentOperators}

In addition to the physical and nuisance operators discussed in Sects.~\ref{sec:PhysicalOperators} and~\ref{sec:NuisanceOperators}, in dimensional regularization in $D=4-2\varepsilon$ space-time dimensions we encounter evanescent operators that disappear in four dimensions. The convention for evanescent operators affects the finite matching coefficients for the physical operators at one loop and hence is part of the scheme definition. In order to avoid a mixing of the unphysical evanescent sector into the physical sector, evanescent operators should be renormalized to have vanishing matrix elements in $D=4$ dimensions~\cite{Buras:1989xd,Dugan:1990df,Herrlich:1994kh}. However, in our case the evanescent operators are generated only at one loop and their renormalization does not affect the matching coefficients of the physical operators at one-loop accuracy.

The definition of the evanescent operators depends on the scheme chosen for $\gamma_5$ in dimensional regularization. The simplest scheme is NDR, which treats $\gamma_5$ to be anticommuting with all Dirac matrices in $D$ space-time dimensions:
\begin{align}
	\{ \gamma_\mu , \gamma_5 \} = 0 \, .
\end{align}
In connection with the NDR scheme, we replace $\tilde\sigma_{\mu\nu}$ in the effective operators by $\tilde\sigma_{\mu\nu}^\mathrm{NDR}$ defined in Eq.~\eqref{eq:SigmaTilde}. As is well known, the NDR scheme leads to algebraic inconsistencies with $\gamma_5$-odd fermion traces~\cite{Jegerlehner:2000dz}. We will use it only for the self-matching of the four-quark operators, since the calculation of the matching of flowed tensor operators $\O^{T1}_{pr}$ and $\O^{T8}_{pr}$ to lower-dimension operators leads to ill-defined Dirac traces.

We will use the original scheme by 't~Hooft and Veltman (HV)~\cite{tHooft:1972tcz,Breitenlohner:1977hr} for the complete matching of the four-quark operators. In the HV scheme, we replace $\tilde\sigma_{\mu\nu}$ in the effective operators by $\tilde\sigma_{\mu\nu}^\mathrm{HV}$ defined in Eq.~\eqref{eq:SigmaTilde}, where the indices of the Levi-Civita symbol only run over four dimensions. We split the metric tensor into two parts that project onto the $4$- and $-2\varepsilon$-dimensional subspaces:
\begin{align}
	\label{eq:HVSubspaces}
	\delta_{\mu\nu} = \bar\delta_{\mu\nu} + \hat\delta_{\mu\nu} \, ,
\end{align}
with $\bar\delta_{\mu\nu} \hat\delta_{\nu\lambda} = 0$, $\bar\delta_{\mu\mu} = 4$, $\hat\delta_{\mu\mu} = -2\varepsilon$, and we define $\bar\gamma_\mu = \bar\delta_{\mu\nu}\gamma_\nu$, $\hat\gamma_\mu = \hat\delta_{\mu\nu}\gamma_\nu$.

In the HV scheme, global chiral symmetry is violated by the regulator. This implies that our renormalized MS operators in general do not fulfill the chiral Ward identities. The Ward identities could be restored by finite renormalizations, as discussed, e.g., in Ref.~\cite{Bhattacharya:2015rsa}. However, in vector-like theories such as QCD, gauge symmetry remains unaffected by the regulator and the scheme remains consistent without symmetry-restoring finite counterterms. Here, we do not perform symmetry-restoring finite renormalizations: this corresponds to a scheme choice and all symmetry-breaking contributions will cancel in relations between observables. A comprehensive treatment of these finite renormalizations will be provided in Ref.~\cite{LEFTHV}.

\subsubsection{Evanescent operators in the NDR scheme}

In the calculation of loop integrals, we encounter higher tensor products of gamma matrices, which in $D=4$ space-time dimensions can be reduced to the tensor structures of the physical four-quark operators. In $D=4-2\varepsilon$ dimensions, such relations do not hold but require the introduction of evanescent structures. For NDR, we choose the same scheme for evanescent four-fermion structures as Ref.~\cite{Dekens:2019ept}, but we translate it from the chiral to the parity basis. Explicitly, we define
\begin{align}
	\gamma_\mu \gamma_\nu \gamma_5 \otimes \gamma_\mu \gamma_\nu &= \bigl(4+ (2\tilde{a}_\mathrm{ev}-1)\varepsilon \bigr)  \gamma_5\otimes\mathds{1} - (2\tilde{a}_\mathrm{ev}+1)\varepsilon \, \mathds{1} \otimes \gamma_5 \nn
		&\quad  - \frac{1}{2} \Bigl[ \sigma_{\mu \nu} \otimes \tilde\sigma_{\mu \nu} + \tilde\sigma_{\mu \nu} \otimes \sigma_{\mu \nu} \Bigr] - E^{(2)} \, , \nn
	\gamma_\mu \gamma_\nu \gamma_\lambda \gamma_\sigma\gamma_5 \otimes \gamma_\mu \gamma_\nu \gamma_\lambda \gamma_\sigma &= 8(5+ 2\tilde{d}_{\mathrm{ev}}\varepsilon) \gamma_5\otimes\mathds{1} + 8(3 - 14\tilde{f}_{\mathrm{ev}}\varepsilon)  \mathds{1} \otimes \gamma_5 \nn
		&\quad - 4(2-\tilde{e}_\mathrm{ev}\varepsilon) \Bigl[ \sigma_{\mu \nu} \otimes \tilde\sigma_{\mu \nu} + \tilde\sigma_{\mu \nu} \otimes \sigma_{\mu \nu} \Bigr] + E^{(4)} \, ,
\end{align}
where our evanescent structures are related to the ones of Ref.~\cite{Dekens:2019ept} by
\begin{align}
	\label{eq:EvanescentStructuresNDR}
	E^{(2)} &= E^{(2)}_{LR}-E^{(2)}_{RL} \, , \nn
	E^{(4)} &= E^{(4)}_{RR} - E^{(4)}_{LL} - E^{(4)}_{LR} + E^{(4)}_{RL} \, .
\end{align}
The evanescent structures depend on the parameters $\tilde a_\mathrm{ev}, \ldots, \tilde f_\mathrm{ev}$, which can be expressed in terms of the evanescent-scheme parameters of Ref.~\cite{Dekens:2019ept} as
\begin{align}
	\tilde a_\mathrm{ev} = a_\mathrm{ev} \, , \quad
	\tilde d_\mathrm{ev} = 4 f_\mathrm{ev} - 3 d_\mathrm{ev} \, , \quad
	\tilde e_\mathrm{ev} = e_\mathrm{ev} \, , \quad 
	\tilde f_\mathrm{ev} = \frac{4}{7} f_\mathrm{ev} + \frac{3}{7} d_\mathrm{ev} \, .
\end{align}
Note that the tensor operators are defined in Eq.~\eqref{eq:FourQuarkParityBasis} in a manifestly symmetric way, since in NDR the following structure is evanescent but non-vanishing in $D$ dimensions:
\begin{align}
	\sigma_{\mu \nu} \otimes \tilde\sigma_{\mu \nu} - \tilde\sigma_{\mu \nu} \otimes \sigma_{\mu \nu} &= 2 (1 + 2 \tilde a_\mathrm{ev})\varepsilon \left[ \gamma_5 \otimes \mathds{1} - \mathds{1}\otimes\gamma_5 \right] - 2 E^{(2)} \, ,
\end{align}
where $\tilde\sigma_{\mu\nu}$ is interpreted as $\tilde\sigma_{\mu\nu}^\mathrm{NDR}$.

Finally, the evanescent four-quark operators are defined by inserting the corresponding evanescent Dirac structures~\eqref{eq:EvanescentStructuresNDR} into two antiquark-quark bilinears. The operators with two different flavors are given by
\begin{align}
	\E^{(2),1}_{pr} &= (1+2 \tilde a_\mathrm{ev})\varepsilon \Big[ (\bar q_p \gamma_5 q_p)(\bar q_r q_r) - (\bar q_r \gamma_5 q_r)(\bar q_p q_p) \Big] \nn
		&\quad + \frac{1}{2} \Big[ (\bar q_p \tilde\sigma_{\mu\nu} q_p)(\bar q_r \sigma_{\mu\nu} q_r) - (\bar q_r \tilde\sigma_{\mu\nu} q_r)(\bar q_p \sigma_{\mu\nu} q_p) \Big] \, , \nn
	\E^{(2),8}_{pr} &= (1+2 \tilde a_\mathrm{ev})\varepsilon \Big[ (\bar q_p \gamma_5 t^a q_p)(\bar q_r t^a q_r) - (\bar q_r \gamma_5 t^a q_r)(\bar q_p t^a q_p) \Big] \nn
		&\quad + \frac{1}{2} \Big[ (\bar q_p \tilde\sigma_{\mu\nu} t^a q_p)(\bar q_r \sigma_{\mu\nu} t^a q_r) - (\bar q_r \tilde\sigma_{\mu\nu} t^a q_r)(\bar q_p \sigma_{\mu\nu} t^a q_p) \Big] \, , \nn
	\E^{(4),1}_{pr} &= (\bar q_p \gamma_\mu \gamma_\nu \gamma_\lambda \gamma_\sigma \gamma_5 q_p)(\bar q_r \gamma_\mu \gamma_\nu \gamma_\lambda \gamma_\sigma q_r) \nn*
		&\quad - 8(5+2 \tilde d_\mathrm{ev}\varepsilon) (\bar q_p \gamma_5 q_p)(\bar q_r q_r) - 8(3-14 \tilde f_\mathrm{ev}\varepsilon) (\bar q_r \gamma_5 q_r)(\bar q_p q_p)  \nn*
		&\quad + 4(2-\tilde e_\mathrm{ev}\varepsilon) \Big[ (\bar q_p \tilde\sigma_{\mu\nu} q_p)(\bar q_r \sigma_{\mu\nu} q_r) + (\bar q_r \tilde\sigma_{\mu\nu} q_r)(\bar q_p \sigma_{\mu\nu} q_p) \Big] \, , \nn
	\E^{(4),8}_{pr} &= (\bar q_p \gamma_\mu \gamma_\nu \gamma_\lambda \gamma_\sigma \gamma_5 t^a q_p)(\bar q_r \gamma_\mu \gamma_\nu \gamma_\lambda \gamma_\sigma t^a q_r) \nn*
		&\quad - 8(5+2 \tilde d_\mathrm{ev}\varepsilon) (\bar q_p \gamma_5 t^a q_p)(\bar q_r t^a q_r) - 8(3-14 \tilde f_\mathrm{ev}\varepsilon) (\bar q_r \gamma_5 t^a q_r)(\bar q_p t^a q_p)  \nn*
		&\quad + 4(2-\tilde e_\mathrm{ev}\varepsilon) \Big[ (\bar q_p \tilde\sigma_{\mu\nu} t^a q_p)(\bar q_r \sigma_{\mu\nu} t^a q_r) + (\bar q_r \tilde\sigma_{\mu\nu} t^a q_r)(\bar q_p \sigma_{\mu\nu} t^a q_p) \Big] \, ,
\end{align}
where $p\neq r$. In the case of a single quark flavor, the evanescent structures~\eqref{eq:EvanescentStructuresNDR} do not show up at one loop.
As mentioned above, tensor operators with identical flavor indices are not included in the physical operator basis, since they can be reduced to scalar operators using the Fierz relations~\eqref{eq:FierzDiracParity} and~\eqref{eq:SUNFierz}. Since the Fierz relations only hold in $D=4$ space-time dimensions, they give rise to the following evanescent operators with single quark flavors:
\begin{align}
	\label{eq:FierzEvanescents}
	\E^{F,1}_{p} &= (\bar q_p \tilde\sigma_{\mu\nu} q_p)(\bar q_p \sigma_{\mu\nu} q_p) - r_{T1,S1} \O_p^{S1} - r_{T1,S8} \O_p^{S8} \, , \nn
	\E^{F,8}_{p} &= (\bar q_p \tilde\sigma_{\mu\nu} t^a q_p)(\bar q_p \sigma_{\mu\nu} t^a q_p) - r_{T8,S1} \O_p^{S1} - r_{T8,S8} \O_p^{S8} \, ,
\end{align}
where
\begin{align}
	\label{eq:FierzCoefficients}
	r_{T1,S1} &= - \frac{4(N_c+2)}{N_c} \, , \quad r_{T1,S8} = 16 \, , \nn
	r_{T8,S1} &= \frac{4(N_c^2-1)}{N_c^2} \, , \quad r_{T8,S8} = - \frac{4(N_c-2)}{N_c} \, .
\end{align}
Fierz-evanescent operators would also show up in the one-loop basis change from Eq.~\eqref{eq:FourQuarkParityBasis} to the LEFT operators in chiral basis.

\subsubsection{Evanescent operators in the HV scheme}

In the HV scheme, we use Eq.~\eqref{eq:HVSubspaces} to split $D$-dimensional Dirac matrices according to
\begin{align}
	\label{eq:GammaSplit}
	\gamma_\mu = \bar\gamma_\mu + \hat\gamma_\mu \, ,
\end{align}
where $\bar\gamma_\mu$ lives in the four-dimensional sub-space with $\mu=1,\ldots,4$, and $\hat\gamma_\mu$ belongs to the evanescent sub-space. Using only the Dirac algebra, we can bring all encountered four-fermion structures to the form
\begin{align}
	\left( \bar\psi_1  \hat\gamma_{\nu_1} \ldots \hat\gamma_{\nu_m} \bar\gamma_{\mu_1} \ldots \bar\gamma_{\mu_n} \gamma^5 \psi_2 \right) \left( \bar\psi_3  \hat\gamma_{\nu_1} \ldots \hat\gamma_{\nu_m} \bar\gamma_{\mu_1} \ldots \bar\gamma_{\mu_n} \psi_4 \right) \, .
\end{align}
For $n>2$, we reduce the number of Dirac matrices in the four-dimensional sub-space using the Chisholm identity
\begin{align}
	\bar\gamma_\mu \bar\gamma_\nu \bar\gamma_\lambda = \bar\gamma_\mu \bar\delta_{\nu\lambda} + \bar\gamma_\lambda \bar\delta_{\mu\nu} - \bar\gamma_\nu \bar\delta_{\mu\lambda} - \bar\gamma_\sigma \gamma_5 \epsilon_{\mu\nu\lambda\sigma} \, .
\end{align}
All structures containing $\hat\gamma_\mu$ matrices are evanescent. In the end, the evanescent operators required at one loop are
\begin{align}
	\E_{p}^{S1} &= (\bar q_p \hat\gamma_\mu \hat\gamma_\nu \gamma_5 q_p)(\bar q_p \hat\gamma_\mu \hat\gamma_\nu q_p) \, , \nn
	\E_{p}^{S8} &= (\bar q_p \hat\gamma_\mu \hat\gamma_\nu \gamma_5 t^a q_p)(\bar q_p \hat\gamma_\mu \hat\gamma_\nu t^a q_p) \, , \nn
	\E_{pr}^{S1} &= (\bar q_p \hat\gamma_\mu \hat\gamma_\nu \gamma_5 q_p)(\bar q_r \hat\gamma_\mu \hat\gamma_\nu q_r) \, , \qquad p \neq r \, , \nn
	\E_{pr}^{S8} &= (\bar q_p \hat\gamma_\mu \hat\gamma_\nu \gamma_5 t^a q_p)(\bar q_r \hat\gamma_\mu \hat\gamma_\nu t^a q_r) \, , \qquad p \neq r \, , \nn
	\E_{pr}^{T1} &= (\bar q_p \hat\gamma_\mu \hat\gamma_\nu \tilde\sigma_{\lambda\sigma} q_p)(\bar q_r \hat\gamma_\mu \hat\gamma_\nu \sigma_{\lambda\sigma}  q_r) \, , \qquad p \neq r \, , \nn
	\E_{pr}^{T8} &= (\bar q_p \hat\gamma_\mu \hat\gamma_\nu \tilde\sigma_{\lambda\sigma} t^a q_p)(\bar q_r \hat\gamma_\mu \hat\gamma_\nu \sigma_{\lambda\sigma} t^a q_r) \, , \qquad p \neq r \, ,
\end{align}
where no implicit sum over flavor indices is understood. Our scheme slightly differs from the evanescent scheme proposed in Ref.~\cite{Dugan:1990df}, where the Lorentz indices of the evanescent Dirac matrices are antisymmetrized. The differences are proportional to physical operators multiplied by $\varepsilon$ and hence result in finite shifts in the one-loop matching.

In the HV scheme, we use the same form of Fierz-evanescent operators as in the NDR scheme, where we replace in Eq.~\eqref{eq:FierzEvanescents} $\tilde\sigma_{\mu\nu}$ by $\tilde\sigma_{\mu\nu}^\mathrm{HV}$.

For the matching to the lower-dimension operators, which we only perform in the HV scheme, we need additional evanescent quark-bilinear operators. A generic classification can be found in Ref.~\cite{Cirigliano:2020msr}. We work with operators of the form
\begin{align}
	\E_p^{(1,F)} &= \bar q_p \gamma_5 \hat\gamma_\mu O_\mu^F q_p \, , \quad
	\E_p^{(2a,F)} = \bar q_p \gamma_5 \hat\gamma_\mu \bar\gamma_\nu O_{\mu\nu}^F q_p \, , \quad
	\E_p^{(2b,F)} = \bar q_p \gamma_5 \hat\gamma_\mu \hat\gamma_\nu O_{\mu\nu}^F q_p \, , \nn
	\E_p^{(3a,F)} &= \bar q_p \gamma_5 \hat\gamma_\mu \bar\gamma_\nu \bar\gamma_\lambda O_{\mu\nu\lambda}^F q_p \, , \quad
	\E_p^{(3b,F)} = \bar q_p \gamma_5 \hat\gamma_\mu \hat\gamma_\nu \bar\gamma_\lambda O_{\mu\nu\lambda}^F q_p \, , \nn
	\E_p^{(3c,F)} &= \bar q_p \gamma_5 \hat\gamma_\mu \hat\gamma_\nu \hat\gamma_\lambda O_{\mu\nu\lambda}^F q_p \, ,
\end{align}
where $O^F_{\mu_1\ldots\mu_n}$ are built out of $\p_\mu$, $G_\mu^a$, $A_\mu$ and color structures. In this scheme, we can easily project to the non-evanescent sector by keeping the momenta and polarization vectors of external photons and gluons in $D=4$ space-time dimensions.

%% file: sections/GradientFlow.tex

\section{Gradient flow}
\label{sec:GradientFlow}

In this section, we briefly review the gradient-flow formalism as established in Refs.~\cite{Luscher:2010iy,Luscher:2013cpa}, largely following the conventions of Ref.~\cite{Mereghetti:2021nkt}. The gradient flow is a gauge-covariant $D+1$-dimensional extension of $D$-dimensional Euclidean QCD. The additional dimension is called flow time $t$ with mass dimension $[t] = -2$. The theory agrees with Euclidean QCD at the boundary $t=0$. Instead of working with the $D+1$-dimensional theory including Lagrange-multiplier fields~\cite{Luscher:2011bx}, we can directly use $D$-dimensional Euclidean QCD and work with flowed quark and gluon fields, defined through the gradient-flow equations.

In the present context, the final goal is to obtain hadronic matrix elements of LEFT operators contributing to the nEDM from lattice QCD. In the first place, we need a non-perturbative and regularization-independent definition of renormalized effective operators, which will be provided by the gradient flow. In a second step, we establish the matching to operators renormalized in the MS scheme, using the short flow-time operator product expansion (SFTE), together with one-loop perturbation theory. Since we are interested in the nEDM, we require matrix elements including an external electromagnetic field, which in the LEFT can couple through the quark electromagnetic current from the renormalizable part of the Lagrangian, or directly through effective operators. Conventionally, the gradient flow is applied to the pure QCD sector of the theory. However, in this case the flow equations explicitly break $U(1)_\mathrm{em}$ invariance. At dimension six, this leads to matching contributions to unphysical MS operators in addition to the ones discussed in Sect.~\ref{sec:Operators}, which are not invariant under $U(1)_\mathrm{em}$. Here, we propose to avoid this problem by using modified quark flow equations that include the (static) external electromagnetic field.

The flowed gluon and quark fields are defined by the flow equations~\cite{Luscher:2010iy,Luscher:2013cpa}
\begin{align}
	\label{eq:FlowEquations}
	\p_t B_\mu &= D_\nu G_{\nu\mu} + \alpha_0 D_\mu \p_\nu B_\nu \, , \nn
	\p_t \chi &= D_\mu D_\mu \chi - \alpha_0 \left( \p_\mu B_\mu \right)\chi \, , \nn
	\p_t \bar\chi &= \bar\chi \overleftarrow{D}_\mu \overleftarrow{D}_\mu + \alpha_0 \bar\chi \left( \p_\mu B_\mu \right) \, ,
\end{align}
and the boundary conditions at vanishing flow time
\begin{align}
	B_\mu(x, t=0) &= G_\mu(x) \, , \nn
	\chi(x, t=0) &= q(x) \, , \nn
	\bar{\chi}(x, t=0) &= \bar q(x) \, .
\end{align}
In Eq.~\eqref{eq:FlowEquations}, $\alpha_0$ denotes a gauge parameter. The gluon flow equation depends on the flowed field-strength tensor, defined by
\begin{align}
	G_{\mu\nu}(x,t) = \p_\mu B_\nu(x,t) - \p_\nu B_\mu(x,t) + \left[B_\mu(x,t), B_\nu(x,t) \right] \, ,
\end{align}
and the flowed covariant derivative, given by
\begin{align}
	 D_{\mu}^{ac} X^c(x,t) = \left( \delta^{ac} \p_{\mu} + f^{abc} B_{\mu}^b(x,t) \right) X^c(x,t)
\end{align}
for a field $X^c(x,t)$ transforming in the adjoint representation of $SU(3)_c$. The flowed covariant derivative appearing in the flow equations for quark and antiquark fields is given by
\begin{align}
	D_{\mu} X(x,t) = \left( \p_{\mu} + B_{\mu}(x,t) + A_\mu(x) \right) X(x,t) \, ,
\end{align}
where $X(x,t)$ denotes a field transforming in the fundamental representation of $SU(3)_c \times U(1)_\mathrm{em}$.
Note that here we include a coupling of the quark fields to the external photon field in order to preserve local $U(1)_\mathrm{em}$ gauge invariance. The external photon field itself remains unflowed.

The gradient flow directly maps the gauge fields to smooth renormalized fields, provided that the $D$-dimensional theory has been renormalized~\cite{Luscher:2011bx}, whereas quark fields require an additional multiplicative renormalization~\cite{Luscher:2013cpa}. At one loop in the MS scheme, the flowed bare (anti-)quark fields $\chi^{(0)}$, $\bar\chi^{(0)}$ are related to renormalized fields $\chi$, $\bar\chi$ by
\begin{align}
	\label{eq:MSQuarkFieldRenormalization}
	\chi^{(0)}(x,t) &= Z_{\chi}^{1/2} \chi(x,t) \, , \quad
	\bar{\chi}^{(0)}(x,t) = Z_{\chi}^{1/2} \bar\chi(x,t) \, , \quad
	Z_{\chi} = 1- \frac{\alpha_s C_F}{4 \pi} \frac{3}{\varepsilon} \, .
\end{align}
A regularization-independent renormalization condition for the quark fields is provided by~\cite{Makino:2014wca,Makino:2014taa}
\begin{align}
	\<0 | \mathring{\bar\chi}(x,t) \overleftrightarrow{\slashed D} \mathring\chi(x,t) | 0 \> = - \frac{2 N_c N_f}{(4\pi)^2 t^2} \, ,
\end{align}
where $\overleftrightarrow D_\mu := D_\mu - \overleftarrow{D}_\mu$. In dimensional regularization, the so-called ringed fields $\mathring\chi$ and $\mathring{\bar\chi}$ differ from renormalized MS fields by a finite renormalization, given at one loop by~\cite{Makino:2014taa,Mereghetti:2021nkt}
\begin{align}
	\label{eq:RingedFields}
	\chi(x,t) &= (8\pi t)^{\varepsilon/2} \zeta_\chi^{1/2} \mathring\chi(x,t) \, , \quad \bar\chi(x,t) = (8\pi t)^{\varepsilon/2} \zeta_\chi^{1/2} \mathring{\bar\chi}(x,t) \, , \nn*
	\zeta_\chi &= 1 - \frac{\alpha_s C_F}{4\pi} \left( 3 \log(8\pi\mu^2t) - \log(432) \right) \, .
\end{align}

If the theory is expressed in terms of renormalized coupling and quark masses, operators built from the flowed gluon and renormalized flowed quark fields are automatically UV finite~\cite{Luscher:2011bx}. Therefore, the gradient flow provides a non-perturbative and regularization-independent definition of renormalized operators.
At short flow times, the flowed operators can be related to MS operators via the SFTE, see Sect.~\ref{sec:SFTE}. If one can find a scale where lattice-QCD computations are not too expensive, but where $\alpha_s$ is still small enough, perturbation theory can be used to determine those coefficients of the SFTE that are not affected by power divergences. Typically, this is the case at an \msbar{} renormalization scale of $\bar\mu \approx 3\GeV$, which is related via
\begin{align}
	\mu = \bar{\mu} \frac{e^{\gamma_E/2}}{(4 \pi)^{1/2}}
\end{align}
to an MS scale of $\mu \approx 1.13\GeV$, where $\gamma_E$ is the Euler constant.
In the case of power divergences, a reliable determination of the matching coefficients must use non-perturbative methods~\cite{Kim:2021qae}. In contrast to the standard approach~\cite{Maiani:1991az}, power divergences show up as $1/t^n$ singularities. Therefore, the gradient flow disentangles power divergences from the continuum limit, which can be taken for any fixed finite flow time.

The differential flow equations have the form of modified heat equations and can be written in integral form~\cite{Luscher:2011bx}. Upon rescaling the gauge field $B_\mu \mapsto g_0 B_\mu$, an expansion of the integral equations in powers of the bare coupling $g_0$ leads to additional Feynman rules compared to perturbative QCD. We use the same conventions as Ref.~\cite{Mereghetti:2021nkt}, apart from the new interaction terms involving the external electromagnetic field that emerge from the quark flow equations. The corresponding Feynman rules are listed in App.~\ref{sec:FeynmanRules}.

%% file: sections/Results.tex

\section{Matching coefficients}
\label{sec:MatchingCoefficients}

\subsection{Short flow-time expansion}
\label{sec:SFTE}

In the following, we apply the SFTE to the case of the four-quark operators given in Eq.~\eqref{eq:FourQuarkParityBasis}. We define flowed renormalized versions of the (physical) operators by replacing the quark fields with the ringed flowed quark fields:
\begin{align}
	\label{eq:FlowedFourQuarkParityBasis}
	\O^{S1}_p(t) &= (\mathring{\bar\chi}_p \gamma_5 \mathring{\chi}_p)(\mathring{\bar\chi}_p \mathring{\chi}_p) \, , \nn
	\O^{S8}_p(t) &= (\mathring{\bar\chi}_p \gamma_5 t^a \mathring{\chi}_p)(\mathring{\bar\chi}_p t^a \mathring{\chi}_p) \, , \nn
	\O^{S1}_{pr}(t) &= (\mathring{\bar\chi}_p \gamma_5 \mathring{\chi}_p)(\mathring{\bar\chi}_r \mathring{\chi}_r) \, , \qquad p \neq r \, , \nn
	\O^{S8}_{pr}(t) &= (\mathring{\bar\chi}_p \gamma_5 t^a \mathring{\chi}_p)(\mathring{\bar\chi}_r t^a \mathring{\chi}_r) \, , \qquad p \neq r \, , \nn
	\O^{T1}_{pr}(t) &= \frac{1}{2} \Bigl[ (\mathring{\bar\chi}_p \tilde{\sigma}_{\mu \nu} \mathring{\chi}_p)(\mathring{\bar\chi}_r \sigma_{\mu \nu} \mathring{\chi}_r) + (\mathring{\bar\chi}_r \tilde{\sigma}_{\mu \nu} \mathring{\chi}_r)(\mathring{\bar\chi}_p \sigma_{\mu \nu} \mathring{\chi}_p) \Bigr] \, , \qquad p \neq r \, , \nn
	\O^{T8}_{pr}(t) &= \frac{1}{2} \Bigl[ (\mathring{\bar\chi}_p \tilde{\sigma}_{\mu \nu} t^a \mathring{\chi}_p)(\mathring{\bar\chi}_r \sigma_{\mu \nu} t^a \mathring{\chi}_r) + (\mathring{\bar\chi}_r \tilde{\sigma}_{\mu \nu} t^a \mathring{\chi}_r)(\mathring{\bar\chi}_p \sigma_{\mu \nu} t^a \mathring{\chi}_p) \Bigr] \, , \qquad p \neq r \, .
\end{align}
The SFTE is an operator-product expansion that relates flowed operators to MS renormalized operators,\footnote{The MS operators are understood as minimally subtracted versions of the bare operators discussed in Sect.~\ref{sec:Operators}, adjusted implicitly by factors $\mu^{n\varepsilon}$ so that in $D$ space-time dimensions the mass dimensions of bare operators differ only by integers, see Ref.~\cite{Mereghetti:2021nkt}.}
\begin{align}
	\label{eq:SFTE}
	\O_i^R(t) = \sum_j c_{ij}(t,\mu) \O_j^\mathrm{MS}(\mu) \, ,
\end{align}
with finite matching coefficients $c_{ij}(t,\mu)$. If these matching coefficients are known, one can obtain the hadronic matrix elements of MS operators from the matrix elements of flowed operators by inverting the SFTE. 
The sum in Eq.~\eqref{eq:SFTE} in principle runs over an infinite tower of operators, but we neglect power corrections from operators beyond dimension six, hence we include all the operators listed in Sect.~\ref{sec:Operators}.
Using the fact that QCD and the flow equations conserve flavor, we write the SFTE for the flowed four-quark operators with two different flavor indices in the form
\begin{align}
	\label{eq:SFTETwoFlavors}
	\O_{pr}^{X}(t) &= 
		c_{X,P}^{pr}(t,\mu)\O^{P}_{p}(\mu) + c_{X,P'}^{pr}(t,\mu)\O^{P}_{r}(\mu) + c_{X,\theta}^{pr}(t,\mu) \O_\theta(\mu) \nn
		&\quad + c_{X,E}^{pr}(t,\mu) \O^{E}_{p}(\mu) + c_{X,E'}^{pr}(t,\mu) \O^{E}_{r}(\mu) \nn
		&\quad + c_{X,CE}^{pr}(t,\mu) \O^{CE}_{p}(\mu) + c_{X,CE'}^{pr}(t,\mu) \O^{CE}_{r}(\mu) \nn
		&\quad + c_{X,\widetilde G}(t,\mu) \O_{\widetilde G}(\mu) \nn
		&\quad + c_{X,S1}(t,\mu) \O_{pr}^{S1,\mathrm{MS}}(\mu) + c_{X,S1'}(t,\mu) \O_{rp}^{S1,\mathrm{MS}}(\mu) \nn
		&\quad + c_{X,S8}(t,\mu)\O_{pr}^{S8,\mathrm{MS}}(\mu) + c_{X,S8'}(t,\mu)\O_{rp}^{S8,\mathrm{MS}}(\mu) \nn
		&\quad + c_{X,T1}(t,\mu) \O_{pr}^{T1,\mathrm{MS}}(\mu) + c_{X,T8}(t,\mu) \O_{pr}^{T8,\mathrm{MS}}(\mu) \nn
		&\quad + \sum_i c_{X,\N_i}^{pr}(t,\mu) \N_i(\mu) + \sum_i c_{X,\E_i}^{pr}(t,\mu) \E_i(\mu) \, ,
\end{align}
where $X \in \{S1, S8, T1, T8\}$ and $p,r$ are fixed indices. The SFTE for the two operators with only one quark flavor can be related to Eq.~\eqref{eq:SFTETwoFlavors} by taking into account the Fierz-evanescent operators~\eqref{eq:FierzEvanescents}, leading to
\begin{align}
	\label{eq:SFTEOneFlavor}
	\O_{p}^{X}(t) &= 
		\left[ c_{X,P}^{pp}(t,\mu) + c_{X,P'}^{pp}(t,\mu) \right] \O^{P}_{p}(\mu) + c_{X,\theta}^{pp}(t,\mu) \O_\theta(\mu) \nn
		&\quad + \left[ c_{X,E}^{pp}(t,\mu) + c_{X,E'}^{pp}(t,\mu) \right] \O^{E}_{p}(\mu) + \left[ c_{X,CE}^{pp}(t,\mu) + c_{X,CE'}^{pp}(t,\mu) \right] \O^{CE}_{p}(\mu) \nn
		&\quad + c_{X,\widetilde G}(t,\mu) \O_{\widetilde G}(\mu) \nn
		&\quad + \left[ c_{X,S1}(t,\mu) + c_{X,S1'}(t,\mu) + c_{X,T1}(t,\mu) r_{T1,S1} + c_{X,T8}(t,\mu) r_{T8,S1} \right] \O_{p}^{S1,\mathrm{MS}}(\mu) \nn
		&\quad + \left[ c_{X,S8}(t,\mu) + c_{X,S8'}(t,\mu) + c_{X,T1}(t,\mu) r_{T1,S8} + c_{X,T8}(t,\mu) r_{T8,S8} \right] \O_{p}^{S8,\mathrm{MS}}(\mu) \nn
		&\quad + \sum_i c_{X,\N_i'}(t,\mu) \N_i(\mu) + \sum_i c_{X,\E_i'}(t,\mu) \E_i(\mu) \, ,
\end{align}
where $X \in \{S1, S8\}$ and the coefficients $r_{X,Y}$ are defined in Eq.~\eqref{eq:FierzCoefficients}. At one loop, there are no diagrams that contribute to purely gluonic Green's functions, hence we disregard $c_{X,\theta}^{pr}$ and $c_{X,\widetilde G}$ in the following. Since the dependence of the matching coefficients on the flavor indices is only through quark masses and because we neglect power corrections beyond dimension six, the coefficients of the dimension-six operators are flavor independent.

As discussed in Sects.~\ref{sec:NuisanceOperators} and~\ref{sec:EvanescentOperators}, the $S$-matrix elements of nuisance and (renormalized) evanescent operators vanish. Therefore, their matching coefficients are of no further interest, as they are not required to obtain the matrix elements of MS four-quark operators. However, we need the unphysical operators in the matching calculation and their definition affects the finite matching coefficients to the physical operators.

\subsection{Loop calculation}

We perform the loop calculation in much the same way as Ref.~\cite{Mereghetti:2021nkt}: we determine the coefficients in the SFTE by calculating suitable amputated Green's functions with the insertion of a flowed operator and by matching the result to the same Green's function with the insertion of the right-hand side of the SFTE, i.e., the insertion of a sum of MS operators. As explained in Ref.~\cite{Mereghetti:2021nkt}, we apply the method of regions~\cite{Beneke:1997zp} to simplify the extraction of the matching coefficients, expanding the integrands before integration in all scales apart from the flow time $t$ of the inserted operator. In this case, the flowed one-loop integrals become very simple single-scale integrals and both momentum and flow-time integrals can be performed with standard methods. No loop integrals with insertions of (unflowed) MS operators need to be calculated, since the expansion of the loops results in scaleless integrals, which vanish in dimensional regularization.

A large degree of automation allows us to perform the calculation efficiently: we use \texttt{qgraf}~\cite{Nogueira:1991ex} for the generation of the Feynman diagrams and our own \texttt{Mathematica} routines for the evaluation of the diagrams, partially making use of \texttt{FeynCalc}~\cite{Mertig:1990an,Shtabovenko:2016sxi,Shtabovenko:2020gxv}.

We perform the whole calculation with generic gauge parameters $\xi$ and $\alpha_0$: their cancellation in the final result for the matching coefficients provides a strong consistency check.

\subsection{Results for the matching coefficients}
\label{sec:Results}

In this section, we present the results of the one-loop matching calculation of flowed four-quark operators to MS operators up to dimension six.
We explicitly show the contribution of the finite renormalization $\zeta_\chi$ that arises from the relation~\eqref{eq:RingedFields} between flowed MS fields and ringed fields. For the matching onto four-quark operators, we present results in both the NDR and HV schemes, whereas for the matching to the lower-dimension operators we only use the HV scheme. For a compact notation, we define
\begin{align}
	\delta_\mathrm{NDR} = \begin{cases}
			1 \, , \quad \text{in the NDR scheme}, \\
			0 \, , \quad \text{in the HV scheme}.
		\end{cases}
\end{align}

The logarithmic dependence of the matching coefficients on the matching scale $\mu$ is predicted by the anomalous dimensions of the MS operators. Applying the change to the chiral basis, we have checked that our results are compatible with the anomalous dimensions of Ref.~\cite{Jenkins:2017dyc}.

\subsubsection{Matching coefficients of the pseudoscalar density}

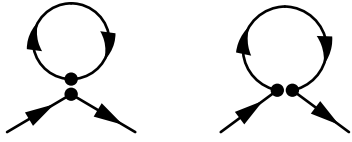
\begin{figure}[t]
	\begin{align*}
		\\[-1cm]
		\begin{gathered}
			\begin{fmfgraph*}(60,60)
				\fmfleft{i1,i2}
				\fmfright{o1,o2}
				\fmftop{t1}
				\fmf{quark,tension=2}{i1,v1,o1}
				\fmf{quark,left}{v2,v3,v2}
				\fmf{phantom, tension=10}{v1,v2}
				\fmf{phantom, tension=5}{t1,v3}
				\fmfdot{v1}
				\fmfdot{v2}
			\end{fmfgraph*}
		\end{gathered}
		\qquad
		\begin{gathered}
			\begin{fmfgraph*}(60,60)
				\fmfleft{i1,i2}
				\fmfright{o1,o2}
				\fmftop{t1}
				\fmf{quark,tension=2}{i1,v1}
				\fmf{quark,tension=2}{v2,o1}
				\fmf{quark,left=0.9}{v1,v3,v2}
				\fmf{phantom, tension=7}{v1,v2}
				\fmf{phantom, tension=5}{t1,v3}
				\fmfdot{v1}
				\fmfdot{v2}
			\end{fmfgraph*}
		\end{gathered}
	\end{align*}
	\caption{Diagrams contributing to the four-quark matching to the pseudoscalar density. In the case of scalar operators, which are not symmetric under exchange of the quark bilinears, there are two different insertions for each diagram.}
	\label{fig:DiagramsTwoPoint}
\end{figure}

The coefficients of the pseudoscalar density in the matching equations are obtained by computing the insertion of flowed four-quark operators into antiquark-quark two-point functions. The diagram topologies are shown in Fig.~\ref{fig:DiagramsTwoPoint}. We find the following results in the HV scheme:
\begin{align}
	\label{eq:MatchingPseudoscalar}
	c_{S1,P}^{pr}(t,\mu) &= -\frac{1}{8\pi^2} \left[ \frac{m_r}{t} + 2 m_r^3 \Big( \log(8\pi\mu^2 t) + 1 \Big) \right] \left( N_c - \frac{1}{2} \delta_{pr}  \right) + \O(\alpha_s,t) \, , \nn*
	c_{S8,P}^{pr}(t,\mu) &= - \frac{C_F}{16\pi^2} \left[ \frac{m_r}{t} + 2 m_r^3 \Big( \log(8\pi\mu^2 t) + 1 \Big) \right] \delta_{pr} + \O(\alpha_s,t) \, , \nn
	c_{S1,P'}^{pr}(t,\mu) &= c_{S8,P'}^{pr}(t,\mu) = \O(\alpha_s,t) \, , \nn
	c_{T1,P}^{pr}(t,\mu) &= c_{T1,P'}^{pr}(t,\mu) = \frac{3}{8\pi^2} \left[ \frac{m_r}{t} + 2 m_r^3 \Big( \log(8\pi\mu^2 t) + 1 \Big) \right] \delta_{pr} + \O(\alpha_s,t) \, , \nn
	c_{T8,P}^{pr}(t,\mu) &= c_{T8,P'}^{pr}(t,\mu) = -\frac{3 C_F}{8\pi^2} \left[ \frac{m_r}{t} + 2 m_r^3 \Big( \log(8\pi\mu^2 t) + 1 \Big) \right] \delta_{pr} + \O(\alpha_s,t) \, .
\end{align}
The matching coefficients of the pseudoscalar density contain at least one mass insertion, which follows from chirality arguments. Since we are matching to a dimension-three operator, the coefficients still contain power divergences in the flow time, which require a non-perturbative subtraction. Therefore, the perturbative result can only be trusted in the case of the $m^3$ contributions. 

The parts of the coefficients containing a Kronecker $\delta_{pr}$ only contribute in the case of single-flavor operators in Eq.~\eqref{eq:SFTEOneFlavor}: since there are no single-flavor tensor operators in our basis, there remains no matching contribution from tensor four-quark operators to the pseudoscalar density. We still list these contributions in Eq.~\eqref{eq:MatchingPseudoscalar}, because this enables the reconstruction of matching coefficients for the case of general flavors: at one loop, a flowed four-quark operator $\O^X_{prst}(t)$ with generic flavor indices matches onto generic pseudoscalar operators $\O^P_{pr}$ with a coefficient proportional to $\delta_{st}$, as well as onto pseudoscalars $\O^P_{pt}$ and $\O^P_{sr}$ with identical coefficients times $\delta_{sr}$ and $\delta_{pt}$, respectively.

\subsubsection{Matching coefficients of dipole operators}

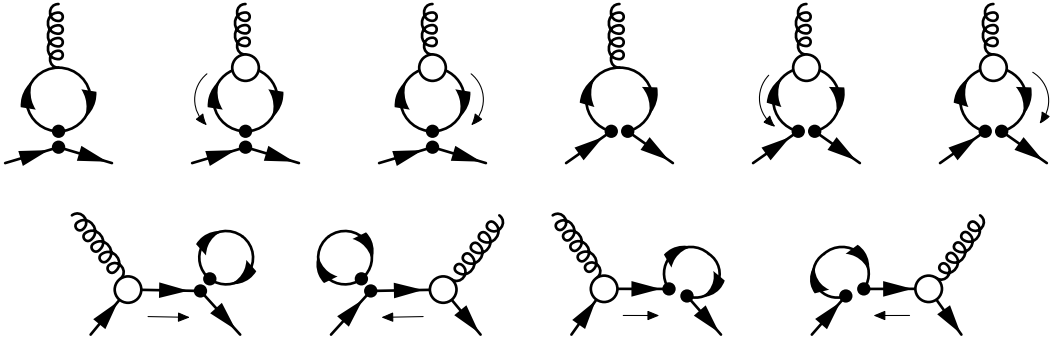
\begin{figure}[t]
	\begin{align*}
		\\[-1cm]
		\begin{gathered}
			\begin{fmfgraph*}(50,60)
				\fmfleft{i1,i2}
				\fmfright{o1,o2}
				\fmftop{t1}
				\fmf{quark,tension=4}{i1,v1,o1}
				\fmf{quark,left}{v2,v3,v2}
				\fmf{phantom, tension=8}{v1,v2}
				\fmf{gluon,tension=2}{t1,v3}
				\fmfdot{v1}
				\fmfdot{v2}
			\end{fmfgraph*}
		\end{gathered}
		\qquad
		\begin{gathered}
			\begin{fmfgraph*}(50,60)
				\fmfleft{i1,i2}
				\fmfright{o1,o2}
				\fmftop{t1}
				\fmf{quark,tension=4}{i1,v1,o1}
				\fmf{quark,left}{v2,v3,v2}
				\fmf{phantom, tension=8}{v1,v2}
				\fmf{gluon,tension=2}{t1,v3}
				\fmfdot{v1}
				\fmfdot{v2}
				\fmfv{decor.shape=circle, decor.filled=empty, decor.size=(3.5mm)}{v3}
				\fmffreeze
				\fmf{mlarrowl,right=0.75}{v3,v2}
			\end{fmfgraph*}
		\end{gathered}
		\qquad
		\begin{gathered}
			\begin{fmfgraph*}(50,60)
				\fmfleft{i1,i2}
				\fmfright{o1,o2}
				\fmftop{t1}
				\fmf{quark,tension=4}{i1,v1,o1}
				\fmf{quark,left}{v2,v3,v2}
				\fmf{phantom, tension=8}{v1,v2}
				\fmf{gluon,tension=2}{t1,v3}
				\fmfdot{v1}
				\fmfdot{v2}
				\fmfv{decor.shape=circle, decor.filled=empty, decor.size=(3.5mm)}{v3}
				\fmffreeze
				\fmf{mlarrowr,left=0.75}{v3,v2}
			\end{fmfgraph*}
		\end{gathered}
		&\qquad
		\begin{gathered}
			\begin{fmfgraph*}(50,60)
				\fmfleft{i1,i2}
				\fmfright{o1,o2}
				\fmftop{t1}
				\fmf{quark,tension=2}{i1,v1}
				\fmf{quark,tension=2}{v2,o1}
				\fmf{quark,left=0.9}{v1,v3,v2}
				\fmf{phantom, tension=5}{v1,v2}
				\fmf{gluon, tension=2}{t1,v3}
				\fmfdot{v1}
				\fmfdot{v2}
			\end{fmfgraph*}
		\end{gathered}
		\qquad
		\begin{gathered}
			\begin{fmfgraph*}(50,60)
				\fmfleft{i1,i2}
				\fmfright{o1,o2}
				\fmftop{t1}
				\fmf{quark,tension=2}{i1,v1}
				\fmf{quark,tension=2}{v2,o1}
				\fmf{quark,left=0.9}{v1,v3,v2}
				\fmf{phantom, tension=5}{v1,v2}
				\fmf{gluon, tension=2}{t1,v3}
				\fmfdot{v1}
				\fmfdot{v2}
				\fmfv{decor.shape=circle, decor.filled=empty, decor.size=(3.5mm)}{v3}
				\fmffreeze
				\fmf{mlarrowl,right=0.75}{v3,v2}
			\end{fmfgraph*}
		\end{gathered}
		\qquad
		\begin{gathered}
			\begin{fmfgraph*}(50,60)
				\fmfleft{i1,i2}
				\fmfright{o1,o2}
				\fmftop{t1}
				\fmf{quark,tension=2}{i1,v1}
				\fmf{quark,tension=2}{v2,o1}
				\fmf{quark,left=0.9}{v1,v3,v2}
				\fmf{phantom, tension=5}{v1,v2}
				\fmf{gluon, tension=2}{t1,v3}
				\fmfdot{v1}
				\fmfdot{v2}
				\fmfv{decor.shape=circle, decor.filled=empty, decor.size=(3.5mm)}{v3}
				\fmffreeze
				\fmf{mlarrowr,left=0.75}{v3,v2}
			\end{fmfgraph*}
		\end{gathered}
		\\[0.25cm]
		\begin{gathered}
			\begin{fmfgraph*}(70,50)
				\fmfleft{i1,i2}
				\fmfright{o1,o2}
				\fmftop{t2,t1}
				\fmf{quark,tension=2}{i1,v4,v1}
				\fmf{quark,tension=2}{v1,o1}
				\fmf{quark,left}{v2,v3,v2}
				\fmf{phantom, tension=7}{v1,v2}
				\fmf{phantom, tension=4}{t1,v3}
				\fmf{gluon, tension=1.25}{v4,t2}
				\fmfdot{v1}
				\fmfdot{v2}
				\fmfv{decor.shape=circle, decor.filled=empty, decor.size=(3.5mm)}{v4}
				\fmffreeze
				\fmf{mlarrowd}{v4,v1}
			\end{fmfgraph*}
		\end{gathered}
		\qquad
		\begin{gathered}
			\begin{fmfgraph*}(70,50)
				\fmfleft{i1,i2}
				\fmfright{o1,o2}
				\fmftop{t1,t2}
				\fmf{quark,tension=2}{i1,v1}
				\fmf{quark,tension=2}{v1,v4,o1}
				\fmf{quark,left}{v2,v3,v2}
				\fmf{phantom, tension=7}{v1,v2}
				\fmf{phantom, tension=4}{t1,v3}
				\fmf{gluon, tension=1.25}{v4,t2}
				\fmfdot{v1}
				\fmfdot{v2}
				\fmfv{decor.shape=circle, decor.filled=empty, decor.size=(3.5mm)}{v4}
				\fmffreeze
				\fmf{mlarrowd}{v4,v1}
			\end{fmfgraph*}
		\end{gathered}
		&\qquad
		\begin{gathered}
			\begin{fmfgraph*}(70,50)
				\fmfleft{i1,i2}
				\fmfright{o1,o2}
				\fmftop{t2,t1}
				\fmf{quark,tension=2}{i1,v4,v1}
				\fmf{quark,tension=2}{v2,o1}
				\fmf{quark,left=0.9}{v1,v3,v2}
				\fmf{phantom, tension=5}{v1,v2}
				\fmf{phantom, tension=2}{t1,v3}
				\fmf{gluon, tension=1.25}{v4,t2}
				\fmfdot{v1}
				\fmfdot{v2}
				\fmfv{decor.shape=circle, decor.filled=empty, decor.size=(3.5mm)}{v4}
				\fmffreeze
				\fmf{mlarrowd}{v4,v1}
			\end{fmfgraph*}
		\end{gathered}
		\qquad
		\begin{gathered}
			\begin{fmfgraph*}(70,50)
				\fmfleft{i1,i2}
				\fmfright{o1,o2}
				\fmftop{t1,t2}
				\fmf{quark,tension=2}{i1,v1}
				\fmf{quark,tension=2}{v2,v4,o1}
				\fmf{quark,left=0.9}{v1,v3,v2}
				\fmf{phantom, tension=5}{v1,v2}
				\fmf{phantom, tension=2}{t1,v3}
				\fmf{gluon, tension=1.25}{v4,t2}
				\fmfdot{v1}
				\fmfdot{v2}
				\fmfv{decor.shape=circle, decor.filled=empty, decor.size=(3.5mm)}{v4}
				\fmffreeze
				\fmf{mlarrowd}{v4,v2}
			\end{fmfgraph*}
		\end{gathered}
	\end{align*}
	\caption{Diagrams contributing to the four-quark matching to the dipole operators. In the case of scalar operators, there are two different insertions for each diagram. With our modified flow equations, the same diagrams appear in the photonic case, with the gluon replaced by a photon, whereas with pure QCD flow equations only the first and fourth diagrams would be present.}
	\label{fig:DiagramsThreePoint}
\end{figure}

The four-quark matching coefficients of the dimension-five dipole operators are obtained by inserting flowed four-quark operators into Green's functions with a quark and antiquark and either a gluon or an external photon, see Fig.~\ref{fig:DiagramsThreePoint}. We only provide results for the HV scheme, since in the NDR scheme the insertion of tensor operators leads to problematic $\gamma_5$-odd fermion traces.

\begin{figure}[t]
	\begin{align*}
		\\[-1cm]
		\begin{gathered}
			\begin{fmfgraph*}(50,45)
				\fmfleft{i1,i2}
				\fmfright{o1,o2}
				\fmftop{t1,t2}
				\fmf{quark,tension=4}{i1,v1,o1}
				\fmf{quark,left=0.2}{v2,v3,v4,v2}
				\fmf{phantom, tension=8}{v1,v2}
				\fmf{gluon,tension=2}{t1,v3}
				\fmf{gluon,tension=2}{t2,v4}
				\fmfdot{v1}
				\fmfdot{v2}
				\fmfv{decor.shape=circle, decor.filled=20, decor.size=(2.5mm)}{v3,v4}
			\end{fmfgraph*}
		\end{gathered}
		\qquad
		\begin{gathered}
			\begin{fmfgraph*}(50,45)
				\fmfleft{i1,i2}
				\fmfright{o1,o2}
				\fmftop{t1,t2}
				\fmf{quark,tension=2}{i1,v1}
				\fmf{quark,tension=2}{v2,o1}
				\fmf{quark,left=0.2}{v1,v3,v4,v2}
				\fmf{phantom, tension=8}{v1,v2}
				\fmf{gluon, tension=2}{t1,v3}
				\fmf{gluon, tension=2}{t2,v4}
				\fmfdot{v1}
				\fmfdot{v2}
				\fmfv{decor.shape=circle, decor.filled=20, decor.size=(2.5mm)}{v3,v4}
			\end{fmfgraph*}
		\end{gathered}
		\qquad
		\begin{gathered}
			\begin{fmfgraph*}(50,45)
				\fmfleft{i1,i2}
				\fmfright{o1,o2}
				\fmftop{t1,t2}
				\fmf{quark,tension=4}{i1,v1,o1}
				\fmf{quark,left}{v2,v3,v2}
				\fmf{phantom, tension=8}{v1,v2}
				\fmf{gluon,tension=2}{t1,v3}
				\fmf{gluon,tension=2}{t2,v3}
				\fmfdot{v1}
				\fmfdot{v2}
				\fmfv{decor.shape=circle, decor.filled=empty, decor.size=(3.5mm)}{v3}
				\fmffreeze
				\fmf{mlarrowr,left=0.75}{v3,v2}
			\end{fmfgraph*}
		\end{gathered}
		&\qquad
		\begin{gathered}
			\begin{fmfgraph*}(50,45)
				\fmfleft{i1,i2}
				\fmfright{o1,o2}
				\fmftop{t1,t2}
				\fmf{quark,tension=2}{i1,v1}
				\fmf{quark,tension=2}{v2,o1}
				\fmf{quark,left=0.9}{v1,v3,v2}
				\fmf{phantom, tension=5}{v1,v2}
				\fmf{gluon, tension=2}{t1,v3}
				\fmf{gluon, tension=2}{t2,v3}
				\fmfdot{v1}
				\fmfdot{v2}
				\fmfv{decor.shape=circle, decor.filled=empty, decor.size=(3.5mm)}{v3}
				\fmffreeze
				\fmf{mlarrowr,left=0.75}{v3,v2}
			\end{fmfgraph*}
		\end{gathered}
		\qquad
		\begin{gathered}
			\begin{fmfgraph*}(50,60)
				\fmfleft{i1,i2}
				\fmfright{o1,o2}
				\fmftop{t1,t2}
				\fmf{quark,tension=4}{i1,v1,o1}
				\fmf{quark,left}{v2,v3,v2}
				\fmf{phantom, tension=6}{v1,v2}
				\fmf{gluon,tension=2}{t1,v4}
				\fmf{gluon,tension=2}{t2,v4}
				\fmf{gluon,tension=1.5}{v4,v3}
				\fmfdot{v1}
				\fmfdot{v2}
				\fmfv{decor.shape=circle, decor.filled=empty, decor.size=(3.5mm)}{v3,v4}
				\fmffreeze
				\fmf{mlarrowr,left=0.75}{v3,v2}
				\fmf{mlarrowr}{v4,v3}
			\end{fmfgraph*}
		\end{gathered}
		\qquad
		\begin{gathered}
			\begin{fmfgraph*}(50,60)
				\fmfleft{i1,i2}
				\fmfright{o1,o2}
				\fmftop{t1,t2}
				\fmf{quark,tension=2}{i1,v1}
				\fmf{quark,tension=2}{v2,o1}
				\fmf{quark,left=0.9}{v1,v3,v2}
				\fmf{phantom, tension=5}{v1,v2}
				\fmf{gluon,tension=2}{t1,v4}
				\fmf{gluon,tension=2}{t2,v4}
				\fmf{gluon,tension=1.5}{v4,v3}
				\fmfdot{v1}
				\fmfdot{v2}
				\fmfv{decor.shape=circle, decor.filled=empty, decor.size=(3.5mm)}{v3,v4}
				\fmffreeze
				\fmf{mlarrowr,left=0.75}{v3,v2}
				\fmf{mlarrowr}{v4,v3}
			\end{fmfgraph*}
		\end{gathered}
		\\[0.25cm]
		\begin{gathered}
			\begin{fmfgraph*}(60,50)
				\fmfleft{i1,i2}
				\fmfright{o1,o2}
				\fmftop{t2,t1}
				\fmf{quark,tension=2}{i1,v4,v1}
				\fmf{quark,tension=2}{v1,o1}
				\fmf{quark,left}{v2,v3,v2}
				\fmf{phantom, tension=6}{v1,v2}
				\fmf{gluon, tension=2}{t1,v3}
				\fmf{gluon, tension=0.75}{v4,t2}
				\fmfdot{v1}
				\fmfdot{v2}
				\fmfv{decor.shape=circle, decor.filled=empty, decor.size=(3.5mm)}{v4}
				\fmfv{decor.shape=circle, decor.filled=20, decor.size=(2.5mm)}{v3}
				\fmffreeze
				\fmf{mlarrowd}{v4,v1}
			\end{fmfgraph*}
		\end{gathered}
		\quad\;
		\begin{gathered}
			\begin{fmfgraph*}(60,50)
				\fmfleft{i1,i2}
				\fmfright{o1,o2}
				\fmftop{t2,t1}
				\fmf{quark,tension=2}{i1,v4,v1}
				\fmf{quark,tension=2}{v2,o1}
				\fmf{quark,left=0.9}{v1,v3,v2}
				\fmf{phantom, tension=5}{v1,v2}
				\fmf{gluon, tension=2}{t1,v3}
				\fmf{gluon, tension=1}{v4,t2}
				\fmfdot{v1}
				\fmfdot{v2}
				\fmfv{decor.shape=circle, decor.filled=empty, decor.size=(3.5mm)}{v4}
				\fmfv{decor.shape=circle, decor.filled=20, decor.size=(2.5mm)}{v3}
				\fmffreeze
				\fmf{mlarrowd}{v4,v1}
			\end{fmfgraph*}
		\end{gathered}
		\quad\;
		\begin{gathered}
			\begin{fmfgraph*}(60,50)
				\fmfleft{i1,i2}
				\fmfright{o1,o2}
				\fmftop{t2,t3,t1}
				\fmf{quark,tension=2}{i1,v4,v1}
				\fmf{quark,tension=2}{v1,o1}
				\fmf{quark,left}{v2,v3,v2}
				\fmf{phantom, tension=6}{v1,v2}
				\fmf{phantom, tension=3}{t1,v3}
				\fmf{gluon, tension=0.5}{v4,t2}
				\fmf{gluon, tension=0.5}{v4,t3}
				\fmfdot{v1}
				\fmfdot{v2}
				\fmfv{decor.shape=circle, decor.filled=empty, decor.size=(3.5mm)}{v4}
				\fmffreeze
				\fmf{mlarrowd}{v4,v1}
			\end{fmfgraph*}
		\end{gathered}
		& \quad\;
		\begin{gathered}
			\begin{fmfgraph*}(60,50)
				\fmfleft{i1,i2}
				\fmfright{o1,o2}
				\fmftop{t2,t3,t1}
				\fmf{quark,tension=2}{i1,v4,v1}
				\fmf{quark,tension=2}{v2,o1}
				\fmf{quark,left=0.9}{v1,v3,v2}
				\fmf{phantom, tension=5}{v1,v2}
				\fmf{phantom, tension=2}{t1,v3}
				\fmf{gluon, tension=0.5}{v4,t2}
				\fmf{gluon, tension=0.5}{v4,t3}
				\fmfdot{v1}
				\fmfdot{v2}
				\fmfv{decor.shape=circle, decor.filled=empty, decor.size=(3.5mm)}{v4}
				\fmffreeze
				\fmf{mlarrowd}{v4,v1}
			\end{fmfgraph*}
		\end{gathered}
		\quad\;
		\begin{gathered}
			\begin{fmfgraph*}(60,50)
				\fmfleft{i1,i2}
				\fmfright{o1,o2}
				\fmftop{t2,t3,t1}
				\fmf{quark,tension=2}{i1,v4,v1}
				\fmf{quark,tension=2}{v1,o1}
				\fmf{quark,left}{v2,v3,v2}
				\fmf{phantom, tension=6}{v1,v2}
				\fmf{phantom, tension=3}{t1,v3}
				\fmf{gluon, tension=1.5}{v4,v5}
				\fmf{gluon, tension=1}{v5,t2}
				\fmf{gluon, tension=1}{v5,t3}
				\fmfdot{v1}
				\fmfdot{v2}
				\fmfv{decor.shape=circle, decor.filled=empty, decor.size=(3.5mm)}{v4,v5}
				\fmffreeze
				\fmf{mlarrowd}{v4,v1}
				\fmf{mlarrowl}{v5,v4}
			\end{fmfgraph*}
		\end{gathered}
		\quad\;
		\begin{gathered}
			\begin{fmfgraph*}(60,50)
				\fmfleft{i1,i2}
				\fmfright{o1,o2}
				\fmftop{t2,t3,t1}
				\fmf{quark,tension=2}{i1,v4,v1}
				\fmf{quark,tension=2}{v2,o1}
				\fmf{quark,left=0.9}{v1,v3,v2}
				\fmf{phantom, tension=5}{v1,v2}
				\fmf{phantom, tension=2}{t1,v3}
				\fmf{gluon, tension=1.5}{v4,v5}
				\fmf{gluon, tension=1}{v5,t2}
				\fmf{gluon, tension=1}{v5,t3}
				\fmfdot{v1}
				\fmfdot{v2}
				\fmfv{decor.shape=circle, decor.filled=empty, decor.size=(3.5mm)}{v4,v5}
				\fmffreeze
				\fmf{mlarrowd}{v4,v1}
				\fmf{mlarrowl}{v5,v4}
			\end{fmfgraph*}
		\end{gathered}
		\\[0.25cm]
		\begin{gathered}
			\begin{fmfgraph*}(80,50)
				\fmfleft{i1,i2}
				\fmfright{o1,o2}
				\fmftop{t2,t3,t1}
				\fmf{quark,tension=2}{i1,v4,v5,v1}
				\fmf{quark,tension=2}{v1,o1}
				\fmf{quark,left}{v2,v3,v2}
				\fmf{phantom, tension=7}{v1,v2}
				\fmf{phantom, tension=3}{t1,v3}
				\fmf{gluon, tension=1}{v4,t2}
				\fmfdot{v1}
				\fmfdot{v2}
				\fmfv{decor.shape=circle, decor.filled=empty, decor.size=(3.5mm)}{v4,v5}
				\fmffreeze
				\fmf{gluon, tension=1}{v5,t3}
				\fmf{mlarrowd}{v4,v5,v1}
			\end{fmfgraph*}
		\end{gathered}
		\qquad
		\begin{gathered}
			\begin{fmfgraph*}(80,50)
				\fmfleft{i1,i2}
				\fmfright{o1,o2}
				\fmftop{t2,t3,t1}
				\fmf{quark,tension=2}{i1,v4,v5}
				\fmf{quark,tension=2.5}{v5,v1}
				\fmf{quark,tension=2}{v2,o1}
				\fmf{quark,left=0.9}{v1,v3,v2}
				\fmf{phantom, tension=5}{v1,v2}
				\fmf{phantom, tension=1.5}{t1,v3}
				\fmf{gluon, tension=1}{v4,t2}
				\fmfdot{v1}
				\fmfdot{v2}
				\fmfv{decor.shape=circle, decor.filled=empty, decor.size=(3.5mm)}{v4,v5}
				\fmffreeze
				\fmf{gluon, tension=1}{v5,t3}
				\fmf{mlarrowd}{v4,v5,v1}
			\end{fmfgraph*}
		\end{gathered}
		&\qquad
		\begin{gathered}
			\begin{fmfgraph*}(80,50)
				\fmfleft{i1,i2}
				\fmfright{o1,o2}
				\fmftop{t2,t1,t3}
				\fmfbottom{b1}
				\fmf{quark,tension=2}{i1,v4,v1}
				\fmf{quark,tension=2}{v1,v5,o1}
				\fmf{quark,left}{v2,v3,v2}
				\fmf{phantom, tension=7}{v1,v2}
				\fmf{phantom, tension=3}{t1,v3}
				\fmf{phantom, tension=2.5}{b1,v1}
				\fmf{gluon, tension=1}{v4,t2}
				\fmf{gluon, tension=1}{v5,t3}
				\fmfdot{v1}
				\fmfdot{v2}
				\fmfv{decor.shape=circle, decor.filled=empty, decor.size=(3.5mm)}{v4,v5}
				\fmffreeze
				\fmf{mlarrowd}{v4,v1}
				\fmf{mlarrowd}{v5,v1}
			\end{fmfgraph*}
		\end{gathered}
		\qquad
		\begin{gathered}
			\begin{fmfgraph*}(80,50)
				\fmfleft{i1,i2}
				\fmfright{o1,o2}
				\fmftop{t2,t1,t3}
				\fmfbottom{b1}
				\fmf{quark,tension=2}{i1,v4,v1}
				\fmf{quark,tension=2}{v2,v5,o1}
				\fmf{quark,left=0.8}{v1,v3,v2}
				\fmf{phantom, tension=5}{v1,v2}
				\fmf{phantom, tension=2}{t1,v3}
				\fmf{phantom, tension=1.25}{b1,v1}
				\fmf{phantom, tension=1.25}{b1,v2}
				\fmf{gluon, tension=1}{v4,t2}
				\fmf{gluon, tension=1}{v5,t3}
				\fmfdot{v1}
				\fmfdot{v2}
				\fmfv{decor.shape=circle, decor.filled=empty, decor.size=(3.5mm)}{v4,v5}
				\fmffreeze
				\fmf{mlarrowd}{v4,v1}
				\fmf{mlarrowd}{v5,v2}
			\end{fmfgraph*}
		\end{gathered}
	\end{align*}
	\caption{Topologies relevant for the $\bar q q GG$ four-point function: we do not show crossed diagrams or diagrams with reversed fermion flow. In diagrams with gray blobs, some quark propagators can be replaced by flow lines, according to the gradient-flow Feynman rules: the gray blobs denote either QCD or flow vertices. In total, there are 76 diagrams.}
	\label{fig:DiagramsqqGG}
\end{figure}
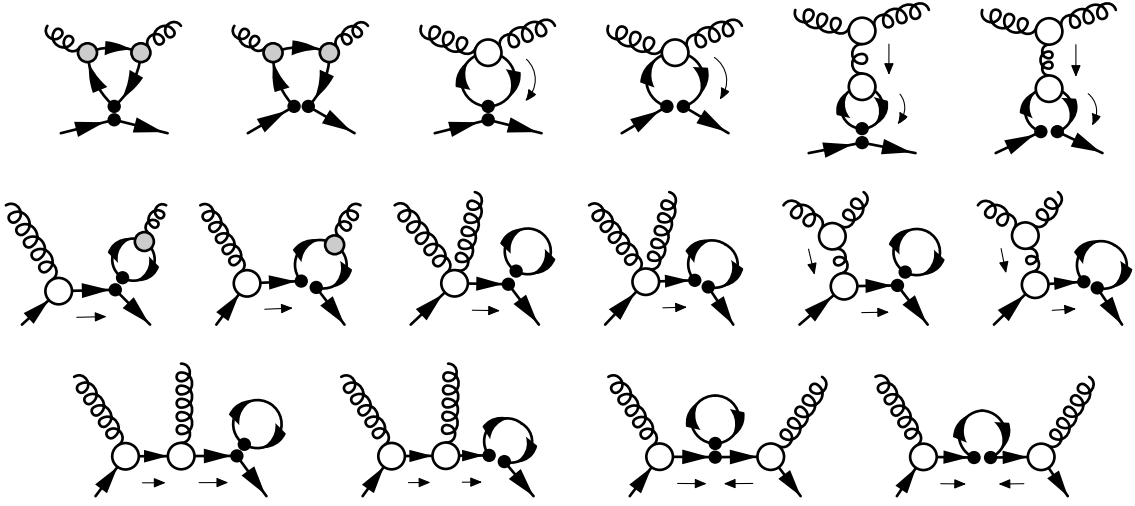

The $\bar q q$ two-point and $\bar q q G$ three-point functions do not fix the coefficients of all the nuisance operators, but they allow us to uniquely determine the coefficient of the physical dipole operator. We checked that in combination with the $\bar q q GG$ four-point function, we obtain a consistent set of matching equations that uniquely determines all coefficients. The topologies for the $\bar q q GG$ four-point function are shown in Fig.~\ref{fig:DiagramsqqGG}.\footnote{We did not consider the $\bar q q GGG$ five-point function, which would provide a further cross check but would require the evaluation of 764 diagrams.}

The results for the matching coefficients of the chromo-electric dipole operator read
\begin{align}
	c_{S1,CE}^{pr}(t,\mu) &= \frac{i m_r}{16\pi^2} \left[ \Big( \log(8\pi\mu^2 t) + 2 \Big) \delta_{pr} - 2 N_c \right] + \O(\alpha_s,t) \, , \nn
	c_{S8,CE}^{pr}(t,\mu) &= \frac{i m_r}{32\pi^2} \left[ \frac{1}{N_c} \Big( \log(8\pi\mu^2 t) + 1 \Big) - 2 C_F \right] \delta_{pr} + \O(\alpha_s,t) \, , \nn
	c_{S1,CE'}^{pr}(t,\mu) &= c_{S8,CE'}^{pr}(t,\mu) = \O(\alpha_s,t) \, , \nn
	c_{T1,CE}^{pr}(t,\mu) &= c_{T1,CE'}^{pr}(t,\mu) = -\frac{i m_r}{8\pi^2} \Big( \log(8\pi\mu^2 t) - 2 \Big) \delta_{pr} + \O(\alpha_s,t) \, , \nn
	c_{T8,CE}^{pr}(t,\mu) &= \frac{i m_r}{8\pi^2} \left[ \Big( \log(8\pi\mu^2 t) + 1 \Big) \left( 1 - \frac{\delta_{pr}}{2N_c} \right) - 3 C_F \delta_{pr} \right] + \O(\alpha_s,t) \, , \nn
	c_{T8,CE'}^{pr}(t,\mu) &= \frac{i m_p}{8\pi^2} \left[ \Big( \log(8\pi\mu^2 t) + 1 \Big) \left( 1 - \frac{\delta_{pr}}{2N_c}  \right) - 3 C_F \delta_{pr} \right] + \O(\alpha_s,t) \, ,
\end{align}
whereas for the coefficients of the electric dipole operator, we obtain
\begin{align}
	c_{S1,E}^{pr}(t,\mu) &= \frac{i m_r}{16\pi^2} \left[ \Big( \log(8\pi\mu^2 t) + 2 \Big) \delta_{pr} - 2 N_c \right] + \O(\alpha_s,t) \, , \nn
	c_{S8,E}^{pr}(t,\mu) &= -\frac{i m_r C_F}{16\pi^2} \Big( \log(8\pi\mu^2 t) + 2 \Big) \delta_{pr} + \O(\alpha_s,t) \, , \nn
	c_{S1,E'}^{pr}(t,\mu) &= c_{S8,E'}^{pr}(t,\mu) = \O(\alpha_s,t) \, , \nn
	c_{T1,E}^{pr}(t,\mu) &= -\frac{i m_r}{8\pi^2} \left[ \Big( \log(8\pi\mu^2 t) + 1 \Big) \left( 2N_c + \delta_{pr} \right) - 3 \delta_{pr} \right] + \O(\alpha_s,t) \, , \nn
	c_{T1,E'}^{pr}(t,\mu) &= -\frac{i m_p}{8\pi^2} \left[ \Big( \log(8\pi\mu^2 t) + 1 \Big) \left( 2N_c + \delta_{pr}  \right) - 3 \delta_{pr} \right] + \O(\alpha_s,t) \, , \nn
	c_{T8,E}^{pr}(t,\mu) &= c_{T8,E'}^{pr}(t,\mu) = \frac{i m_r C_F}{8\pi^2} \Big( \log(8\pi\mu^2 t) - 2 \Big) \delta_{pr} + \O(\alpha_s,t) \, .
\end{align}
The coefficients of the dipole operators are always proportional to a mass insertion, which again follows from chirality. Therefore, these matching coefficients do not contain any power divergences.

As in the case of the pseudoscalar density, we list contributions that only appear for single-flavor operators (and hence are of no relevance in the case of tensor operators), in order to enable the reconstruction of the case of generic flavors: as before, a flowed four-quark operator $\O^X_{prst}(t)$ matches onto generic dipole operators $\O^{E,CE}_{pr}$ with a coefficient proportional to $\delta_{st}$, onto dipole operators $\O^{E,CE}_{st}$ with a coefficient proportional to $\delta_{pr}$, as well as onto dipoles $\O^{E,CE}_{pt}$ and $\O^{E,CE}_{sr}$ with identical coefficients times $\delta_{sr}$ and $\delta_{pt}$, respectively.
We note that in the case of general flavors, a gluon EOM operator as in Eq.~\eqref{eq:GluonEOMOperator} is generated, leading to a penguin contribution to the four-fermion matching, which is absent for flavor-conserving operators relevant to the nEDM.

\subsubsection{Matching coefficients of four-quark operators}

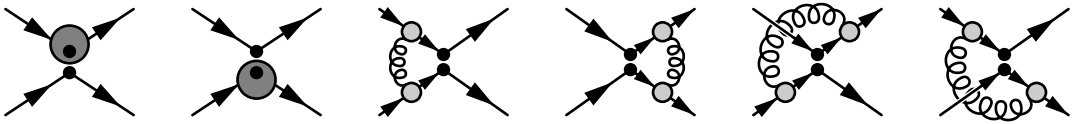
\begin{figure}[t]
	\begin{align*}
		\\[-1cm]
		\begin{gathered}
			\begin{fmfgraph*}(60,40)
				\fmfleft{i1,i2}
				\fmfright{o1,o2}
				\fmf{phantom}{i1,v3,o1}
				\fmf{phantom}{i2,v4,o2}
				\fmf{phantom, tension=2}{v3,v4}
				\fmf{quark,tension=1}{i1,v1,o1}
				\fmf{quark,tension=1}{i2,v2,o2}
				\fmf{phantom, tension=4}{v1,v2}
				\fmfv{decor.shape=circle, decor.filled=50, decor.size=(5mm)}{v4}
				\fmfdot{v1}
				\fmfdot{v2}
			\end{fmfgraph*}
		\end{gathered}
		\quad
		\begin{gathered}
			\begin{fmfgraph*}(60,40)
				\fmfleft{i1,i2}
				\fmfright{o1,o2}
				\fmf{phantom}{i1,v3,o1}
				\fmf{phantom}{i2,v4,o2}
				\fmf{phantom, tension=2}{v3,v4}
				\fmf{quark,tension=1}{i1,v1,o1}
				\fmf{quark,tension=1}{i2,v2,o2}
				\fmf{phantom, tension=4}{v1,v2}
				\fmfv{decor.shape=circle, decor.filled=50, decor.size=(5mm)}{v3}
				\fmfdot{v1}
				\fmfdot{v2}
			\end{fmfgraph*}
		\end{gathered}
		\quad
		\begin{gathered}
			\begin{fmfgraph*}(60,40)
				\fmfleft{i1,i2}
				\fmfright{o1,o2}
				\fmf{quark,tension=2}{i1,v3,v1}
				\fmf{quark,tension=1}{v1,o1}
				\fmf{quark,tension=2}{i2,v4,v2}
				\fmf{quark,tension=1}{v2,o2}
				\fmf{phantom, tension=6}{v1,v2}
				\fmffreeze
				\fmf{gluon,right=0.35}{v4,v3}
				\fmfv{decor.shape=circle, decor.filled=20, decor.size=(2.5mm)}{v3,v4}
				\fmfdot{v1}
				\fmfdot{v2}
			\end{fmfgraph*}
		\end{gathered}
		\quad
		\begin{gathered}
			\begin{fmfgraph*}(60,40)
				\fmfleft{i1,i2}
				\fmfright{o1,o2}
				\fmf{quark,tension=1}{i1,v1}
				\fmf{quark,tension=2}{v1,v4,o1}
				\fmf{quark,tension=1}{i2,v2}
				\fmf{quark,tension=2}{v2,v3,o2}
				\fmf{phantom, tension=6}{v1,v2}
				\fmffreeze
				\fmf{gluon,right=0.35}{v4,v3}
				\fmfv{decor.shape=circle, decor.filled=20, decor.size=(2.5mm)}{v3,v4}
				\fmfdot{v1}
				\fmfdot{v2}
			\end{fmfgraph*}
		\end{gathered}
		\quad
		\begin{gathered}
			\begin{fmfgraph*}(60,40)
				\fmfleft{i1,i2}
				\fmfright{o1,o2}
				\fmf{quark,tension=2}{i1,v3,v1}
				\fmf{quark,tension=1}{v1,o1}
				\fmf{phantom,tension=1}{i2,v2}
				\fmf{quark,tension=2}{v2,v4,o2}
				\fmf{phantom, tension=6}{v1,v2}
				\fmffreeze
				\fmf{gluon,right=1}{v4,v3}
				\fmf{plain,tension=1,rubout}{i2,v6}
				\fmf{phantom,tension=2}{v6,v2}
				\fmf{phantom,tension=1}{i2,v5}
				\fmf{quark,tension=1.5}{v5,v2}
				\fmfv{decor.shape=circle, decor.filled=20, decor.size=(2.5mm)}{v3,v4}
				\fmfdot{v1}
				\fmfdot{v2}
			\end{fmfgraph*}
		\end{gathered}
		\quad
		\begin{gathered}
			\begin{fmfgraph*}(60,40)
				\fmfleft{i2,i1}
				\fmfright{o2,o1}
				\fmf{quark,tension=2}{i1,v3,v1}
				\fmf{quark,tension=1}{v1,o1}
				\fmf{phantom,tension=1}{i2,v2}
				\fmf{quark,tension=2}{v2,v4,o2}
				\fmf{phantom, tension=6}{v1,v2}
				\fmffreeze
				\fmf{gluon,right=1}{v3,v4}
				\fmf{plain,tension=1,rubout}{i2,v6}
				\fmf{phantom,tension=2}{v6,v2}
				\fmf{phantom,tension=1}{i2,v5}
				\fmf{quark,tension=1.5}{v5,v2}
				\fmfv{decor.shape=circle, decor.filled=20, decor.size=(2.5mm)}{v3,v4}
				\fmfdot{v1}
				\fmfdot{v2}
			\end{fmfgraph*}
		\end{gathered}
	\end{align*}
	\caption{Topologies relevant for the $qq\to qq$ four-point function: in the first two diagrams, the dark-gray blob denotes the sum of all sub-diagrams contributing to the insertion of a flowed quark bilinear, shown in Fig.~\ref{fig:DiagramsBilinears}. In the remaining diagrams, some quark propagators can be replaced by flow lines, according to the gradient-flow Feynman rules: the gray blobs denote either QCD or flow vertices. We do not show crossed diagrams. In total, there are 72 diagrams.}
	\label{fig:Diagrams4Quark}
\end{figure}

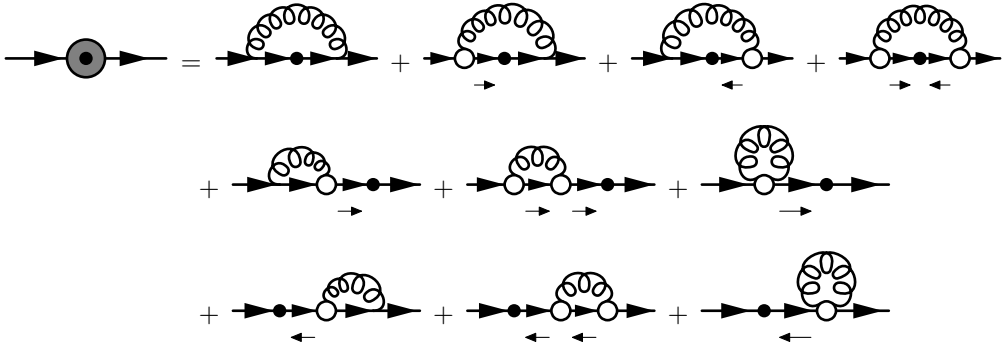
\begin{figure}[t]
	\begin{align*}
		\begin{gathered}
			\begin{fmfgraph*}(60,40)
				\fmfleft{l1} \fmfright{r1}
				\fmf{phantom}{l1,v1,r1}
				\fmf{quark}{l1,v2,r1}
				\fmffreeze
				\fmfv{decor.shape=circle, decor.filled=50, decor.size=(5mm)}{v1}
				\fmfdot{v2}
			\end{fmfgraph*}
		\end{gathered}
		&\;=\;
		\begin{gathered}
			\begin{fmfgraph*}(60,40)
				\fmfleft{l1} \fmfright{r1}
				\fmf{quark}{l1,v1,v2,v3,r1}
				\fmffreeze
				\fmf{gluon,right}{v3,v1}
				\fmfdot{v2}
			\end{fmfgraph*}
		\end{gathered}
		\;+\;
		\begin{gathered}
			\begin{fmfgraph*}(60,40)
				\fmfleft{l1} \fmfright{r1}
				\fmf{quark}{l1,v1,v2,v3,r1}
				\fmffreeze
				\fmf{gluon,right}{v3,v1}
				\fmf{phantom}{l1,v4,v5,v6,r1}
				\fmffreeze
				\fmf{marrowd}{v4,v5}
				\fmfdot{v2}
				\fmfv{decor.shape=circle, decor.filled=empty, decor.size=(2.5mm)}{v1}
			\end{fmfgraph*}
		\end{gathered}
		\;+\;
		\begin{gathered}
			\begin{fmfgraph*}(60,40)
				\fmfleft{l1} \fmfright{r1}
				\fmf{quark}{l1,v1,v2,v3,r1}
				\fmffreeze
				\fmf{gluon,right}{v3,v1}
				\fmf{phantom}{l1,v4,v5,v6,r1}
				\fmffreeze
				\fmf{marrowd}{v6,v5}
				\fmfdot{v2}
				\fmfv{decor.shape=circle, decor.filled=empty, decor.size=(2.5mm)}{v3}
			\end{fmfgraph*}
		\end{gathered}
		\;+\;
		\begin{gathered}
			\begin{fmfgraph*}(60,40)
				\fmfleft{l1} \fmfright{r1}
				\fmf{quark}{l1,v1,v2,v3,r1}
				\fmffreeze
				\fmf{gluon,right}{v3,v1}
				\fmf{phantom}{l1,v4,v5,v6,r1}
				\fmffreeze
				\fmf{marrowd}{v4,v5}
				\fmf{marrowd}{v6,v5}
				\fmfdot{v2}
				\fmfv{decor.shape=circle, decor.filled=empty, decor.size=(2.5mm)}{v1,v3}
			\end{fmfgraph*}
		\end{gathered}
		\\
		&\quad+\;
		\begin{gathered}
			\begin{fmfgraph*}(70,40)
				\fmfleft{l1} \fmfright{r1}
				\fmf{quark}{l1,v1,v2,v3,r1}
				\fmffreeze
				\fmf{gluon,right}{v2,v1}
				\fmf{phantom}{l1,v4,v5,v6,r1}
				\fmffreeze
				\fmf{marrowd}{v5,v6}
				\fmfdot{v3}
				\fmfv{decor.shape=circle, decor.filled=empty, decor.size=(2.5mm)}{v2}
			\end{fmfgraph*}
		\end{gathered}
		\;+\;
		\begin{gathered}
			\begin{fmfgraph*}(70,40)
				\fmfleft{l1} \fmfright{r1}
				\fmf{quark}{l1,v1,v2,v3,r1}
				\fmffreeze
				\fmf{gluon,right}{v2,v1}
				\fmf{phantom}{l1,v4,v5,v6,r1}
				\fmffreeze
				\fmf{marrowd}{v4,v5,v6}
				\fmfdot{v3}
				\fmfv{decor.shape=circle, decor.filled=empty, decor.size=(2.5mm)}{v1,v2}
			\end{fmfgraph*}
		\end{gathered}
		\;+\;
		\begin{gathered}
			\begin{fmfgraph*}(70,40)
				\fmfleft{l1} \fmfright{r1}
				\fmf{quark}{l1,v1,v2,r1}
				\fmffreeze
				\fmf{gluon,right}{v1,v1}
				\fmf{phantom}{l1,v4,v5,r1}
				\fmffreeze
				\fmf{marrowd}{v4,v5}
				\fmfdot{v2}
				\fmfv{decor.shape=circle, decor.filled=empty, decor.size=(2.5mm)}{v1}
			\end{fmfgraph*}
		\end{gathered}
		\\
		&\quad+\;
		\begin{gathered}
			\begin{fmfgraph*}(70,40)
				\fmfleft{l1} \fmfright{r1}
				\fmf{quark}{l1,v1,v2,v3,r1}
				\fmffreeze
				\fmf{gluon,right}{v3,v2}
				\fmf{phantom}{l1,v4,v5,v6,r1}
				\fmffreeze
				\fmf{marrowd}{v5,v4}
				\fmfdot{v1}
				\fmfv{decor.shape=circle, decor.filled=empty, decor.size=(2.5mm)}{v2}
			\end{fmfgraph*}
		\end{gathered}
		\;+\;
		\begin{gathered}
			\begin{fmfgraph*}(70,40)
				\fmfleft{l1} \fmfright{r1}
				\fmf{quark}{l1,v1,v2,v3,r1}
				\fmffreeze
				\fmf{gluon,right}{v3,v2}
				\fmf{phantom}{l1,v4,v5,v6,r1}
				\fmffreeze
				\fmf{marrowd}{v6,v5,v4}
				\fmfdot{v1}
				\fmfv{decor.shape=circle, decor.filled=empty, decor.size=(2.5mm)}{v2,v3}
			\end{fmfgraph*}
		\end{gathered}
		\;+\;
		\begin{gathered}
			\begin{fmfgraph*}(70,40)
				\fmfleft{l1} \fmfright{r1}
				\fmf{quark}{l1,v1,v2,r1}
				\fmffreeze
				\fmf{gluon,right}{v2,v2}
				\fmf{phantom}{l1,v4,v5,r1}
				\fmffreeze
				\fmf{marrowd}{v5,v4}
				\fmfdot{v1}
				\fmfv{decor.shape=circle, decor.filled=empty, decor.size=(2.5mm)}{v2}
			\end{fmfgraph*}
		\end{gathered} \\[-1cm]
	\end{align*}
	\caption{Gradient-flow diagrams for the matching of quark bilinears.}
	\label{fig:DiagramsBilinears}
\end{figure}

The $q_p q_r \to q_p q_r$ four-point function allows us to extract the coefficients of the MS four-quark operators in the SFTE. In the matching between operators of equal mass dimension, the expansion in the IR scales amounts to setting external momenta and quark masses to zero. The matching of the four-point function involves 72 diagrams with insertions of the flowed operators, shown in Fig.~\ref{fig:Diagrams4Quark} (in the case of scalar operators, there are two different insertions in each diagram). A subset of the diagrams can be related to the matching of flowed quark-bilinear operators, shown in Fig.~\ref{fig:DiagramsBilinears}. Collecting the results for the NDR and HV schemes in one expression, we obtain for the matching coefficients in the SFTE of the flowed scalar singlet operator
\begin{align}
	\label{eq:cijS1}
	c_{S1,S1}(t,\mu) &= \zeta_\chi^{-2} - \frac{\alpha_s C_F}{\pi} \left[ 3\log(8 \pi \mu^2 t) + 3 - 2 \delta_\mathrm{NDR} \right] + \O(\alpha_s^2,t) \, , \nn
	c_{S1,S1'}(t,\mu) &= \O(\alpha_s^2,t) \, , \nn
	c_{S1,S8}(t,\mu) &= - \frac{ \alpha_s}{4 \pi} \left[ 2 - \delta_\mathrm{NDR} (1-2 \tilde a_\mathrm{ev}) \right] + \O(\alpha_s^2,t) \, , \nn
	c_{S1,S8'}(t,\mu) &= \frac{\alpha_s}{4 \pi} \delta_\mathrm{NDR} (1+2 \tilde a_\mathrm{ev}) + \O(\alpha_s^2,t) \, , \nn
	c_{S1,T1}(t,\mu) &= \O(\alpha_s^2,t) \, , \nn
	c_{S1,T8}(t,\mu) &= \frac{\alpha_s}{8 \pi} \left[ 2\log(8 \pi \mu^2 t) + 3 \right] + \O(\alpha_s^2,t) \, .
\end{align}
The matching coefficients in the SFTE of the flowed scalar octet operator are given by
\begin{align}
	\label{eq:cijS8}
	c_{S8,S1}(t,\mu) &= - \frac{\alpha_s}{8 \pi} \frac{C_F}{N_c} \left[ 2 - \delta_\mathrm{NDR} \left(1 - 2 \tilde a_\mathrm{ev}\right) \right] + \O(\alpha_s^2,t) \, , \nn
	c_{S8,S1'}(t,\mu) &= \frac{\alpha_s}{8 \pi} \frac{C_F}{N_c} \delta_\mathrm{NDR} \left(1 + 2\tilde a_\mathrm{ev}\right) + \O(\alpha_s^2,t) \, , \nn
	c_{S8,S8}(t,\mu) &= \zeta_\chi^{-2} - \frac{\alpha_s}{4\pi N_c} \left[ 3 (N_c^2-2)\log(8 \pi \mu^2 t) + \frac{N_c^2-8}{2} + \delta_{\mathrm{NDR}}\left(\frac{N_c^2}{4}(1-2\tilde{a}_{\mathrm{ev}}) + 3+2\tilde{a}_{\mathrm{ev}} \right) \right] \nn*
		&\quad + \O(\alpha_s^2,t) \, , \nn
	c_{S8,S8'}(t,\mu) &= - \frac{\alpha_s}{16 \pi} \frac{N_c^2-4}{N_c} \delta_\mathrm{NDR} (1+2\tilde a_\mathrm{ev}) + \O(\alpha_s^2,t) \, , \nn
	c_{S8,T1}(t,\mu) &= \frac{\alpha_s}{16 \pi} \frac{C_F}{N_c} \left[ 2 \log(8 \pi \mu^2 t) + 3 \right] + \O(\alpha_s^2,t) \, , \nn
	c_{S8,T8}(t,\mu) &= - \frac{\alpha_s}{32 \pi} \frac{N_c^2-4}{N_c} \left[ 2 \log(8 \pi \mu^2 t) + 3 \right] + \O(\alpha_s^2,t) \, .
\end{align}
For the flowed tensor singlet operator, we find
\begin{align}
	\label{eq:cijT1}
	c_{T1,S1}(t,\mu) &= c_{T1,S1'}(t,\mu) = \O(\alpha_s^2,t) \, , \nn
	c_{T1,S8}(t,\mu) &= c_{T1,S8'}(t,\mu) = \frac{\alpha_s }{\pi} \left[ 3 (2 \log(8 \pi \mu^2 t) + 3) + 2\delta_\mathrm{NDR}(\tilde d_\mathrm{ev} - 7 \tilde f_\mathrm{ev} + 1) \right] + \O(\alpha_s^2,t) \, , \nn
	c_{T1,T1}(t,\mu) &= \zeta_\chi^{-2} - \frac{\alpha_s C_F }{\pi} \log(8 \pi \mu^2 t) + \O(\alpha_s^2,t) \, , \nn*
	c_{T1,T8}(t,\mu) &= - \frac{\alpha_s}{2 \pi} \left[ 1 + \delta_{\mathrm{NDR}} (5 - 4 \tilde{e}_{\mathrm{ev}}) \right] + \O(\alpha_s^2,t) \, ,
\end{align}
whereas for the coefficients in the SFTE of the flowed tensor octet operator, we obtain
\begin{align}
	\label{eq:cijT8}
	c_{T8,S1}(t,\mu) &= c_{T8,S1'}(t,\mu) = \frac{\alpha_s}{2 \pi} \frac{C_F}{N_c} \left[ 3  (2\log(8 \pi \mu^2 t) + 3) + 2\delta_\mathrm{NDR}(\tilde d_\mathrm{ev} - 7 \tilde f_\mathrm{ev} + 1)  \right] + \O(\alpha_s^2,t) \, , \nn
	c_{T8,S8}(t,\mu) &= c_{T8,S8'}(t,\mu) = - \frac{\alpha_s}{4 \pi} \frac{N_c^2-4}{N_c} \left[ 3  (2\log(8 \pi \mu^2 t) + 3) + 2\delta_\mathrm{NDR}(\tilde d_\mathrm{ev} - 7 \tilde f_\mathrm{ev} + 1)  \right] \nn
		&\qquad\qquad\qquad\qquad + \O(\alpha_s^2,t) \, , \nn
	c_{T8,T1}(t,\mu) &= - \frac{\alpha_s}{4 \pi}\frac{C_F}{N_c} \left[ 1 + \delta_\mathrm{NDR}(5 - 4 \tilde e_\mathrm{ev}) \right]  + \O(\alpha_s^2,t) \, , \nn
	c_{T8,T8}(t,\mu) &= \zeta_\chi^{-2} - \frac{\alpha_s}{8 \pi N_c} \left[2(5 N_c^2-2) \log(8 \pi \mu^2 t) + 7N_c^2+4 + \delta_\mathrm{NDR} ( 4 \tilde e_\mathrm{ev} (N_c^2 - 4) - 9 N_c^2 + 20 ) \right] \nn
		&\quad + \O(\alpha_s^2,t) \, .
\end{align}
At one-loop order, the finite renormalization $\zeta_\chi$ only affects the diagonal matching coefficients, which are equal to 1 at tree level.

Due to the $SU(N_c)$ Fierz relation~\eqref{eq:SUNFierz}, our results only hold for the color gauge group $SU(N_c)$: we use Eq.~\eqref{eq:QuadraticCasimirs} to write the results in a compact form.\footnote{This does not imply that $C_F$ arises in all places directly from the product of two generators $-t^a t^a$.}

\subsection{Anomalous axial rotations}

As noted above, the pseudoscalar density~\eqref{eq:PseudoscalarDensity} can be removed by an anomalous axial field redefinition. This amounts to replacing the pseudoscalar density by an EOM operator~\cite{Bhattacharya:2015rsa}. In the HV scheme, one finds the following relation of bare operators:
\begin{align}
	\N_p' := \bar q_{Ep} \gamma_5 q_p + \bar q_p \gamma_5 q_{Ep} &= 2 m_p \O^P_p + \bar q_p \overleftrightarrow{\hat{\slashed D}} \gamma_5 q_p - \p_\mu \left( \bar q_p \bar\gamma_\mu \gamma_5 q_p \right) \, .
\end{align}
Our matching results are given in the basis specified in Sect.~\ref{sec:Operators}, i.e., we do not include the EOM operator $\N_p'$. Instead, one could eliminate the pseudoscalar density in favor of the EOM operator~$\N_p'$. This would result in a shift in the evanescent anomaly operator $\bar q_p \overleftrightarrow{\hat{\slashed D}} \gamma_5 q_p$, which is only visible in the HV scheme. The requirement that renormalized evanescent operators have vanishing matrix elements induces then a finite renormalization of the theta term. This is a manifestation of the anomaly in dimensional regularization, where the determinant of field redefinitions in the path integral is always trivial.

In the presence of $CP$-odd operators that also violate chiral symmetry, physical $CP$ violation arises only if complex phases remain in the Lagrangian after vacuum alignment~\cite{Dashen:1970et}. Here, we assume that the dominant source of chiral symmetry breaking is given by non-vanishing mass matrices. A detailed discussion of vacuum alignment can be found in Ref.~\cite{Bhattacharya:2015rsa}.

\subsection{Impact of higher-order effects}

\begin{figure}
	\centering
	\footnotesize
	\scalebox{0.8}{
	\input{images/matchingCoeffS1HV}
	\input{images/matchingCoeffS8HV}}
	\\[-0.2cm]
	\scalebox{0.8}{
	\input{images/matchingCoeffT1HV}
	\input{images/matchingCoeffT8HV}}
	\\[-0.4cm]
	\normalsize
	\input{images/label}
	\\[-0.4cm]
	\caption{Scale dependence of the diagonal four-quark matching coefficients in the HV scheme, evaluated at $\mu_0$. The solid blue curves are for a fixed value of $\alpha_s(\bar{\mu}_0)$ while the dashed red curves are for $\alpha_s$ evaluated according to Eq.~\eqref{eq:alphaScaleDependence}. Flowed operators are defined in terms of ringed fields. As input for $\alpha_s$ we use the solution of the QCD $\beta$ function at two loops.}
	\label{fig:ScalePlotsHV}
\end{figure}
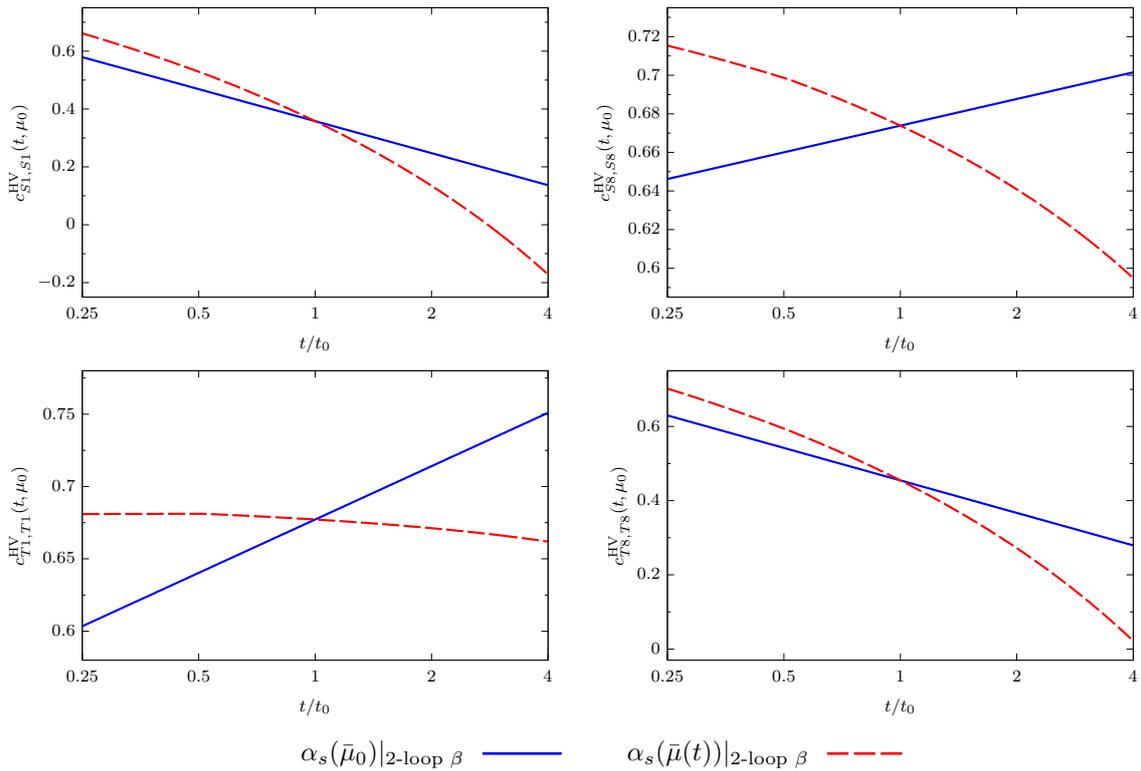

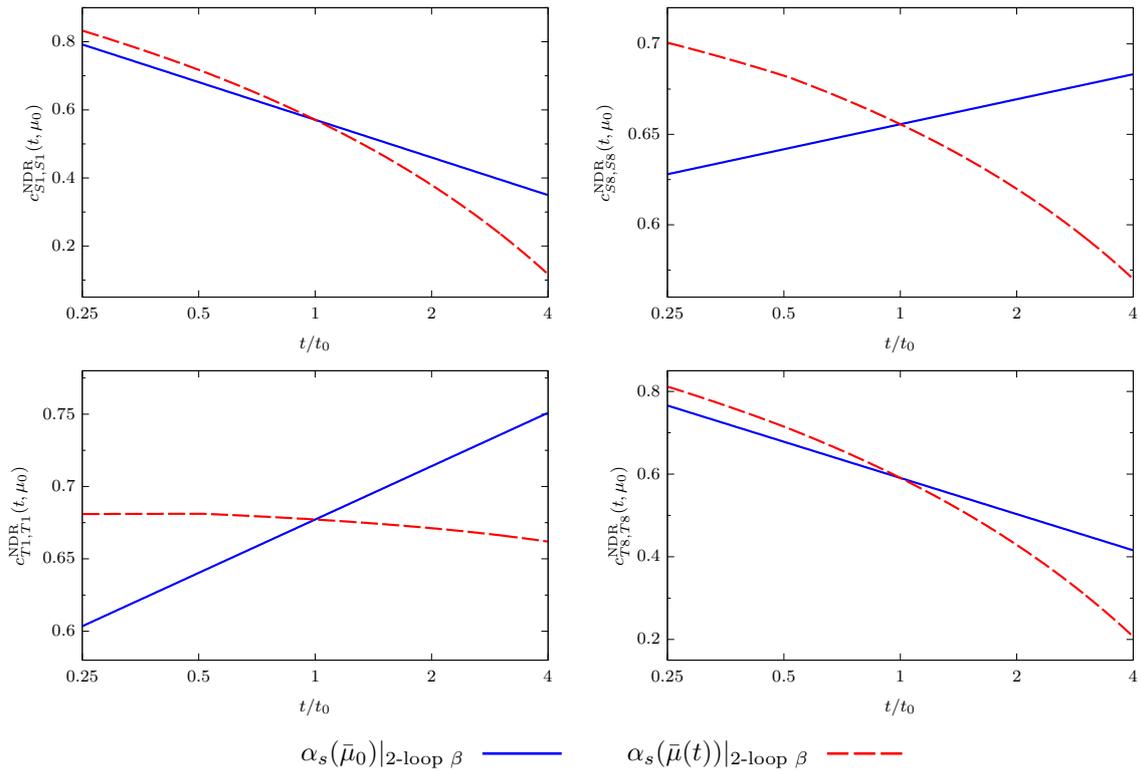
\begin{figure}
	\centering
	\footnotesize
	\scalebox{0.8}{
	\input{images/matchingCoeffS1NDR}
	\input{images/matchingCoeffS8NDR}}
	\\[-0.2cm]
	\scalebox{0.8}{
	\input{images/matchingCoeffT1NDR}
	\input{images/matchingCoeffT8NDR}}
	\\[-0.4cm]
	\normalsize
	\input{images/label}
	\\[-0.4cm]
	\caption{Scale dependence of the diagonal four-quark matching coefficients, evaluated at $\mu_0$ and in the NDR scheme for $\tilde{a}_{\mathrm{ev}}=\ldots =\tilde{f}_{\mathrm{ev}}=1$.}
	\label{fig:ScalePlotsNDR}
\end{figure}

In the following, we evaluate the matching coefficients numerically and estimate their perturbative uncertainty. 

The matching coefficients depend on the flow time $t$ and the MS renormalization scale $\mu$ through logarithms $\log(8 \pi \mu^2 t)$, which are dictated by the anomalous dimensions of the operators. For a given flow time $t_0$, the matching should be performed at an MS scale close to
\begin{align}
	\label{eq:MatchingScale}
	\mu_0 = \frac{1}{\sqrt{8 \pi t_0}} \, ,
\end{align}
in order to avoid large logarithms in the matching. As pointed out in Sect.~\ref{sec:GradientFlow}, this typically corresponds to a reference scale of $\bar\mu_0 \approx 3\GeV$ in the \msbar{} scheme, or $\mu_0 \approx1.13\GeV$ in the MS scheme.

An additional scale dependence of the matching coefficients arises through the strong coupling constant $\alpha_s(\bar\mu)$, which however is beyond the control of our one-loop calculation. As in Ref.~\cite{Mereghetti:2021nkt}, we use this residual scale dependence as an estimate of higher-order corrections to the matching coefficients. To this end, we evaluate the matching coefficients at $\mu_0 = 1.13\GeV$ and around $t_0$ in the range $t \in \left[ (1/4) t_0, 4t_0\right]$. The results for the diagonal matching coefficients $c_{X,X}(t,\mu_0)$ are shown in Figs.~\ref{fig:ScalePlotsHV} and~\ref{fig:ScalePlotsNDR} for the HV and NDR schemes, respectively. The blue curves are for fixed $\alpha_s=\alpha_s(\bar{\mu}_0)$, whereas for the red curves the coupling is evaluated at
\begin{align}
	\label{eq:alphaScaleDependence}
	\alpha_s(\bar{\mu}(t)) = \alpha_s \left( \frac{1}{\sqrt{8 \pi t}} \frac{(4 \pi)^{1/2}}{e^{\gamma_E/2}} \right) \, .
\end{align}
The scale dependence of the coupling is determined with the two-loop QCD $\beta$-function~\cite{Caswell:1974gg,Jones:1974mm,Egorian:1978zx,Tarasov:1980au,Larin:1993tp} and the input value at the weak scale $\alpha_s(M_Z) = 0.1179$~\cite{ParticleDataGroup:2020ssz}. The maximal difference between blue and red curves in the considered range illustrates the residual scale dependence and gives a very rough estimate of higher-order corrections to the matching coefficients: we expect the one-loop contribution to the matching coefficients to have a relative perturbative uncertainty of
\begin{align}
	\label{eq:PerturbativeUncertaintyEstimate}
	1 - \frac{\alpha_s(\bar{\mu}(4 t_0))}{\alpha_s(\bar{\mu}_0)} \approx 40\% \, .
\end{align}
This estimate is based solely on the logarithmic terms of higher order in $\alpha_s$ that arise from the RG evolution of the coupling. For the matching coefficient, one is mostly interested in the finite contributions, as all logarithmic contributions vanish at the matching scale~\eqref{eq:MatchingScale}. Obtaining the actual higher-order matching corrections requires a proper two-loop matching calculation, which is beyond the scope of this paper.

We observe that compared to the tree-level matching the one-loop contributions provide corrections of the order of $30\% - 60\%$ to the diagonal matching coefficients, while the off-diagonal matching coefficients of course only start at one loop. The numerical values for the four-quark matching coefficients are listed in Table~\ref{tab:FourQuarkMatchingCoefficientsNumerics}, using the two-loop QCD $\beta$ function. We note that in some cases the HV scheme leads to much larger off-diagonal contributions than the NDR scheme. It would be interesting to see if this effect can be traced back to the spurious breaking of chiral symmetry by the regulator~\cite{LEFTHV}.

\begin{table}
\begin{center}
\scalebox{0.85}{
\begin{tabular}{c|cccccc}
\toprule
$c_{ij}^{\mathrm{tree+1L}}$ & $j=S1$ & $j=S1'$ & $j=S8$ & $j=S8'$ & $j=T1$ & $j=T8$ \\ 
\midrule
\midrule
& \multicolumn{6}{c}{HV scheme} \\
\midrule
$i=S1$	&  $0.4(2)$ & $0$ & $-0.04(1)$	& $0$ & $0$ & $0.03(1)$  \\
$i=S8$	&  $-0.009(3)$ & $0$ & $0.7(1)$ & $0$ & $0.007(2)$ &   $-0.013(4)$  \\
$i=T1$	& $0$ & $0$ & $0.7(3)$ & $0.7(3)$ & $0.7(1)$ & $-0.04(1)$ \\
$i=T8$	& $0.16(6)$ & $0.16(6)$ & $-0.3(1)$ & $-0.3(1)$ & $-0.009(3)$ & $0.5(2)$ \\
\midrule
& \multicolumn{6}{c}{NDR scheme} \\
\midrule
$i=S1$	&  $0.6(2)$ & $0$ & $-0.06(2)$	& $0.06(2)$ & $0$ & $0.03(1)$  \\
$i=S8$	&  $-0.013(5)$ & $0.013(5)$ & $0.7(1)$ & $-0.03(1)$ & $0.007(2)$ &   $-0.012(4)$  \\
$i=T1$	& $0$ & $0$ & $-0.08(3)$ & $-0.08(3)$ & $0.7(1)$ & $-0.08(3)$ \\
$i=T8$	& $-0.018(6)$ & $-0.018(6)$ & $0.03(1)$ & $0.03(1)$ & $-0.018(6)$ & $0.6(1)$ \\
\bottomrule
\end{tabular}}
\end{center}
\caption{Numerical values of the four-quark matching coefficients in the HV and NDR schemes, evaluated at $\mu_0, t_0$ and using $\left. \alpha_s(\bar{\mu}_0) \right|_\text{2-loop}$. The errors correspond to the (higher-order) scale dependence of the coupling for $t \in \left[(1/4)t_0, 4 t_0 \right]$. The index $i$ runs over the flowed operators (in terms of ringed fields), while $j$ runs over the MS operators.}
\label{tab:FourQuarkMatchingCoefficientsNumerics}
\end{table}

%% file: images/matchingCoeffS1HV.tex
\begingroup
  \makeatletter
  \providecommand\color[2][]{%
    \GenericError{(gnuplot) \space\space\space\@spaces}{%
      Package color not loaded in conjunction with
      terminal option `colourtext'%
    }{See the gnuplot documentation for explanation.%
    }{Either use 'blacktext' in gnuplot or load the package
      color.sty in LaTeX.}%
    \renewcommand\color[2][]{}%
  }%
  \providecommand\includegraphics[2][]{%
    \GenericError{(gnuplot) \space\space\space\@spaces}{%
      Package graphicx or graphics not loaded%
    }{See the gnuplot documentation for explanation.%
    }{The gnuplot epslatex terminal needs graphicx.sty or graphics.sty.}%
    \renewcommand\includegraphics[2][]{}%
  }%
  \providecommand\rotatebox[2]{#2}%
  \@ifundefined{ifGPcolor}{%
    \newif\ifGPcolor
    \GPcolorfalse
  }{}%
  \@ifundefined{ifGPblacktext}{%
    \newif\ifGPblacktext
    \GPblacktexttrue
  }{}%
  \let\gplgaddtomacro\g@addto@macro
  \gdef\gplbacktext{}%
  \gdef\gplfronttext{}%
  \makeatother
  \ifGPblacktext
    \def\colorrgb#1{}%
    \def\colorgray#1{}%
  \else
    \ifGPcolor
      \def\colorrgb#1{\color[rgb]{#1}}%
      \def\colorgray#1{\color[gray]{#1}}%
      \expandafter\def\csname LTw\endcsname{\color{white}}%
      \expandafter\def\csname LTb\endcsname{\color{black}}%
      \expandafter\def\csname LTa\endcsname{\color{black}}%
      \expandafter\def\csname LT0\endcsname{\color[rgb]{1,0,0}}%
      \expandafter\def\csname LT1\endcsname{\color[rgb]{0,1,0}}%
      \expandafter\def\csname LT2\endcsname{\color[rgb]{0,0,1}}%
      \expandafter\def\csname LT3\endcsname{\color[rgb]{1,0,1}}%
      \expandafter\def\csname LT4\endcsname{\color[rgb]{0,1,1}}%
      \expandafter\def\csname LT5\endcsname{\color[rgb]{1,1,0}}%
      \expandafter\def\csname LT6\endcsname{\color[rgb]{0,0,0}}%
      \expandafter\def\csname LT7\endcsname{\color[rgb]{1,0.3,0}}%
      \expandafter\def\csname LT8\endcsname{\color[rgb]{0.5,0.5,0.5}}%
    \else
      \def\colorrgb#1{\color{black}}%
      \def\colorgray#1{\color[gray]{#1}}%
      \expandafter\def\csname LTw\endcsname{\color{white}}%
      \expandafter\def\csname LTb\endcsname{\color{black}}%
      \expandafter\def\csname LTa\endcsname{\color{black}}%
      \expandafter\def\csname LT0\endcsname{\color{black}}%
      \expandafter\def\csname LT1\endcsname{\color{black}}%
      \expandafter\def\csname LT2\endcsname{\color{black}}%
      \expandafter\def\csname LT3\endcsname{\color{black}}%
      \expandafter\def\csname LT4\endcsname{\color{black}}%
      \expandafter\def\csname LT5\endcsname{\color{black}}%
      \expandafter\def\csname LT6\endcsname{\color{black}}%
      \expandafter\def\csname LT7\endcsname{\color{black}}%
      \expandafter\def\csname LT8\endcsname{\color{black}}%
    \fi
  \fi
    \setlength{\unitlength}{0.0500bp}%
    \ifx\gptboxheight\undefined%
      \newlength{\gptboxheight}%
      \newlength{\gptboxwidth}%
      \newsavebox{\gptboxtext}%
    \fi%
    \setlength{\fboxrule}{0.5pt}%
    \setlength{\fboxsep}{1pt}%
    \definecolor{tbcol}{rgb}{1,1,1}%
\begin{picture}(5400.00,3528.00)%
    \gplgaddtomacro\gplbacktext{%
      \csname LTb\endcsname
      \put(594,730){\makebox(0,0)[r]{\strut{}$-0.2$}}%
      \put(594,1272){\makebox(0,0)[r]{\strut{}$0$}}%
      \put(594,1815){\makebox(0,0)[r]{\strut{}$0.2$}}%
      \put(594,2357){\makebox(0,0)[r]{\strut{}$0.4$}}%
      \put(594,2900){\makebox(0,0)[r]{\strut{}$0.6$}}%
      \put(660,429){\makebox(0,0){\strut{}$0.25$}}%
      \put(1746,429){\makebox(0,0){\strut{}$0.5$}}%
      \put(2832,429){\makebox(0,0){\strut{}$1$}}%
      \put(3917,429){\makebox(0,0){\strut{}$2$}}%
      \put(5003,429){\makebox(0,0){\strut{}$4$}}%
    }%
    \gplgaddtomacro\gplfronttext{%
      \csname LTb\endcsname
      \put(88,1950){\rotatebox{-270}{\makebox(0,0){\strut{}$c_{S1,S1}^\mathrm{HV}(t,\mu_0)$}}}%
      \put(2831,154){\makebox(0,0){\strut{}$t/t_0$}}%
    }%
    \gplbacktext
    \put(0,0){\includegraphics[width={270.00bp},height={176.40bp}]{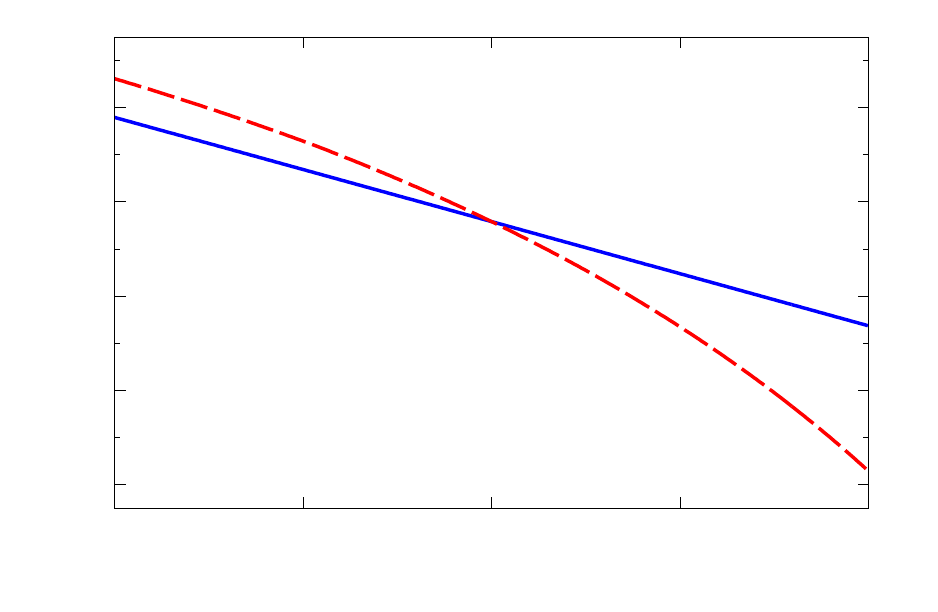}}%
    \gplfronttext
  \end{picture}%
\endgroup

%% file: images/matchingCoeffS8HV.tex
\begingroup
  \makeatletter
  \providecommand\color[2][]{%
    \GenericError{(gnuplot) \space\space\space\@spaces}{%
      Package color not loaded in conjunction with
      terminal option `colourtext'%
    }{See the gnuplot documentation for explanation.%
    }{Either use 'blacktext' in gnuplot or load the package
      color.sty in LaTeX.}%
    \renewcommand\color[2][]{}%
  }%
  \providecommand\includegraphics[2][]{%
    \GenericError{(gnuplot) \space\space\space\@spaces}{%
      Package graphicx or graphics not loaded%
    }{See the gnuplot documentation for explanation.%
    }{The gnuplot epslatex terminal needs graphicx.sty or graphics.sty.}%
    \renewcommand\includegraphics[2][]{}%
  }%
  \providecommand\rotatebox[2]{#2}%
  \@ifundefined{ifGPcolor}{%
    \newif\ifGPcolor
    \GPcolorfalse
  }{}%
  \@ifundefined{ifGPblacktext}{%
    \newif\ifGPblacktext
    \GPblacktexttrue
  }{}%
  \let\gplgaddtomacro\g@addto@macro
  \gdef\gplbacktext{}%
  \gdef\gplfronttext{}%
  \makeatother
  \ifGPblacktext
    \def\colorrgb#1{}%
    \def\colorgray#1{}%
  \else
    \ifGPcolor
      \def\colorrgb#1{\color[rgb]{#1}}%
      \def\colorgray#1{\color[gray]{#1}}%
      \expandafter\def\csname LTw\endcsname{\color{white}}%
      \expandafter\def\csname LTb\endcsname{\color{black}}%
      \expandafter\def\csname LTa\endcsname{\color{black}}%
      \expandafter\def\csname LT0\endcsname{\color[rgb]{1,0,0}}%
      \expandafter\def\csname LT1\endcsname{\color[rgb]{0,1,0}}%
      \expandafter\def\csname LT2\endcsname{\color[rgb]{0,0,1}}%
      \expandafter\def\csname LT3\endcsname{\color[rgb]{1,0,1}}%
      \expandafter\def\csname LT4\endcsname{\color[rgb]{0,1,1}}%
      \expandafter\def\csname LT5\endcsname{\color[rgb]{1,1,0}}%
      \expandafter\def\csname LT6\endcsname{\color[rgb]{0,0,0}}%
      \expandafter\def\csname LT7\endcsname{\color[rgb]{1,0.3,0}}%
      \expandafter\def\csname LT8\endcsname{\color[rgb]{0.5,0.5,0.5}}%
    \else
      \def\colorrgb#1{\color{black}}%
      \def\colorgray#1{\color[gray]{#1}}%
      \expandafter\def\csname LTw\endcsname{\color{white}}%
      \expandafter\def\csname LTb\endcsname{\color{black}}%
      \expandafter\def\csname LTa\endcsname{\color{black}}%
      \expandafter\def\csname LT0\endcsname{\color{black}}%
      \expandafter\def\csname LT1\endcsname{\color{black}}%
      \expandafter\def\csname LT2\endcsname{\color{black}}%
      \expandafter\def\csname LT3\endcsname{\color{black}}%
      \expandafter\def\csname LT4\endcsname{\color{black}}%
      \expandafter\def\csname LT5\endcsname{\color{black}}%
      \expandafter\def\csname LT6\endcsname{\color{black}}%
      \expandafter\def\csname LT7\endcsname{\color{black}}%
      \expandafter\def\csname LT8\endcsname{\color{black}}%
    \fi
  \fi
    \setlength{\unitlength}{0.0500bp}%
    \ifx\gptboxheight\undefined%
      \newlength{\gptboxheight}%
      \newlength{\gptboxwidth}%
      \newsavebox{\gptboxtext}%
    \fi%
    \setlength{\fboxrule}{0.5pt}%
    \setlength{\fboxsep}{1pt}%
    \definecolor{tbcol}{rgb}{1,1,1}%
\begin{picture}(5400.00,3528.00)%
    \gplgaddtomacro\gplbacktext{%
      \csname LTb\endcsname
      \put(594,865){\makebox(0,0)[r]{\strut{}$0.6$}}%
      \put(594,1227){\makebox(0,0)[r]{\strut{}$0.62$}}%
      \put(594,1589){\makebox(0,0)[r]{\strut{}$0.64$}}%
      \put(594,1951){\makebox(0,0)[r]{\strut{}$0.66$}}%
      \put(594,2312){\makebox(0,0)[r]{\strut{}$0.68$}}%
      \put(594,2674){\makebox(0,0)[r]{\strut{}$0.7$}}%
      \put(594,3036){\makebox(0,0)[r]{\strut{}$0.72$}}%
      \put(660,429){\makebox(0,0){\strut{}$0.25$}}%
      \put(1746,429){\makebox(0,0){\strut{}$0.5$}}%
      \put(2832,429){\makebox(0,0){\strut{}$1$}}%
      \put(3917,429){\makebox(0,0){\strut{}$2$}}%
      \put(5003,429){\makebox(0,0){\strut{}$4$}}%
    }%
    \gplgaddtomacro\gplfronttext{%
      \csname LTb\endcsname
      \put(88,1950){\rotatebox{-270}{\makebox(0,0){\strut{}$c_{S8,S8}^\mathrm{HV}(t,\mu_0)$}}}%
      \put(2831,154){\makebox(0,0){\strut{}$t/t_0$}}%
    }%
    \gplbacktext
    \put(0,0){\includegraphics[width={270.00bp},height={176.40bp}]{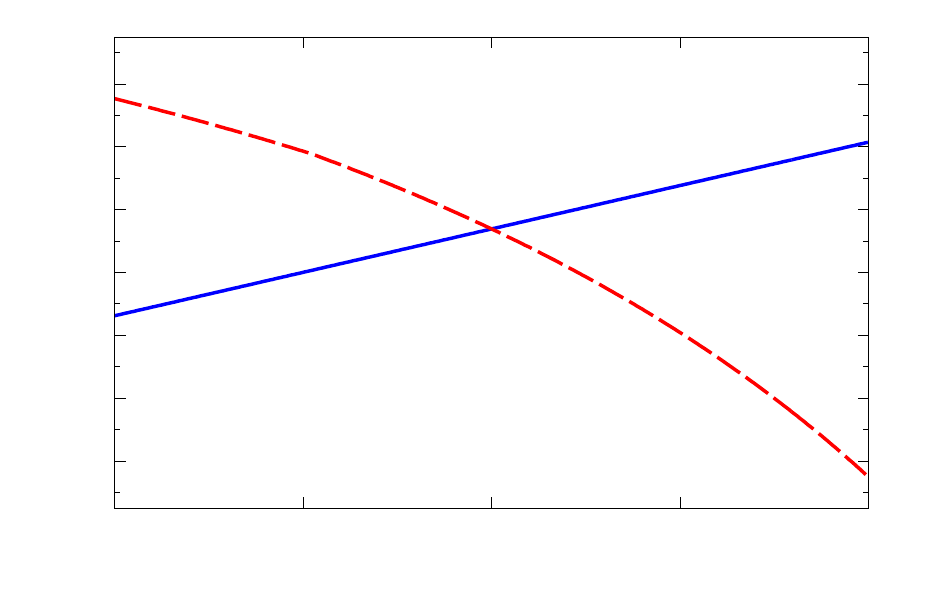}}%
    \gplfronttext
  \end{picture}%
\endgroup

%% file: images/matchingCoeffT1HV.tex
\begingroup
  \makeatletter
  \providecommand\color[2][]{%
    \GenericError{(gnuplot) \space\space\space\@spaces}{%
      Package color not loaded in conjunction with
      terminal option `colourtext'%
    }{See the gnuplot documentation for explanation.%
    }{Either use 'blacktext' in gnuplot or load the package
      color.sty in LaTeX.}%
    \renewcommand\color[2][]{}%
  }%
  \providecommand\includegraphics[2][]{%
    \GenericError{(gnuplot) \space\space\space\@spaces}{%
      Package graphicx or graphics not loaded%
    }{See the gnuplot documentation for explanation.%
    }{The gnuplot epslatex terminal needs graphicx.sty or graphics.sty.}%
    \renewcommand\includegraphics[2][]{}%
  }%
  \providecommand\rotatebox[2]{#2}%
  \@ifundefined{ifGPcolor}{%
    \newif\ifGPcolor
    \GPcolorfalse
  }{}%
  \@ifundefined{ifGPblacktext}{%
    \newif\ifGPblacktext
    \GPblacktexttrue
  }{}%
  \let\gplgaddtomacro\g@addto@macro
  \gdef\gplbacktext{}%
  \gdef\gplfronttext{}%
  \makeatother
  \ifGPblacktext
    \def\colorrgb#1{}%
    \def\colorgray#1{}%
  \else
    \ifGPcolor
      \def\colorrgb#1{\color[rgb]{#1}}%
      \def\colorgray#1{\color[gray]{#1}}%
      \expandafter\def\csname LTw\endcsname{\color{white}}%
      \expandafter\def\csname LTb\endcsname{\color{black}}%
      \expandafter\def\csname LTa\endcsname{\color{black}}%
      \expandafter\def\csname LT0\endcsname{\color[rgb]{1,0,0}}%
      \expandafter\def\csname LT1\endcsname{\color[rgb]{0,1,0}}%
      \expandafter\def\csname LT2\endcsname{\color[rgb]{0,0,1}}%
      \expandafter\def\csname LT3\endcsname{\color[rgb]{1,0,1}}%
      \expandafter\def\csname LT4\endcsname{\color[rgb]{0,1,1}}%
      \expandafter\def\csname LT5\endcsname{\color[rgb]{1,1,0}}%
      \expandafter\def\csname LT6\endcsname{\color[rgb]{0,0,0}}%
      \expandafter\def\csname LT7\endcsname{\color[rgb]{1,0.3,0}}%
      \expandafter\def\csname LT8\endcsname{\color[rgb]{0.5,0.5,0.5}}%
    \else
      \def\colorrgb#1{\color{black}}%
      \def\colorgray#1{\color[gray]{#1}}%
      \expandafter\def\csname LTw\endcsname{\color{white}}%
      \expandafter\def\csname LTb\endcsname{\color{black}}%
      \expandafter\def\csname LTa\endcsname{\color{black}}%
      \expandafter\def\csname LT0\endcsname{\color{black}}%
      \expandafter\def\csname LT1\endcsname{\color{black}}%
      \expandafter\def\csname LT2\endcsname{\color{black}}%
      \expandafter\def\csname LT3\endcsname{\color{black}}%
      \expandafter\def\csname LT4\endcsname{\color{black}}%
      \expandafter\def\csname LT5\endcsname{\color{black}}%
      \expandafter\def\csname LT6\endcsname{\color{black}}%
      \expandafter\def\csname LT7\endcsname{\color{black}}%
      \expandafter\def\csname LT8\endcsname{\color{black}}%
    \fi
  \fi
    \setlength{\unitlength}{0.0500bp}%
    \ifx\gptboxheight\undefined%
      \newlength{\gptboxheight}%
      \newlength{\gptboxwidth}%
      \newsavebox{\gptboxtext}%
    \fi%
    \setlength{\fboxrule}{0.5pt}%
    \setlength{\fboxsep}{1pt}%
    \definecolor{tbcol}{rgb}{1,1,1}%
\begin{picture}(5400.00,3528.00)%
    \gplgaddtomacro\gplbacktext{%
      \csname LTb\endcsname
      \put(594,865){\makebox(0,0)[r]{\strut{}$0.6$}}%
      \put(594,1544){\makebox(0,0)[r]{\strut{}$0.65$}}%
      \put(594,2222){\makebox(0,0)[r]{\strut{}$0.7$}}%
      \put(594,2900){\makebox(0,0)[r]{\strut{}$0.75$}}%
      \put(660,429){\makebox(0,0){\strut{}$0.25$}}%
      \put(1746,429){\makebox(0,0){\strut{}$0.5$}}%
      \put(2832,429){\makebox(0,0){\strut{}$1$}}%
      \put(3917,429){\makebox(0,0){\strut{}$2$}}%
      \put(5003,429){\makebox(0,0){\strut{}$4$}}%
    }%
    \gplgaddtomacro\gplfronttext{%
      \csname LTb\endcsname
      \put(88,1950){\rotatebox{-270}{\makebox(0,0){\strut{}$c_{T1,T1}^\mathrm{HV}(t,\mu_0)$}}}%
      \put(2831,154){\makebox(0,0){\strut{}$t/t_0$}}%
    }%
    \gplbacktext
    \put(0,0){\includegraphics[width={270.00bp},height={176.40bp}]{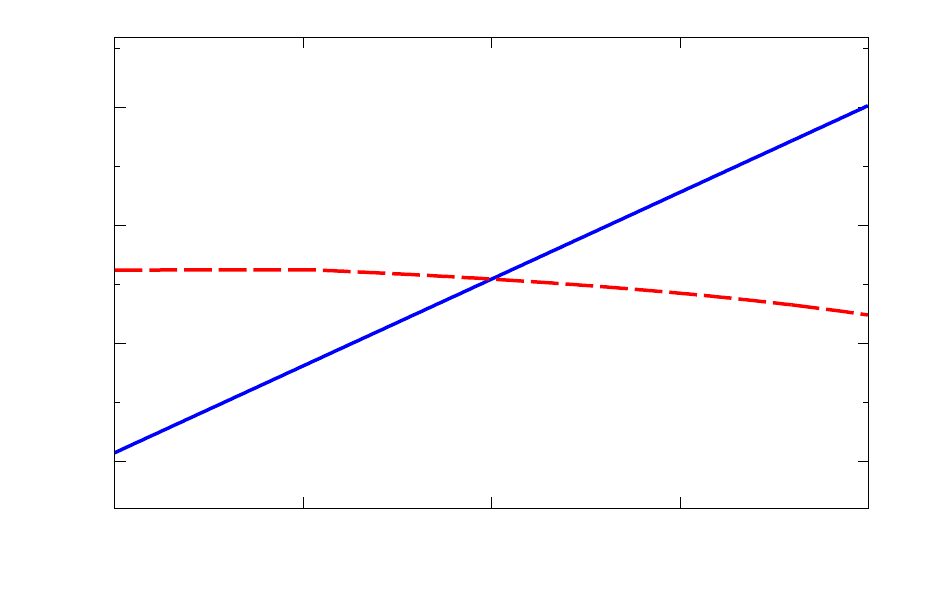}}%
    \gplfronttext
  \end{picture}%
\endgroup

%% file: images/matchingCoeffT8HV.tex
\begingroup
  \makeatletter
  \providecommand\color[2][]{%
    \GenericError{(gnuplot) \space\space\space\@spaces}{%
      Package color not loaded in conjunction with
      terminal option `colourtext'%
    }{See the gnuplot documentation for explanation.%
    }{Either use 'blacktext' in gnuplot or load the package
      color.sty in LaTeX.}%
    \renewcommand\color[2][]{}%
  }%
  \providecommand\includegraphics[2][]{%
    \GenericError{(gnuplot) \space\space\space\@spaces}{%
      Package graphicx or graphics not loaded%
    }{See the gnuplot documentation for explanation.%
    }{The gnuplot epslatex terminal needs graphicx.sty or graphics.sty.}%
    \renewcommand\includegraphics[2][]{}%
  }%
  \providecommand\rotatebox[2]{#2}%
  \@ifundefined{ifGPcolor}{%
    \newif\ifGPcolor
    \GPcolorfalse
  }{}%
  \@ifundefined{ifGPblacktext}{%
    \newif\ifGPblacktext
    \GPblacktexttrue
  }{}%
  \let\gplgaddtomacro\g@addto@macro
  \gdef\gplbacktext{}%
  \gdef\gplfronttext{}%
  \makeatother
  \ifGPblacktext
    \def\colorrgb#1{}%
    \def\colorgray#1{}%
  \else
    \ifGPcolor
      \def\colorrgb#1{\color[rgb]{#1}}%
      \def\colorgray#1{\color[gray]{#1}}%
      \expandafter\def\csname LTw\endcsname{\color{white}}%
      \expandafter\def\csname LTb\endcsname{\color{black}}%
      \expandafter\def\csname LTa\endcsname{\color{black}}%
      \expandafter\def\csname LT0\endcsname{\color[rgb]{1,0,0}}%
      \expandafter\def\csname LT1\endcsname{\color[rgb]{0,1,0}}%
      \expandafter\def\csname LT2\endcsname{\color[rgb]{0,0,1}}%
      \expandafter\def\csname LT3\endcsname{\color[rgb]{1,0,1}}%
      \expandafter\def\csname LT4\endcsname{\color[rgb]{0,1,1}}%
      \expandafter\def\csname LT5\endcsname{\color[rgb]{1,1,0}}%
      \expandafter\def\csname LT6\endcsname{\color[rgb]{0,0,0}}%
      \expandafter\def\csname LT7\endcsname{\color[rgb]{1,0.3,0}}%
      \expandafter\def\csname LT8\endcsname{\color[rgb]{0.5,0.5,0.5}}%
    \else
      \def\colorrgb#1{\color{black}}%
      \def\colorgray#1{\color[gray]{#1}}%
      \expandafter\def\csname LTw\endcsname{\color{white}}%
      \expandafter\def\csname LTb\endcsname{\color{black}}%
      \expandafter\def\csname LTa\endcsname{\color{black}}%
      \expandafter\def\csname LT0\endcsname{\color{black}}%
      \expandafter\def\csname LT1\endcsname{\color{black}}%
      \expandafter\def\csname LT2\endcsname{\color{black}}%
      \expandafter\def\csname LT3\endcsname{\color{black}}%
      \expandafter\def\csname LT4\endcsname{\color{black}}%
      \expandafter\def\csname LT5\endcsname{\color{black}}%
      \expandafter\def\csname LT6\endcsname{\color{black}}%
      \expandafter\def\csname LT7\endcsname{\color{black}}%
      \expandafter\def\csname LT8\endcsname{\color{black}}%
    \fi
  \fi
    \setlength{\unitlength}{0.0500bp}%
    \ifx\gptboxheight\undefined%
      \newlength{\gptboxheight}%
      \newlength{\gptboxwidth}%
      \newsavebox{\gptboxtext}%
    \fi%
    \setlength{\fboxrule}{0.5pt}%
    \setlength{\fboxsep}{1pt}%
    \definecolor{tbcol}{rgb}{1,1,1}%
\begin{picture}(5400.00,3528.00)%
    \gplgaddtomacro\gplbacktext{%
      \csname LTb\endcsname
      \put(594,698){\makebox(0,0)[r]{\strut{}$0$}}%
      \put(594,1394){\makebox(0,0)[r]{\strut{}$0.2$}}%
      \put(594,2090){\makebox(0,0)[r]{\strut{}$0.4$}}%
      \put(594,2785){\makebox(0,0)[r]{\strut{}$0.6$}}%
      \put(660,429){\makebox(0,0){\strut{}$0.25$}}%
      \put(1746,429){\makebox(0,0){\strut{}$0.5$}}%
      \put(2832,429){\makebox(0,0){\strut{}$1$}}%
      \put(3917,429){\makebox(0,0){\strut{}$2$}}%
      \put(5003,429){\makebox(0,0){\strut{}$4$}}%
    }%
    \gplgaddtomacro\gplfronttext{%
      \csname LTb\endcsname
      \put(220,1950){\rotatebox{-270}{\makebox(0,0){\strut{}$c_{T8,T8}^\mathrm{HV}(t,\mu_0)$}}}%
      \put(2831,154){\makebox(0,0){\strut{}$t/t_0$}}%
    }%
    \gplbacktext
    \put(0,0){\includegraphics[width={270.00bp},height={176.40bp}]{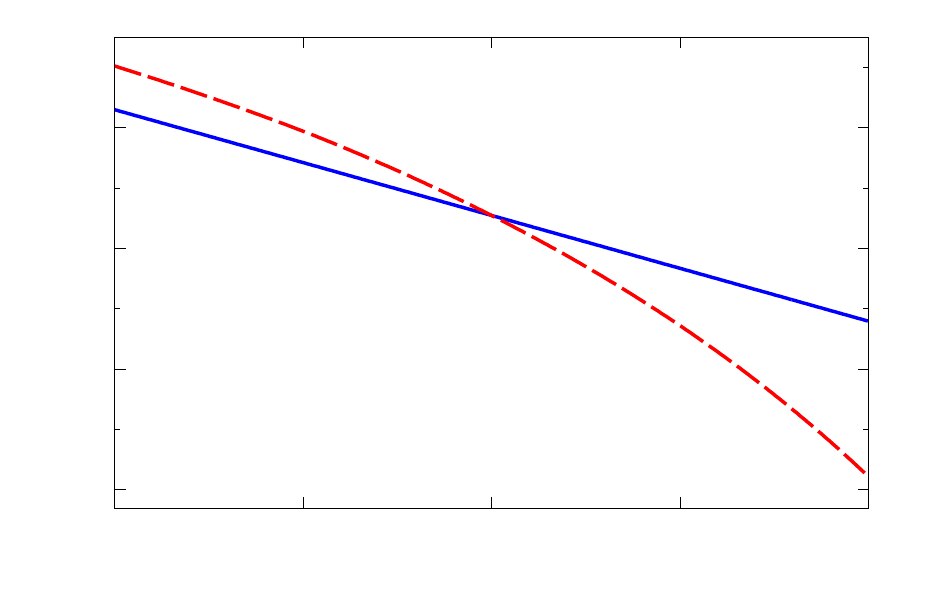}}%
    \gplfronttext
  \end{picture}%
\endgroup

%% file: images/label.tex
\begingroup
  \makeatletter
  \providecommand\color[2][]{%
    \GenericError{(gnuplot) \space\space\space\@spaces}{%
      Package color not loaded in conjunction with
      terminal option `colourtext'%
    }{See the gnuplot documentation for explanation.%
    }{Either use 'blacktext' in gnuplot or load the package
      color.sty in LaTeX.}%
    \renewcommand\color[2][]{}%
  }%
  \providecommand\includegraphics[2][]{%
    \GenericError{(gnuplot) \space\space\space\@spaces}{%
      Package graphicx or graphics not loaded%
    }{See the gnuplot documentation for explanation.%
    }{The gnuplot epslatex terminal needs graphicx.sty or graphics.sty.}%
    \renewcommand\includegraphics[2][]{}%
  }%
  \providecommand\rotatebox[2]{#2}%
  \@ifundefined{ifGPcolor}{%
    \newif\ifGPcolor
    \GPcolorfalse
  }{}%
  \@ifundefined{ifGPblacktext}{%
    \newif\ifGPblacktext
    \GPblacktexttrue
  }{}%
  \let\gplgaddtomacro\g@addto@macro
  \gdef\gplbacktext{}%
  \gdef\gplfronttext{}%
  \makeatother
  \ifGPblacktext
    \def\colorrgb#1{}%
    \def\colorgray#1{}%
  \else
    \ifGPcolor
      \def\colorrgb#1{\color[rgb]{#1}}%
      \def\colorgray#1{\color[gray]{#1}}%
      \expandafter\def\csname LTw\endcsname{\color{white}}%
      \expandafter\def\csname LTb\endcsname{\color{black}}%
      \expandafter\def\csname LTa\endcsname{\color{black}}%
      \expandafter\def\csname LT0\endcsname{\color[rgb]{1,0,0}}%
      \expandafter\def\csname LT1\endcsname{\color[rgb]{0,1,0}}%
      \expandafter\def\csname LT2\endcsname{\color[rgb]{0,0,1}}%
      \expandafter\def\csname LT3\endcsname{\color[rgb]{1,0,1}}%
      \expandafter\def\csname LT4\endcsname{\color[rgb]{0,1,1}}%
      \expandafter\def\csname LT5\endcsname{\color[rgb]{1,1,0}}%
      \expandafter\def\csname LT6\endcsname{\color[rgb]{0,0,0}}%
      \expandafter\def\csname LT7\endcsname{\color[rgb]{1,0.3,0}}%
      \expandafter\def\csname LT8\endcsname{\color[rgb]{0.5,0.5,0.5}}%
    \else
      \def\colorrgb#1{\color{black}}%
      \def\colorgray#1{\color[gray]{#1}}%
      \expandafter\def\csname LTw\endcsname{\color{white}}%
      \expandafter\def\csname LTb\endcsname{\color{black}}%
      \expandafter\def\csname LTa\endcsname{\color{black}}%
      \expandafter\def\csname LT0\endcsname{\color{black}}%
      \expandafter\def\csname LT1\endcsname{\color{black}}%
      \expandafter\def\csname LT2\endcsname{\color{black}}%
      \expandafter\def\csname LT3\endcsname{\color{black}}%
      \expandafter\def\csname LT4\endcsname{\color{black}}%
      \expandafter\def\csname LT5\endcsname{\color{black}}%
      \expandafter\def\csname LT6\endcsname{\color{black}}%
      \expandafter\def\csname LT7\endcsname{\color{black}}%
      \expandafter\def\csname LT8\endcsname{\color{black}}%
    \fi
  \fi
    \setlength{\unitlength}{0.0500bp}%
    \ifx\gptboxheight\undefined%
      \newlength{\gptboxheight}%
      \newlength{\gptboxwidth}%
      \newsavebox{\gptboxtext}%
    \fi%
    \setlength{\fboxrule}{0.5pt}%
    \setlength{\fboxsep}{1pt}%
    \definecolor{tbcol}{rgb}{1,1,1}%
\begin{picture}(7200.00,907.20)%
    \gplgaddtomacro\gplbacktext{%
    }%
    \gplgaddtomacro\gplfronttext{%
      \csname LTb\endcsname
      \put(2740,471){\makebox(0,0)[r]{\strut{}$\alpha_s(\bar\mu_0)|_\text{2-loop $\beta$}$}}%
      \csname LTb\endcsname
      \put(5311,471){\makebox(0,0)[r]{\strut{}$\alpha_s(\bar\mu(t))|_\text{2-loop $\beta$}$}}%
    }%
    \gplbacktext
    \put(0,0){\includegraphics[width={360.00bp},height={45.36bp}]{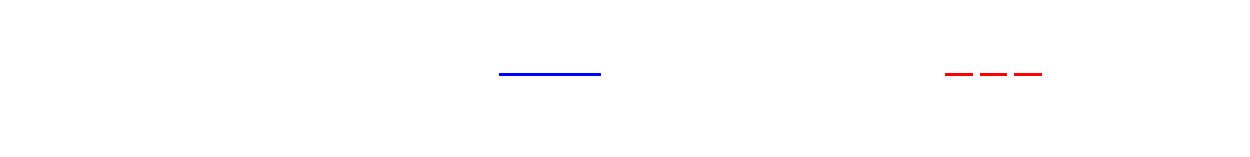}}%
    \gplfronttext
  \end{picture}%
\endgroup

%% file: images/matchingCoeffS1NDR.tex
\begingroup
  \makeatletter
  \providecommand\color[2][]{%
    \GenericError{(gnuplot) \space\space\space\@spaces}{%
      Package color not loaded in conjunction with
      terminal option `colourtext'%
    }{See the gnuplot documentation for explanation.%
    }{Either use 'blacktext' in gnuplot or load the package
      color.sty in LaTeX.}%
    \renewcommand\color[2][]{}%
  }%
  \providecommand\includegraphics[2][]{%
    \GenericError{(gnuplot) \space\space\space\@spaces}{%
      Package graphicx or graphics not loaded%
    }{See the gnuplot documentation for explanation.%
    }{The gnuplot epslatex terminal needs graphicx.sty or graphics.sty.}%
    \renewcommand\includegraphics[2][]{}%
  }%
  \providecommand\rotatebox[2]{#2}%
  \@ifundefined{ifGPcolor}{%
    \newif\ifGPcolor
    \GPcolorfalse
  }{}%
  \@ifundefined{ifGPblacktext}{%
    \newif\ifGPblacktext
    \GPblacktexttrue
  }{}%
  \let\gplgaddtomacro\g@addto@macro
  \gdef\gplbacktext{}%
  \gdef\gplfronttext{}%
  \makeatother
  \ifGPblacktext
    \def\colorrgb#1{}%
    \def\colorgray#1{}%
  \else
    \ifGPcolor
      \def\colorrgb#1{\color[rgb]{#1}}%
      \def\colorgray#1{\color[gray]{#1}}%
      \expandafter\def\csname LTw\endcsname{\color{white}}%
      \expandafter\def\csname LTb\endcsname{\color{black}}%
      \expandafter\def\csname LTa\endcsname{\color{black}}%
      \expandafter\def\csname LT0\endcsname{\color[rgb]{1,0,0}}%
      \expandafter\def\csname LT1\endcsname{\color[rgb]{0,1,0}}%
      \expandafter\def\csname LT2\endcsname{\color[rgb]{0,0,1}}%
      \expandafter\def\csname LT3\endcsname{\color[rgb]{1,0,1}}%
      \expandafter\def\csname LT4\endcsname{\color[rgb]{0,1,1}}%
      \expandafter\def\csname LT5\endcsname{\color[rgb]{1,1,0}}%
      \expandafter\def\csname LT6\endcsname{\color[rgb]{0,0,0}}%
      \expandafter\def\csname LT7\endcsname{\color[rgb]{1,0.3,0}}%
      \expandafter\def\csname LT8\endcsname{\color[rgb]{0.5,0.5,0.5}}%
    \else
      \def\colorrgb#1{\color{black}}%
      \def\colorgray#1{\color[gray]{#1}}%
      \expandafter\def\csname LTw\endcsname{\color{white}}%
      \expandafter\def\csname LTb\endcsname{\color{black}}%
      \expandafter\def\csname LTa\endcsname{\color{black}}%
      \expandafter\def\csname LT0\endcsname{\color{black}}%
      \expandafter\def\csname LT1\endcsname{\color{black}}%
      \expandafter\def\csname LT2\endcsname{\color{black}}%
      \expandafter\def\csname LT3\endcsname{\color{black}}%
      \expandafter\def\csname LT4\endcsname{\color{black}}%
      \expandafter\def\csname LT5\endcsname{\color{black}}%
      \expandafter\def\csname LT6\endcsname{\color{black}}%
      \expandafter\def\csname LT7\endcsname{\color{black}}%
      \expandafter\def\csname LT8\endcsname{\color{black}}%
    \fi
  \fi
    \setlength{\unitlength}{0.0500bp}%
    \ifx\gptboxheight\undefined%
      \newlength{\gptboxheight}%
      \newlength{\gptboxwidth}%
      \newsavebox{\gptboxtext}%
    \fi%
    \setlength{\fboxrule}{0.5pt}%
    \setlength{\fboxsep}{1pt}%
    \definecolor{tbcol}{rgb}{1,1,1}%
\begin{picture}(5400.00,3528.00)%
    \gplgaddtomacro\gplbacktext{%
      \csname LTb\endcsname
      \put(594,1073){\makebox(0,0)[r]{\strut{}$0.2$}}%
      \put(594,1711){\makebox(0,0)[r]{\strut{}$0.4$}}%
      \put(594,2349){\makebox(0,0)[r]{\strut{}$0.6$}}%
      \put(594,2988){\makebox(0,0)[r]{\strut{}$0.8$}}%
      \put(660,429){\makebox(0,0){\strut{}$0.25$}}%
      \put(1746,429){\makebox(0,0){\strut{}$0.5$}}%
      \put(2832,429){\makebox(0,0){\strut{}$1$}}%
      \put(3917,429){\makebox(0,0){\strut{}$2$}}%
      \put(5003,429){\makebox(0,0){\strut{}$4$}}%
    }%
    \gplgaddtomacro\gplfronttext{%
      \csname LTb\endcsname
      \put(220,1950){\rotatebox{-270}{\makebox(0,0){\strut{}$c_{S1,S1}^\mathrm{NDR}(t,\mu_0)$}}}%
      \put(2831,154){\makebox(0,0){\strut{}$t/t_0$}}%
    }%
    \gplbacktext
    \put(0,0){\includegraphics[width={270.00bp},height={176.40bp}]{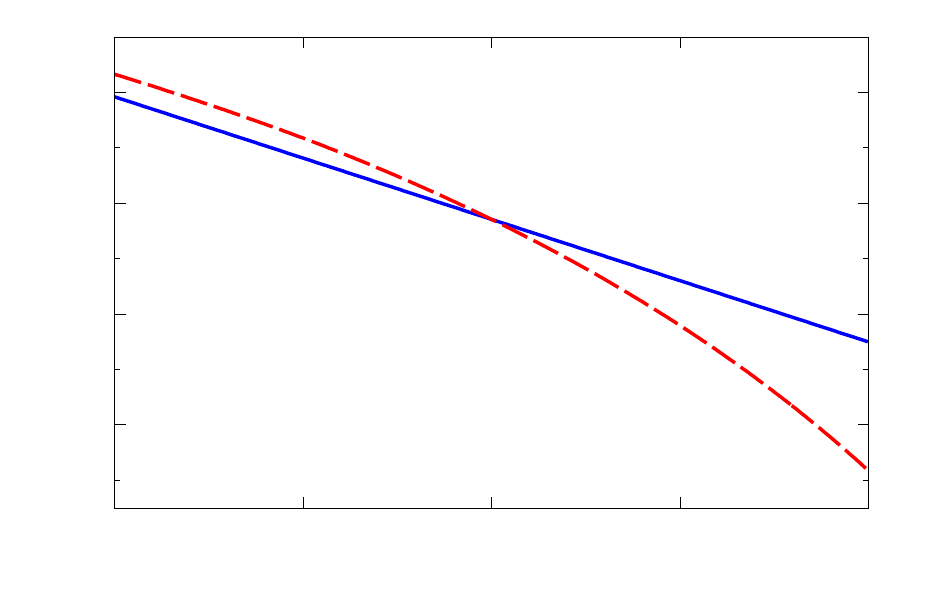}}%
    \gplfronttext
  \end{picture}%
\endgroup

%% file: images/matchingCoeffS8NDR.tex
\begingroup
  \makeatletter
  \providecommand\color[2][]{%
    \GenericError{(gnuplot) \space\space\space\@spaces}{%
      Package color not loaded in conjunction with
      terminal option `colourtext'%
    }{See the gnuplot documentation for explanation.%
    }{Either use 'blacktext' in gnuplot or load the package
      color.sty in LaTeX.}%
    \renewcommand\color[2][]{}%
  }%
  \providecommand\includegraphics[2][]{%
    \GenericError{(gnuplot) \space\space\space\@spaces}{%
      Package graphicx or graphics not loaded%
    }{See the gnuplot documentation for explanation.%
    }{The gnuplot epslatex terminal needs graphicx.sty or graphics.sty.}%
    \renewcommand\includegraphics[2][]{}%
  }%
  \providecommand\rotatebox[2]{#2}%
  \@ifundefined{ifGPcolor}{%
    \newif\ifGPcolor
    \GPcolorfalse
  }{}%
  \@ifundefined{ifGPblacktext}{%
    \newif\ifGPblacktext
    \GPblacktexttrue
  }{}%
  \let\gplgaddtomacro\g@addto@macro
  \gdef\gplbacktext{}%
  \gdef\gplfronttext{}%
  \makeatother
  \ifGPblacktext
    \def\colorrgb#1{}%
    \def\colorgray#1{}%
  \else
    \ifGPcolor
      \def\colorrgb#1{\color[rgb]{#1}}%
      \def\colorgray#1{\color[gray]{#1}}%
      \expandafter\def\csname LTw\endcsname{\color{white}}%
      \expandafter\def\csname LTb\endcsname{\color{black}}%
      \expandafter\def\csname LTa\endcsname{\color{black}}%
      \expandafter\def\csname LT0\endcsname{\color[rgb]{1,0,0}}%
      \expandafter\def\csname LT1\endcsname{\color[rgb]{0,1,0}}%
      \expandafter\def\csname LT2\endcsname{\color[rgb]{0,0,1}}%
      \expandafter\def\csname LT3\endcsname{\color[rgb]{1,0,1}}%
      \expandafter\def\csname LT4\endcsname{\color[rgb]{0,1,1}}%
      \expandafter\def\csname LT5\endcsname{\color[rgb]{1,1,0}}%
      \expandafter\def\csname LT6\endcsname{\color[rgb]{0,0,0}}%
      \expandafter\def\csname LT7\endcsname{\color[rgb]{1,0.3,0}}%
      \expandafter\def\csname LT8\endcsname{\color[rgb]{0.5,0.5,0.5}}%
    \else
      \def\colorrgb#1{\color{black}}%
      \def\colorgray#1{\color[gray]{#1}}%
      \expandafter\def\csname LTw\endcsname{\color{white}}%
      \expandafter\def\csname LTb\endcsname{\color{black}}%
      \expandafter\def\csname LTa\endcsname{\color{black}}%
      \expandafter\def\csname LT0\endcsname{\color{black}}%
      \expandafter\def\csname LT1\endcsname{\color{black}}%
      \expandafter\def\csname LT2\endcsname{\color{black}}%
      \expandafter\def\csname LT3\endcsname{\color{black}}%
      \expandafter\def\csname LT4\endcsname{\color{black}}%
      \expandafter\def\csname LT5\endcsname{\color{black}}%
      \expandafter\def\csname LT6\endcsname{\color{black}}%
      \expandafter\def\csname LT7\endcsname{\color{black}}%
      \expandafter\def\csname LT8\endcsname{\color{black}}%
    \fi
  \fi
    \setlength{\unitlength}{0.0500bp}%
    \ifx\gptboxheight\undefined%
      \newlength{\gptboxheight}%
      \newlength{\gptboxwidth}%
      \newsavebox{\gptboxtext}%
    \fi%
    \setlength{\fboxrule}{0.5pt}%
    \setlength{\fboxsep}{1pt}%
    \definecolor{tbcol}{rgb}{1,1,1}%
\begin{picture}(5400.00,3528.00)%
    \gplgaddtomacro\gplbacktext{%
      \csname LTb\endcsname
      \put(594,1272){\makebox(0,0)[r]{\strut{}$0.6$}}%
      \put(594,2120){\makebox(0,0)[r]{\strut{}$0.65$}}%
      \put(594,2968){\makebox(0,0)[r]{\strut{}$0.7$}}%
      \put(660,429){\makebox(0,0){\strut{}$0.25$}}%
      \put(1746,429){\makebox(0,0){\strut{}$0.5$}}%
      \put(2832,429){\makebox(0,0){\strut{}$1$}}%
      \put(3917,429){\makebox(0,0){\strut{}$2$}}%
      \put(5003,429){\makebox(0,0){\strut{}$4$}}%
    }%
    \gplgaddtomacro\gplfronttext{%
      \csname LTb\endcsname
      \put(88,1950){\rotatebox{-270}{\makebox(0,0){\strut{}$c_{S8,S8}^\mathrm{NDR}(t,\mu_0)$}}}%
      \put(2831,154){\makebox(0,0){\strut{}$t/t_0$}}%
    }%
    \gplbacktext
    \put(0,0){\includegraphics[width={270.00bp},height={176.40bp}]{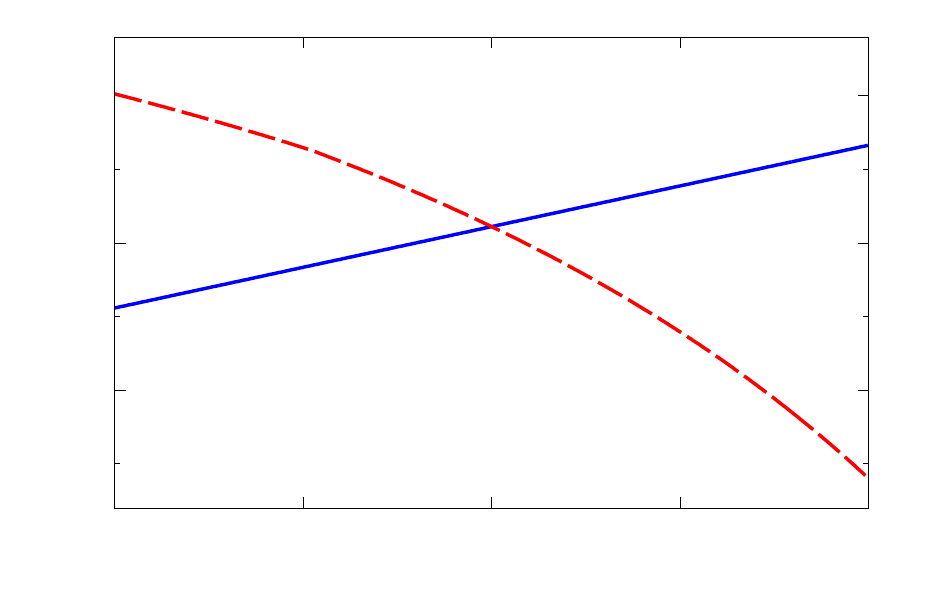}}%
    \gplfronttext
  \end{picture}%
\endgroup

%% file: images/matchingCoeffT1NDR.tex
\begingroup
  \makeatletter
  \providecommand\color[2][]{%
    \GenericError{(gnuplot) \space\space\space\@spaces}{%
      Package color not loaded in conjunction with
      terminal option `colourtext'%
    }{See the gnuplot documentation for explanation.%
    }{Either use 'blacktext' in gnuplot or load the package
      color.sty in LaTeX.}%
    \renewcommand\color[2][]{}%
  }%
  \providecommand\includegraphics[2][]{%
    \GenericError{(gnuplot) \space\space\space\@spaces}{%
      Package graphicx or graphics not loaded%
    }{See the gnuplot documentation for explanation.%
    }{The gnuplot epslatex terminal needs graphicx.sty or graphics.sty.}%
    \renewcommand\includegraphics[2][]{}%
  }%
  \providecommand\rotatebox[2]{#2}%
  \@ifundefined{ifGPcolor}{%
    \newif\ifGPcolor
    \GPcolorfalse
  }{}%
  \@ifundefined{ifGPblacktext}{%
    \newif\ifGPblacktext
    \GPblacktexttrue
  }{}%
  \let\gplgaddtomacro\g@addto@macro
  \gdef\gplbacktext{}%
  \gdef\gplfronttext{}%
  \makeatother
  \ifGPblacktext
    \def\colorrgb#1{}%
    \def\colorgray#1{}%
  \else
    \ifGPcolor
      \def\colorrgb#1{\color[rgb]{#1}}%
      \def\colorgray#1{\color[gray]{#1}}%
      \expandafter\def\csname LTw\endcsname{\color{white}}%
      \expandafter\def\csname LTb\endcsname{\color{black}}%
      \expandafter\def\csname LTa\endcsname{\color{black}}%
      \expandafter\def\csname LT0\endcsname{\color[rgb]{1,0,0}}%
      \expandafter\def\csname LT1\endcsname{\color[rgb]{0,1,0}}%
      \expandafter\def\csname LT2\endcsname{\color[rgb]{0,0,1}}%
      \expandafter\def\csname LT3\endcsname{\color[rgb]{1,0,1}}%
      \expandafter\def\csname LT4\endcsname{\color[rgb]{0,1,1}}%
      \expandafter\def\csname LT5\endcsname{\color[rgb]{1,1,0}}%
      \expandafter\def\csname LT6\endcsname{\color[rgb]{0,0,0}}%
      \expandafter\def\csname LT7\endcsname{\color[rgb]{1,0.3,0}}%
      \expandafter\def\csname LT8\endcsname{\color[rgb]{0.5,0.5,0.5}}%
    \else
      \def\colorrgb#1{\color{black}}%
      \def\colorgray#1{\color[gray]{#1}}%
      \expandafter\def\csname LTw\endcsname{\color{white}}%
      \expandafter\def\csname LTb\endcsname{\color{black}}%
      \expandafter\def\csname LTa\endcsname{\color{black}}%
      \expandafter\def\csname LT0\endcsname{\color{black}}%
      \expandafter\def\csname LT1\endcsname{\color{black}}%
      \expandafter\def\csname LT2\endcsname{\color{black}}%
      \expandafter\def\csname LT3\endcsname{\color{black}}%
      \expandafter\def\csname LT4\endcsname{\color{black}}%
      \expandafter\def\csname LT5\endcsname{\color{black}}%
      \expandafter\def\csname LT6\endcsname{\color{black}}%
      \expandafter\def\csname LT7\endcsname{\color{black}}%
      \expandafter\def\csname LT8\endcsname{\color{black}}%
    \fi
  \fi
    \setlength{\unitlength}{0.0500bp}%
    \ifx\gptboxheight\undefined%
      \newlength{\gptboxheight}%
      \newlength{\gptboxwidth}%
      \newsavebox{\gptboxtext}%
    \fi%
    \setlength{\fboxrule}{0.5pt}%
    \setlength{\fboxsep}{1pt}%
    \definecolor{tbcol}{rgb}{1,1,1}%
\begin{picture}(5400.00,3528.00)%
    \gplgaddtomacro\gplbacktext{%
      \csname LTb\endcsname
      \put(594,865){\makebox(0,0)[r]{\strut{}$0.6$}}%
      \put(594,1544){\makebox(0,0)[r]{\strut{}$0.65$}}%
      \put(594,2222){\makebox(0,0)[r]{\strut{}$0.7$}}%
      \put(594,2900){\makebox(0,0)[r]{\strut{}$0.75$}}%
      \put(660,429){\makebox(0,0){\strut{}$0.25$}}%
      \put(1746,429){\makebox(0,0){\strut{}$0.5$}}%
      \put(2832,429){\makebox(0,0){\strut{}$1$}}%
      \put(3917,429){\makebox(0,0){\strut{}$2$}}%
      \put(5003,429){\makebox(0,0){\strut{}$4$}}%
    }%
    \gplgaddtomacro\gplfronttext{%
      \csname LTb\endcsname
      \put(88,1950){\rotatebox{-270}{\makebox(0,0){\strut{}$c_{T1,T1}^\mathrm{NDR}(t,\mu_0)$}}}%
      \put(2831,154){\makebox(0,0){\strut{}$t/t_0$}}%
    }%
    \gplbacktext
    \put(0,0){\includegraphics[width={270.00bp},height={176.40bp}]{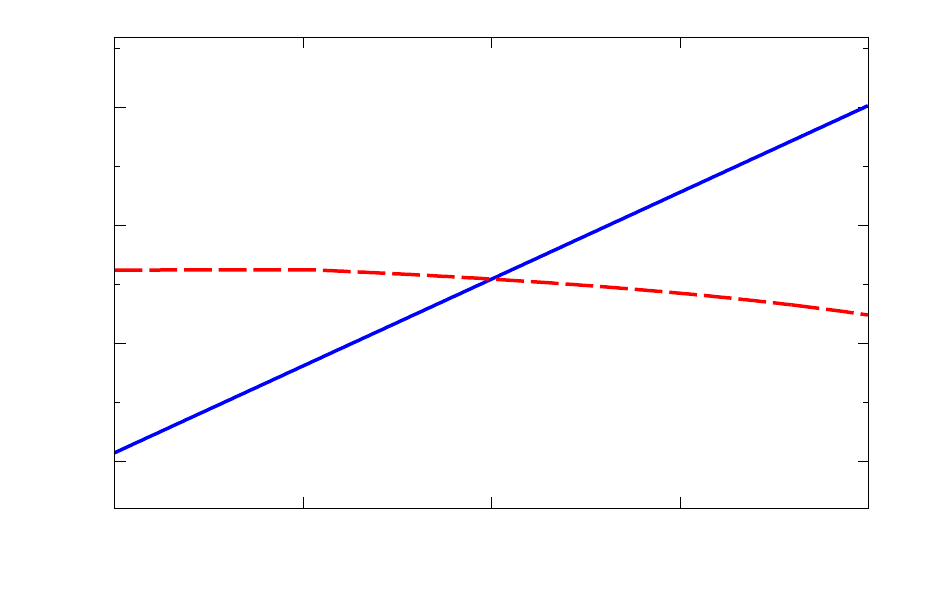}}%
    \gplfronttext
  \end{picture}%
\endgroup

%% file: images/matchingCoeffT8NDR.tex
\begingroup
  \makeatletter
  \providecommand\color[2][]{%
    \GenericError{(gnuplot) \space\space\space\@spaces}{%
      Package color not loaded in conjunction with
      terminal option `colourtext'%
    }{See the gnuplot documentation for explanation.%
    }{Either use 'blacktext' in gnuplot or load the package
      color.sty in LaTeX.}%
    \renewcommand\color[2][]{}%
  }%
  \providecommand\includegraphics[2][]{%
    \GenericError{(gnuplot) \space\space\space\@spaces}{%
      Package graphicx or graphics not loaded%
    }{See the gnuplot documentation for explanation.%
    }{The gnuplot epslatex terminal needs graphicx.sty or graphics.sty.}%
    \renewcommand\includegraphics[2][]{}%
  }%
  \providecommand\rotatebox[2]{#2}%
  \@ifundefined{ifGPcolor}{%
    \newif\ifGPcolor
    \GPcolorfalse
  }{}%
  \@ifundefined{ifGPblacktext}{%
    \newif\ifGPblacktext
    \GPblacktexttrue
  }{}%
  \let\gplgaddtomacro\g@addto@macro
  \gdef\gplbacktext{}%
  \gdef\gplfronttext{}%
  \makeatother
  \ifGPblacktext
    \def\colorrgb#1{}%
    \def\colorgray#1{}%
  \else
    \ifGPcolor
      \def\colorrgb#1{\color[rgb]{#1}}%
      \def\colorgray#1{\color[gray]{#1}}%
      \expandafter\def\csname LTw\endcsname{\color{white}}%
      \expandafter\def\csname LTb\endcsname{\color{black}}%
      \expandafter\def\csname LTa\endcsname{\color{black}}%
      \expandafter\def\csname LT0\endcsname{\color[rgb]{1,0,0}}%
      \expandafter\def\csname LT1\endcsname{\color[rgb]{0,1,0}}%
      \expandafter\def\csname LT2\endcsname{\color[rgb]{0,0,1}}%
      \expandafter\def\csname LT3\endcsname{\color[rgb]{1,0,1}}%
      \expandafter\def\csname LT4\endcsname{\color[rgb]{0,1,1}}%
      \expandafter\def\csname LT5\endcsname{\color[rgb]{1,1,0}}%
      \expandafter\def\csname LT6\endcsname{\color[rgb]{0,0,0}}%
      \expandafter\def\csname LT7\endcsname{\color[rgb]{1,0.3,0}}%
      \expandafter\def\csname LT8\endcsname{\color[rgb]{0.5,0.5,0.5}}%
    \else
      \def\colorrgb#1{\color{black}}%
      \def\colorgray#1{\color[gray]{#1}}%
      \expandafter\def\csname LTw\endcsname{\color{white}}%
      \expandafter\def\csname LTb\endcsname{\color{black}}%
      \expandafter\def\csname LTa\endcsname{\color{black}}%
      \expandafter\def\csname LT0\endcsname{\color{black}}%
      \expandafter\def\csname LT1\endcsname{\color{black}}%
      \expandafter\def\csname LT2\endcsname{\color{black}}%
      \expandafter\def\csname LT3\endcsname{\color{black}}%
      \expandafter\def\csname LT4\endcsname{\color{black}}%
      \expandafter\def\csname LT5\endcsname{\color{black}}%
      \expandafter\def\csname LT6\endcsname{\color{black}}%
      \expandafter\def\csname LT7\endcsname{\color{black}}%
      \expandafter\def\csname LT8\endcsname{\color{black}}%
    \fi
  \fi
    \setlength{\unitlength}{0.0500bp}%
    \ifx\gptboxheight\undefined%
      \newlength{\gptboxheight}%
      \newlength{\gptboxwidth}%
      \newsavebox{\gptboxtext}%
    \fi%
    \setlength{\fboxrule}{0.5pt}%
    \setlength{\fboxsep}{1pt}%
    \definecolor{tbcol}{rgb}{1,1,1}%
\begin{picture}(5400.00,3528.00)%
    \gplgaddtomacro\gplbacktext{%
      \csname LTb\endcsname
      \put(594,788){\makebox(0,0)[r]{\strut{}$0.2$}}%
      \put(594,1563){\makebox(0,0)[r]{\strut{}$0.4$}}%
      \put(594,2338){\makebox(0,0)[r]{\strut{}$0.6$}}%
      \put(594,3113){\makebox(0,0)[r]{\strut{}$0.8$}}%
      \put(660,429){\makebox(0,0){\strut{}$0.25$}}%
      \put(1746,429){\makebox(0,0){\strut{}$0.5$}}%
      \put(2832,429){\makebox(0,0){\strut{}$1$}}%
      \put(3917,429){\makebox(0,0){\strut{}$2$}}%
      \put(5003,429){\makebox(0,0){\strut{}$4$}}%
    }%
    \gplgaddtomacro\gplfronttext{%
      \csname LTb\endcsname
      \put(220,1950){\rotatebox{-270}{\makebox(0,0){\strut{}$c_{T8,T8}^\mathrm{NDR}(t,\mu_0)$}}}%
      \put(2831,154){\makebox(0,0){\strut{}$t/t_0$}}%
    }%
    \gplbacktext
    \put(0,0){\includegraphics[width={270.00bp},height={176.40bp}]{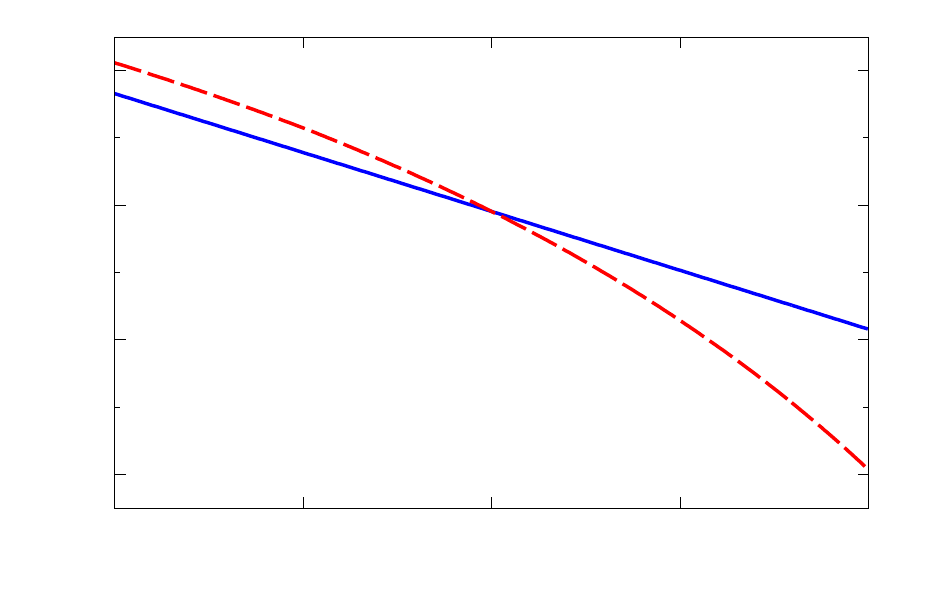}}%
    \gplfronttext
  \end{picture}%
\endgroup

%% file: sections/Conclusions.tex

\section{Conclusions}
\label{sec:Conclusions}

In the present paper, we have calculated the one-loop matching for flavor-neutral $CP$-odd four-quark operators between the gradient-flow scheme and the more familiar MS scheme used in EFT analyses. The matching coefficients are obtained by inserting flowed four-quark operators into two-, three-, and four-point Green's functions and applying the method of regions to extract the coefficients of the MS operators. We provide the coefficients of four-quark operators both in the NDR and HV schemes. For the calculation of the coefficients of lower-dimension operators, we have only used the HV scheme in order to avoid problematic $\gamma_5$-odd traces in NDR. The matching coefficients to lower-dimension operators are provided in a way that allows one to reconstruct the case of generic quark flavors.

For the matching to the dimension-five EDM operator, we have employed modified quark-flow equations that involve the static external electromagnetic field and manifestly respect $U(1)_\mathrm{em}$ invariance. Working instead with pure QCD flow equations would result in matching contributions to unphysical dimension-six operators that are not $U(1)_\mathrm{em}$ gauge invariant.

The gradient flow is a promising scheme for the treatment of $CP$-odd operators that contribute to the neutron EDM, providing a regularization-independent definition of renormalized operators. In lattice-QCD implementations, the gradient flow disentangles the power-divergent mixing with lower-dimension operators from the continuum limit, which can be taken for any fixed non-vanishing flow time. Our results extend previous work on the gradient-flow matching of dimension-five operators~\cite{Mereghetti:2021nkt} and they provide a necessary ingredient for future lattice-QCD computations of the contribution of four-quark operators to the neutron EDM. With a forthcoming study of the $CP$-odd three-gluon operator~\cite{CP3GOMatching}, the gradient-flow matching for the complete set of operators up to dimension six that contribute to the neutron EDM will be available at one loop. As our studies show, in some cases the desired accuracy goal motivates the calculation of the matching at two loops~\cite{Harlander:2022tgk,Harlander:2022vgf}. 

%% file: sections/Conventions.tex

\section{Conventions}
\label{sec:Conventions}

We adopt the same conventions as Ref.~\cite{Mereghetti:2021nkt}, which are briefly summarized in the following.

\subsection[$SU(3)$]{\boldmath $SU(3)$}

The anti-Hermitian $SU(3)_c$ generators $t^a$ are defined in terms of the Gell-Mann matrices $\lambda_a$ as
\begin{align}
	t^a = -i \frac{\lambda_a}{2} \, .
\end{align}
The generators, $SU(N_c)$ structure constants $f^{abc}$, and the totally symmetric tensor $d^{abc}$ fulfill
\begin{align}
	\left[t^a, t^b \right] &= f^{abc}t^c,\quad \left\lbrace t^a, t^b \right\rbrace = -\frac{1}{N_c}\delta^{ab} - i d^{abc}t^c, \quad \mathrm{Tr}\left[t^a, t^b \right] = -\frac{1}{2}\delta^{ab}, \quad t^a t^a = -C_F \, , \nn
	f^{abc}f^{abd} &= C_A \delta^{cd}, \quad d^{abc}d^{abd} = \frac{N_c^2 - 4}{N_c} \delta^{cd} \, ,
\end{align}
with the quadratic Casimir invariants
\begin{align}
	\label{eq:QuadraticCasimirs}
	C_A = N_c, \quad C_F = \frac{N_c^2 - 1}{2N_c}.
\end{align}
For the reduction of expressions with two and more generators, the following identities are useful:
\begin{align}
	\left( t^a t^b \right)_{\alpha \beta} &= \frac{1}{2} \left[t^a, t^b \right]_{\alpha \beta} + \frac{1}{2} \left\lbrace t^a, t^b \right\rbrace_{\alpha \beta} = \frac{1}{2} \left(f^{abc}- i d^{abc} \right)t^c_{\alpha \beta} - \frac{1}{2 N_c}\delta^{ab}\delta_{\alpha \beta} \, , \nn
	\left( t^b t^a t^b \right)_{\alpha \beta} &= \frac{1}{2N_c} t^a_{\alpha \beta} \, , \nn
	\left( t^c t^a t^b t^c \right)_{\alpha \beta} &= \frac{1}{4} \delta^{ab} \delta_{\alpha \beta} + \frac{1}{2N_c} \left( t^a t^b \right)_{\alpha \beta} \, .
\end{align}

\subsection{Dirac algebra}

Our Hermitian Dirac matrices can be related to the more commonly used Dirac matrices in Minkowski conventions by
\begin{align}
	\gamma_4 &:= \gamma_0^M = \gamma^0_M \, , \quad
	\gamma_k := i \gamma_k^M = - i \gamma^k_M \, , \quad \text{for } k=1,2,3.
\end{align}
In dimensional regularization in $D = 4 - 2 \varepsilon$ Euclidean space-time dimensions, the Dirac algebra is defined by
\begin{equation}
	\{ \gamma_{\mu}, \gamma_{\nu} \} = 2\delta_{\mu \nu} \, .
\end{equation}
Our convention for $\gamma_5$ is
\begin{equation}
\label{gamma5E}
	\gamma_5 = \gamma_1 \gamma_2 \gamma_3 \gamma_4 = \frac{1}{4!} \epsilon_{\mu \nu \lambda \sigma} \gamma_\mu \gamma_\nu \gamma_\lambda \gamma_\sigma
\end{equation}
with the Levi-Civita tensor normalized to $\epsilon_{1 2 3 4}=+1$. The matrix $\gamma_5$ is Hermitian and fulfills $\gamma_5^2 = \mathds{1}$. In the NDR scheme, we use
\begin{align}
	\{ \gamma_5, \gamma_\mu \} = 0
\end{align}
for all Dirac matrices $\gamma_\mu$. As is well known, this prescription leads to problematic $\gamma_5$-odd traces, since together with the cyclicity of the trace it implies $\Tr(\gamma_5 \gamma_\mu \gamma_\nu \gamma_\lambda \gamma_\sigma) = 0$ for $D \neq 4$. In the HV scheme, the Dirac matrices projected with Eq.~\eqref{eq:HVSubspaces} to the four-dimensional and evanescent subspaces are treated differently and fulfill
\begin{align}
	\{ \gamma_5, \bar\gamma_\mu \} = [ \gamma_5, \hat\gamma_\mu ] = 0 \, .
\end{align}
In the HV scheme, the Lorentz indices of the Levi-Civita symbol only run over four space-time dimensions.

%% file: sections/FeynmanRules.tex

\section{Feynman rules}
\label{sec:FeynmanRules}

Our Feynman rules for Euclidean QCD and the perturbative solution of the flow equations largely agree with the ones given in App.~B of Ref.~\cite{Mereghetti:2021nkt}. The only exception are the additional flow vertices listed in App.~\ref{sec:PhotonicVertices}, which involve the external electromagnetic field and emerge from our modified quark flow equation. In App.~\ref{sec:OperatorInsertionRules}, we list the Feynman rules for the operator insertions.

\subsection{Photonic flow vertices}
\label{sec:PhotonicVertices}

We suppress flavor indices, because all flow vertices are flavor conserving. Flow lines are marked by a solid adjacent arrow, whereas a dashed adjacent arrow indicates that the line is either a flow line or a flowed propagator. Note that all momenta $q_i$ are chosen to be outgoing.
\begin{align}
	\nn
	\begin{gathered}
	\begin{fmfgraph*}(60,40)
		\fmfleft{i1,i2}
		\fmfright{o1}
		\fmflabel{$q_2,\beta$}{i1}
		\fmflabel{$\mu, q_1$}{i2}
		\fmflabel{$q_3,\alpha$}{o1}
		\fmf{quark}{i1,v1}
		\fmf{photon}{i2,v1}
		\fmf{quark}{v1,o1}
		\fmfv{d.sh=circle, d.f=empty, label=$t$,l.a=30, l.d=10}{v1}
		\fmffreeze
		\fmf{dlarrowdr}{i1,v1}
		\fmf{marrowd}{v1,o1}
	\end{fmfgraph*}
	\end{gathered}
	\qquad \quad &= \; -i \delta_{\alpha \beta} \int_0^\infty dt \left( {q_1}_\mu + 2 {q_2}_\mu \right) \, , \\[1cm]
	\begin{gathered}
		\begin{fmfgraph*}(60,40)
			\fmfleft{i1,i2}
			\fmfright{o1}
			\fmflabel{$q_2,\beta$}{i1}
			\fmflabel{$\mu, q_1$}{i2}
			\fmflabel{$q_3,\alpha$}{o1}
			\fmf{quark}{i1,v1}
			\fmf{photon}{i2,v1}
			\fmf{quark}{v1,o1}
			\fmfv{d.sh=circle, d.f=empty, label=$t$,l.a=30, l.d=10}{v1}
			\fmffreeze
			\fmf{marrowdr}{v1,i1}
			\fmf{darrowd}{o1,v1}
		\end{fmfgraph*}
	\end{gathered}
	\qquad \quad &= \; i \delta_{\alpha \beta} \int_0^\infty dt \left( {q_1}_\mu + 2{q_3}_\mu \right) \, , \\[1cm]
	\begin{gathered}
		\begin{fmfgraph*}(70,70)
			\fmfleft{i1}
			\fmftop{t1}
			\fmfright{o1}
			\fmfbottom{b1}
			\fmflabel{$q_2,\beta$}{i1}
			\fmflabel{$\nu, q_3$}{b1}
			\fmflabel{$\mu, q_1$}{t1}
			\fmflabel{$q_4,\alpha$}{o1}
			\fmf{quark}{i1,v1}
			\fmf{photon}{t1,v1}
			\fmf{photon}{v1,b1}
			\fmf{quark}{v1,o1}
			\fmfv{d.sh=circle, d.f=empty, label=$t$,l.a=30, l.d=10}{v1}
			\fmffreeze
			\fmf{marrowd}{v1,o1}
			\fmf{darrowd}{i1,v1}
		\end{fmfgraph*}
	\end{gathered}
	\qquad \quad &= \qquad \quad
	\begin{gathered}
		\begin{fmfgraph*}(70,70)
			\fmfleft{i1}
			\fmftop{t1}
			\fmfright{o1}
			\fmfbottom{b1}
			\fmflabel{$q_2,\beta$}{i1}
			\fmflabel{$\nu, q_3$}{b1}
			\fmflabel{$\mu, q_1$}{t1}
			\fmflabel{$q_4,\alpha$}{o1}
			\fmf{quark}{i1,v1}
			\fmf{photon}{t1,v1}
			\fmf{photon}{v1,b1}
			\fmf{quark}{v1,o1}
			\fmfv{d.sh=circle, d.f=empty, label=$t$,l.a=30, l.d=10}{v1}
			\fmffreeze
			\fmf{darrowd}{o1,v1}
			\fmf{marrowd}{v1,i1}
		\end{fmfgraph*}
	\end{gathered}
	\qquad \quad = \; 2 \delta_{\mu \nu} \delta_{\alpha \beta}  \int_0^\infty dt \, , \\[1cm]
	\begin{gathered}
		\begin{fmfgraph*}(70,70)
			\fmfleft{i1}
			\fmftop{t1}
			\fmfright{o1}
			\fmfbottom{b1}
			\fmflabel{$q_2,\beta$}{i1}
			\fmflabel{$a,\nu, q_3$}{b1}
			\fmflabel{$\mu, q_1$}{t1}
			\fmflabel{$q_4,\alpha$}{o1}
			\fmf{quark}{i1,v1}
			\fmf{photon}{t1,v1}
			\fmf{gluon}{v1,b1}
			\fmf{quark}{v1,o1}
			\fmfv{d.sh=circle, d.f=empty, label=$t$,l.a=30, l.d=10}{v1}
			\fmffreeze
			\fmf{marrowd}{v1,o1}
			\fmf{darrowd}{i1,v1}
			\fmf{darrowl}{b1,v1}
		\end{fmfgraph*}
	\end{gathered}
	\qquad \quad &= \qquad \quad
	\begin{gathered}
		\begin{fmfgraph*}(70,70)
			\fmfleft{i1}
			\fmftop{t1}
			\fmfright{o1}
			\fmfbottom{b1}
			\fmflabel{$q_2,\beta$}{i1}
			\fmflabel{$a, \nu, q_3$}{b1}
			\fmflabel{$\mu, q_1$}{t1}
			\fmflabel{$q_4,\alpha$}{o1}
			\fmf{quark}{i1,v1}
			\fmf{photon}{t1,v1}
			\fmf{gluon}{v1,b1}
			\fmf{quark}{v1,o1}
			\fmfv{d.sh=circle, d.f=empty, label=$t$,l.a=30, l.d=10}{v1}
			\fmffreeze
			\fmf{darrowd}{o1,v1}
			\fmf{marrowd}{v1,i1}
			\fmf{darrowl}{b1,v1}
		\end{fmfgraph*}
	\end{gathered}
	\qquad \quad = \; 2 \delta_{\mu \nu} t^a_{\alpha \beta}  \int_0^\infty dt \, . \\\nonumber
\end{align} 

\subsection{Operator insertions}
\label{sec:OperatorInsertionRules}

Here, we list all the required vertex rules for effective operators. In contrast to the convention in Ref.~\cite{Mereghetti:2021nkt}, we regard the operators as part of the Lagrangian~\eqref{eq:ApproximateLEFTLagrangian} and include in the Feynman rules both the Wilson coefficient and the minus sign from the exponential in the generating functional
\begin{align}
	Z_E[J] = \int \! \D G \, \D \bar q \, \D q \, \D \bar c \, \D c \, e^{-S_E[J]} \, ,
\end{align}
where $S_E[J] = \int d^D x \, \L[J]$ denotes the Euclidean action including external sources $J$.
Flavor indices are denoted by $p,r,s,t$, Dirac indices by $i,j,k,l$, and color indices by $\alpha, \beta,\gamma, \delta$.

The Feynman rule for the pseudoscalar density reads
\begin{align}
	\begin{gathered}
		\begin{fmfgraph*}(60,20)
			\fmfleft{i1}
			\fmfright{o1}
			\fmf{quark}{i1,v1}
			\fmf{quark}{v1,o1}
			\fmfv{d.sh=circle, d.f=shaded, decor.size=(4mm)}{v1}
			\fmflabel{$p, i, \alpha$}{o1}
			\fmflabel{$r, j, \beta$}{i1}
		\end{fmfgraph*}
	\end{gathered}
	\qquad \qquad = \; - L^P_{pr} (\gamma_5)_{ij} \delta_{\alpha\beta} \, .
\end{align}
The rules for the quark electric and chromo-electric dipole operators are given by (all momenta $q_i$ are outgoing)
\begin{align}
	\nn
	\begin{gathered}
		\begin{fmfgraph*}(70,60)
			\fmfleft{l1,l2} \fmfright{r1}
			\fmf{quark,tension=2}{l1,v1}
			\fmf{photon,tension=2}{v1,l2}
			\fmf{quark,tension=2}{v1,r1}
			\fmflabel{$q_1,\mu$}{l2}
			\fmflabel{$r,j,\beta$}{l1}
			\fmflabel{$p,i,\alpha$}{r1}
			\fmfv{decor.shape=circle, decor.filled=shaded, decor.size=(4mm)}{v1}
		\end{fmfgraph*}
	\end{gathered}
	\qquad\qquad &= \; - L^E_{pr} \, 2i \delta_{\alpha\beta} (\tilde\sigma_{\mu\nu})_{ij} {q_1}_\nu \, , \\[1cm]
	\begin{gathered}
		\begin{fmfgraph*}(70,60)
			\fmfleft{l1,l2} \fmfright{r1}
			\fmf{quark,tension=2}{l1,v1}
			\fmf{gluon,tension=2}{v1,l2}
			\fmf{quark,tension=2}{v1,r1}
			\fmfv{decor.shape=circle, decor.filled=shaded, decor.size=(4mm)}{v1}
			\fmflabel{$q_1,\mu,a$}{l2}
			\fmflabel{$r,j,\beta$}{l1}
			\fmflabel{$p,i,\alpha$}{r1}
		\end{fmfgraph*}
	\end{gathered}
	\qquad\qquad &= \; - L^{CE}_{pr} \, 2i (t^a)_{\alpha\beta} (\tilde\sigma_{\mu\nu})_{ij} {q_1}_\nu \, , \\[1cm]
	\begin{gathered}
		\begin{fmfgraph*}(80,70)
			\fmfleft{l1} \fmftop{t1} \fmfbottom{b1} \fmfright{r1}
			\fmf{gluon,tension=2}{v1,b1}
			\fmf{quark,tension=2}{l1,v1}
			\fmf{gluon,tension=2}{t1,v1}
			\fmf{quark,tension=2}{v1,r1}
			\fmfv{decor.shape=circle, decor.filled=shaded, decor.size=(4mm)}{v1}
			\fmflabel{$\mu,a$}{t1}
			\fmflabel{$\nu,b$}{b1}
			\fmflabel{$p,i,\alpha$}{r1}
			\fmflabel{$r,j,\beta$}{l1}
		\end{fmfgraph*}
	\end{gathered}
	\qquad\qquad &= \; - L^{CE}_{pr} \, 2 (\tilde\sigma_{\mu\nu})_{ij} f^{abc} (t^c)_{\alpha\beta} \, . \\\nonumber
\end{align}

The Feynman rule for color-octet four-quark operators reads
\begin{align}
	\nn
	\begin{gathered}
		\begin{fmfgraph*}(60,30)
			\fmfleft{i1,i2}
			\fmfright{o1,o2}
			\fmf{quark}{i1,v1,o1}
			\fmf{quark}{i2,v2,o2}
			\fmf{phantom, tension=5}{v1,v2}
			\fmfdot{v1}
			\fmfdot{v2}
			\fmflabel{$p, i, \alpha$}{o1}
			\fmflabel{$r, j, \beta$}{i1}
			\fmflabel{$s, k, \gamma$}{o2}
			\fmflabel{$t, l, \delta$}{i2}
			\fmfv{label=$\Gamma_1 t^a$,label.dist=10}{v1}
			\fmfv{label=$\Gamma_2 t^a$,label.dist=10}{v2}
		\end{fmfgraph*}
	\end{gathered}
	\qquad \quad = \; -L_{prst}(\Gamma_1)_{ij}(\Gamma_2)_{kl} (t^a)_{\alpha\beta} (t^a)_{\gamma\delta} \, , \\\nonumber
\end{align}
where $\Gamma_{1,2}$ are the Dirac structures of the two bilinears. For the color-singlet operators, the $SU(3)_c$ generators are replaced by the identity. We explicitly distinguish the two bilinears and hence in general require four instead of two separate insertions of the four-quark vertex into each topology, which doubles the number of diagrams mentioned in the main text.